\documentclass[11pt,a4paper]{article}
\pdfoutput=1
\usepackage{jheppub}
\usepackage{amsmath}
\def\th{\theta}

\def\cob{\delta}

\def\ka{\kappa}

\DeclareMathOperator{\Tr}{Tr}

\newcommand{\hf}{\frac{1}{2}}
\newcommand{\qu}{\frac{1}{4}}

\def\til#1{\widetilde{#1}}

\def\b#1{\overline{#1}}
\def\del{\partial}

\def\bra{\langle}
\def\ket{\rangle}
\def\lf{\left}
\def\ri{\right}

\def\lrya{\leftrightarrow}

\def\la{\lambda}

\def\h#1{\widehat{#1}}

\def\bt{\beta}
\def\ga{\gamma}
\def\Ga{\Gamma}

\def\rt#1{\sqrt{#1}}
\def\sitarel#1#2{\mathrel{\mathop{\kern0pt #1}\limits_{#2}}}

\title{Orientifolding of the ABJ Fermi gas}

\author{Kazumi Okuyama}

\affiliation{Department of Physics, Shinshu University, Matsumoto 390-8621, Japan}

\emailAdd{kazumi@azusa.shinshu-u.ac.jp}

\abstract{
The grand partition functions of ABJ theory
can be factorized into even and odd parts
under the reflection of fermion coordinate
in the Fermi gas approach.
In some cases, the even/odd part of ABJ grand partition function
is equal to that of $\mathcal{N}=5$
$O(n)\times USp(n')$ theory,
hence it is natural to think of the even/odd projection of
grand partition function as an orientifolding
of ABJ Fermi gas system.
By a systematic WKB analysis,
we determine the coefficients in the perturbative part of grand potential
of such orientifold ABJ theory. We also 
find the exact form of the first few ``half-instanton'' corrections coming from
the twisted sector of the reflection of fermion coordinate.
For the Chern-Simons level
$k=2,4,8$ we find closed form expressions of the
grand partition functions of orientifold ABJ theory,
and for $k=2,4$ we prove the functional relations among
the grand partition functions
conjectured in arXiv:1410.7658.
}

\begin{document}
\maketitle

\section{Introduction}\label{sec:intro}
The Fermi gas approach, first introduced in \cite{MP},
to study the partition function of certain
$3d$ Chern-Simons-matter theories on $S^3$ enables us
to extract the valuable information of the
non-perturbative effects in the holographically dual
M-theory side.
In particular, in the case of  $\mathcal{N}=6$
 $U(N+M)_k \times U(N)_{-k}$
 ABJ(M) theory,
which is dual to M-theory on $AdS_4\times S^7/\mathbb{Z}_k$,
we have a very detailed understanding
of the instanton effects coming from
the M2-branes wrapping some 3-cycles, thanks to the relation to
the (refined) topological string on local $\mathbb{P}^1\times \mathbb{P}^1$ \cite{HMMO} (see \cite{ABJM-review} for a review). 
Remarkably,
for the special values of the Chern-Simons level $k=1,2,4$,
 it is found that the grand partition functions of ABJ(M) theories
 can be written explicitly in closed forms
 in terms of the Jacobi theta functions \cite{CGM,GHM}.

 The grand partition functions of ABJ(M) theories can be naturally factorized into the even part and the odd part under the reflection
 $x\to -x$ of the fermion coordinate,
 and it is conjectured in \cite{GHM} that the factorized 
 grand partition functions of ABJ theories
 enjoy certain functional relations,
 which are reminiscent of the quantum Wronskian relation obeyed by the spectral determinant of some quantum mechanical system \cite{Dorey:1998pt,Voros:2012se}.
 Recently, it is realized that 
 such projection to the even/odd sectors under the reflection
 of Fermi gas system has a close relation to the orientifolding
 on the bulk M-theory side \cite{Mezei:2013gqa,Moriyama:2015jsa,Assel:2015hsa,MS,Okuyama:2015auc,Honda2}.
 In this paper, we will consider the grand partition functions of the even/odd sectors of 
the ABJ Fermi gas under the reflection, which we call the ``orientifold ABJ theory''.
 For some special values of parameter $M$,
 it is found that 
 such orientifold ABJ theory is related to the Fermi gas
 representation of the $\mathcal{N}=5$
 $O(n)\times USp(n')$ Chern-Simons-matter theory \cite{MS,Honda2}, which  
 is obtained by introducing an orientifold plane
 into the brane construction of ABJ theory \cite{ABJ}.
 Thus, the ``orientifold ABJ theory'' in such cases are indeed
 related to the orientifolding in the bulk M-theory side.

 In this paper, we will study the (grand) partition function
of orientifold ABJ theory in detail. We develop a systematic method to compute the small $k$ expansion (WKB expansion) of the grand potential of orientifold ABJ theory, and determine the coefficients
 $C,B$ and $A$ in the perturbative part of grand potential.
 We also study the coefficients of ``half-instanton'' corrections to the grand potential, which have half-weight $\mathcal{O}(e^{-\mu})$
 of the membrane instanton $\mathcal{O}(e^{-2\mu})$ in ABJ theory.
 This type of half-instanton corrections has also appeared in
 other theories \cite{MS,Okuyama:2015auc} which have  dual
description as  M-theory on certain
 orientifolds, and it is natural to identify
 such half-instantons as the effect coming from orientifold planes.

We also compute the exact values of the canonical partition functions 
of orientifold ABJ theory up to some $N=N_\text{max}$ for various values of $k,M$.
Using this data of exact values of partition functions, one can extract the 
instanton corrections to the grand potential which are summarized in Appendix \ref{app:Jpm-inst}.
For the case of $k=2,4,8$ in Appendix \ref{app:Jpm-inst},
by looking at the first few coefficients of this instanton expansion, 
one can easily guess the all-order resummation of instanton corrections and
write down the closed form expression of the grand potential. 
Once we know the exact grand potential, we can reconstruct the grand partition function
by performing the ``periodic sum'' in \eqref{periodic-sum} \cite{HMO2}. 
In this way, we find
the closed form expressions of the grand partition functions of orientifold ABJ theory 
with $k=2,4,8$. It turns out that 
 the grand potential at $k=2,4$
 are essentially determined by the genus-zero and genus-one free energies of the (refined) topological string on local $\mathbb{P}^1\times \mathbb{P}^1$,
 and the exact grand partition functions
of orientifold ABJ theory 
for $k=2,4$ are proportional to the Jacobi theta function
$\vartheta_3(v,\tau)$, as in the case of ABJ(M) theory before
orientifolding \cite{CGM,GHM}. 
We find that the exact forms of grand partition functions with
$k=2,4$ we obtained indeed satisfy the functional relations conjectured in \cite{GHM}.
On the other hand,  the exact grand partition function
for $k=8$ is qualitatively different from the $k=2,4$ cases, 
due to the all-genus corrections in the worldsheet instanton part
of grand potential.

 This paper is organized as follows. In section \ref{sec:exactZ},
 we compute the exact values of the canonical partition functions of
 orientifold ABJ theories for various $k$ and $M$, and determine
 the instanton corrections to the grand potential, which are summarized
 in Appendix \ref{app:Jpm-inst}.
 In section \ref{sec:WKB}, we study the WKB expansion of
 the grand potential of orientifold ABJ theory, and find the
 coefficients $C,B$ and $A$
 in the perturbative part of grand potential in closed forms. We also find the closed form expression of the first few coefficients
of ``half-instantons'' coming from the twisted sector of the reflection of fermion coordinate.
 In section \ref{sec:k=2}, \ref{sec:k=4}, and \ref{sec:k=8}, we
 write down the exact grand partition functions of orientifold ABJ
 theory at $k=2$, $k=4$, and $k=8$, respectively.
 In section \ref{sec:qw-rel}, for the $k=2,4$ cases
 we prove the functional relations among grand partition functions conjectured in \cite{GHM}.
 We conclude in section \ref{sec:conclusion}.
 Some useful properties of theta functions are summarized in
 Appendix \ref{app:theta}.
 
 \section{Exact partition functions of orientifold ABJ theory}
 \label{sec:exactZ}
 In this section, we will compute the exact values of
 the canonical partition functions $Z_\pm(N,k,M)$
 of orientifold ABJ theory for various integral $k,M$ up to some
 $N=N_\text{max}$.
 Using the exact data of $Z_\pm(N,k,M)$, we determine
 the instanton corrections to the grand potential
 of orientifold ABJ theory.

 \subsection{Review of ABJ Fermi gas}
 Before going to the orientifold ABJ theory,
let us first review the Fermi gas approach
of $U(N_1)_{k}\times U(N_2)_{-k}$ ABJ theory.
It is convenient to parametrize the theory by $(N,k,M)$, where
$k$ is the Chern-Simons level, $N$ is the rank of the
second gauge group,
and $M$ is the difference of the ranks of
the first and the second gauge groups:
\begin{equation}
 N=N_2,\quad M=N_1-N_2.
\end{equation}
Without loss of generality, we can assume $k>0$ and $M\geq0$.
Note that the ABJM theory is a special case of ABJ theory
with $M=0$.
We are interested in the grand partition function
of ABJ theory on $S^3$ obtained from the canonical one by summing over
$N$ with fixed $k,M$ 
\begin{equation}
 \Xi(\mu,k,M)=1+\sum_{N=1}^\infty Z(N,k,M)e^{N\mu},
  \label{ABJ-Xi}
\end{equation}
where $\mu$ is the chemical
potential conjugate to $N$.

In the Fermi gas approach \cite{MP},
the grand partition function
of ABJ theory \eqref{ABJ-Xi} is written as a system of fermions
with no multi-body interactions (ideal Fermi gas)
\begin{equation}
 \Xi(\mu,k,M)=\det(1+e^\mu\rho).
\end{equation}
The density matrix $\rho$
of ABJ Fermi gas is given by \cite{AHS,Honda1,HO}\footnote{In \cite{MaMo}, 
a different expression of ABJ grand partition function is considered.}
\begin{equation}
 \rho(x,y)=\frac{\rt{E(x)E(y)}}{2\cosh\frac{x-y}{2k}},\qquad
E(x)=\frac{1}{2\pi k}\frac{1}{e^{\frac{x}{2}}+(-1)^M e^{-\frac{x}{2}}}
\prod_{s=-\frac{M-1}{2}}^{\frac{M-1}{2}}\tanh\frac{x+2\pi is}{2k}.
\label{rho-ABJ}
\end{equation}

As discussed in \cite{HMO2},
the large $\mu$ expansion of the {\it total grand potential} 
$\log\Xi(\mu,k,M)$
consists of two parts:
the oscillatory part and the non-oscillatory part.
We will focus on the latter part, known as the \emph{modified grand potential} $J(\mu,k,M)$.
The grand partition function can be reconstructed
from
the modified grand potential $J(\mu,k,M)$ by the following
{\it periodic sum} \cite{HMO2}
\begin{equation}
\Xi(\mu,k,M)= \sum_{n\in\mathbb{Z}}e^{J(\mu+2\pi in,k,M)}.
\label{periodic-sum}
\end{equation}
This summation over the $2\pi i$-shift of $\mu$ recovers the periodicity of grand partition function
\begin{equation}
 \Xi(\mu+2\pi i,k,M)=\Xi(\mu,k,M),
\end{equation}
which is required from the very definition of
 $\Xi(\mu,k,M)$ in \eqref{ABJ-Xi}.
Moreover, as discussed in \cite{HMO2}
one can obtain the
canonical partition function at fixed $N$
from the integral transform
of the modified grand potential
\begin{equation}
 Z(N,k,M)=\int_{-\pi i}^{\pi i}\frac{d\mu}{2\pi i}\Xi(\mu,k,M)
  e^{-N\mu}=\int_{\mathcal{C}}\frac{d\mu}{2\pi i}e^{J(\mu,k,M)-N\mu},
  \label{Z-int}
\end{equation}
 where $\mathcal{C}$ is a contour on the complex $\mu$-plane
 running from $e^{-\frac{\pi i}{3}}\infty$ to $e^{+\frac{\pi i}{3}}\infty$.

 According to the conjecture of \cite{HMMO},
 the modified grand potential
 of ABJ theory is completely determined
 by the refined topological string on local $\mathbb{P}^1\times \mathbb{P}^1$
 \begin{equation}
  J(\mu,k,M)=F_\text{top}(T_1,T_2,g_s)+\frac{1}{2\pi i}
   \frac{\del}{\del g_s}\left[g_s F_\text{NS}
			 \left(\frac{T_1}{g_s},\frac{T_2}{g_s},\frac{1}{g_s}\right)\right],
   \label{Jconjecture}
 \end{equation}
 where $g_s=2/k$ denotes the string coupling
 and $F_\text{top}$ and $F_\text{NS}$ represent the free energy
 of the standard topological string and the refined topological string
 in the Nekrasov-Shatashvili limit, respectively.
 The K{\"a}hler parameters $T_1,T_2$ of local $\mathbb{P}^1\times \mathbb{P}^1$ are related to the chemical potential $\mu$
 of ABJ Fermi gas by
 \begin{equation}
  T_{1,2}=\frac{4\mu_\text{eff}}{k}\pm \pi i(1-2b).
   \label{Tvsmueff}
 \end{equation}
Here we introduced the notation
\begin{equation}
 b=\frac{M}{k},
\label{b-def}
\end{equation} 
and the ``effective'' chemical potential $\mu_\text{eff}$ in \eqref{Tvsmueff}
is related to $\mu$ by the so-called quantum mirror map \cite{Aganagic:2011mi}.
See \cite{HMMO} for more details.
In \cite{HO}, it is shown that this conjecture \eqref{Jconjecture}
is in complete agreement with the exact values
of the canonical partition functions of ABJ theory.

It is useful to decompose the modified grand potential
\eqref{Jconjecture} into two parts
\begin{equation}
 J(\mu,k,M)=J^\text{pert}(\mu,k,M)+J^\text{np}(\mu,k,M),
\label{J-decomp}
\end{equation}
where $J^\text{pert}(\mu,k,M)$ is the perturbative part 
\begin{equation}
 J^\text{pert}(\mu,k,M)=\frac{C(k)}{3}\mu^3+B(k,M)\mu+A(k,M)
\label{Jpert-ABJ}
\end{equation}
and $J^\text{np}(\mu,k,M)$ is the non-perturbative part 
which is exponentially suppressed in the large $\mu$ limit.
The coefficients $C(k)$ and $B(k,M)$ in the perturbative part \eqref{Jpert-ABJ}
are given by \cite{MaMo,HO}
\begin{equation}
 C(k)=\frac{2}{\pi^2k},\qquad B(k,M)=\frac{1}{3k}-\frac{k}{12}+\frac{k}{2}\left(\hf-b\right)^2,
\label{CB-ABJ}
\end{equation}
and the constant $A(k,M)$ in \eqref{Jpert-ABJ} is 
given by \cite{HO}
\begin{equation}
 A(k,M)=A_c(k)+F_\text{coni}(k,M),
\label{A-ABJ}
\end{equation}
where $A_c(k)$ is the constant term in the ABJM theory,
which is closely related to a resummation of the constant map contributions in the topological string
\cite{KEK,HO1,HO2}
\begin{equation}
 A_c(k)=-\frac{k^2\zeta(3)}{8\pi^2}+4\int_0^\infty dx\frac{x}{e^{2\pi x}-1}
  \log\left(2\sinh\frac{2\pi x}{k}\right).
  \label{Acon}
\end{equation}
The second term in \eqref{A-ABJ}
denotes the free energy of the $\mathcal{N}=2$
$U(M)_k$ pure Chern-Simons theory on $S^3$ \footnote{Note that in \eqref{Fconi}
there is no shift of the Chern-Simons level $k$ in a regularization preserving supersymmetry.
See \cite{Aharony:2015mjs} for a discussion on this point.}
\begin{equation}
 F_\text{coni}(k,M)=-\log Z_\text{CS}(U(M)_k)
=-\log\left[k^{-\frac{M}{2}}\prod_{s=1}^{M-1}\Bigl(2\sin\frac{\pi s}{k}\Bigr)^{M-s}\right].
\label{Fconi}
\end{equation}
Via large $N$ duality, this is equal to the free energy
of the topological string on the resolved conifold \cite{Gopakumar:1998ki}.
As discussed in \cite{HO2},
by a certain resummation of the genus expansion, we can write down a useful integral representation
of \eqref{Fconi}
\begin{equation}
 F_\text{coni}(k,M)= -\frac{k^2}{4\pi^2}F_0(b)+\int_0^\infty
dx \frac{x}{e^{2\pi x}-1}\log\left(\frac{\cosh \frac{2\pi x}{k}-\cos2\pi b}{\cosh \frac{2\pi x}{k}-1}\right),
\label{Fconi-int}
\end{equation}
where $F_0(b)$ is the genus-zero free energy
of resolved conifold
\begin{equation}
 F_0(b)=\text{Re}[\text{Li}_3(e^{2\pi ib})]-\zeta(3).
\end{equation}
Here we assumed $k,b\in\mathbb{R}$, and
$ \text{Li}_s(z)=\sum_{n=1}^\infty \frac{z^n}{n^s}$ denotes
the polylogarithm.
This integral representation \eqref{Fconi-int}
defines a natural analytic continuation of $k,M$ to non-integer values.

Plugging the decomposition \eqref{J-decomp}
of the modified grand potential
into \eqref{Z-int}
and expanding the non-perturbative part, we find
that the canonical partition function $Z(N)$
at fixed $N$ can be also decomposed into perturbative and
non-perturbative parts:
\begin{equation}
 Z(N)=Z_\text{pert}(N)+Z_\text{np}(N),
\label{Zexpand}
\end{equation}
where  the perturbative part $Z_\text{pert}(N)$
is given by the Airy function \cite{MP,FHM}
\begin{equation}
 Z_\text{pert}(N)=C^{-\frac{1}{3}}e^A\text{Ai}\Big[C^{-\frac{1}{3}}(N-B)\Big],
\label{Zpert}
\end{equation}
and the non-perturbative correction
$Z_\text{np}(N)$ is 
given by a sum of the derivatives of the Airy functions \cite{HMO2}.

We should mention one important property of the grand partition function of ABJ
theory. Due to the Seiberg-like duality \cite{ABJ}, 
$\Xi(\mu,k,M)$ is invariant under $M\to k-M$
\begin{equation}
 \Xi(\mu,k,k-M)=\Xi(\mu,k,M).
\label{sei-dual}
\end{equation}
In terms of the parameter $b$ in \eqref{b-def},
the grand partition function is invariant under $b\to 1-b$,
which corresponds to the exchange of two K{\"a}hler parameters
$T_{1}\lrya T_2$ in
\eqref{Tvsmueff}.
For the physical ABJ theory with integer $k$ and $M$,
it is argued that
the supersymmetry is spontaneously broken when $M>k$ \cite{ABJ}.
Therefore, for a fixed $k$, the independent values of $M$
are $M=0,1,\cdots,k$.
Taking account of the Seiberg-like duality \eqref{sei-dual},
the physically independent values of $M$ are further reduced to
\begin{equation}
 M=0,1,\cdots, [k/2].
\end{equation}
\subsection{Orientifold ABJ theory}\label{subsec:Zpm}
In this subsection, we will consider the even/odd
projection of the ABJ Fermi gas under the reflection $R$
of the fermion coordinate
\begin{equation}
 R:x\to -x.
\end{equation}
This kind of projection
of Fermi gas system into $R=\pm1$ sectors appeared previously in some examples
\cite{MS,Honda2,Okuyama:2015auc}, and those examples have
holographically dual description as the
M-theory on certain orientifolds.
We expect that the projection of ABJ Fermi gas into even/odd sectors is 
also holographically dual to
the M-theory on some orientifolds.

Since the density matrix $\rho(x,y)$ of ABJ theory is invariant under the reflection of coordinates
\begin{equation}
 \rho(-x,-y)=\rho(x,y),
\end{equation}
we can consider the projection of $\rho$ into the even/odd parts
$\rho_\pm$
\begin{equation}
 \rho_\pm(x,y)=\frac{\rho(x,y)\pm\rho(x,-y)}{2},
\end{equation}
which, in the operator language, is simply written as
\begin{equation}
 \rho_\pm=\rho\frac{1\pm R}{2}.
\end{equation}
Then the grand partition function of ABJ theory is
naturally factorized into the even/odd part
\begin{equation}
 \Xi(\mu,k,M)=\Xi_{+}(\mu,k,M)\Xi_{-}(\mu,k,M)
\end{equation}
where $\Xi_\pm(\mu,k,M)$ are the
Fredholm determinant of $\rho_\pm$
\begin{equation}
 \Xi_{\pm}(\mu,k,M)=\det(1+e^\mu\rho_{\pm}).
\end{equation}
From this grand partition function $\Xi_{\pm}(\mu,k,M)$,
we can read off the canonical partition function
of orientifold ABJ theory
   $Z_\pm(N,k,M)$ with fixed $N$ by expanding
   in $e^\mu$
   \begin{equation}
   \Xi_{\pm}(\mu,k,M)= 1+\sum_{N=1}^\infty Z_\pm(N,k,M) e^{N\mu}.
   \end{equation}
Note that the Seiberg-like duality \eqref{sei-dual}
holds for each sector
\begin{equation}
 \Xi_{\pm}(\mu,k,k-M)=\Xi_{\pm}(\mu,k,M).
\end{equation}

From \eqref{rho-ABJ}, one can see that $\rho_{+}(x,y)$
can be written as
\begin{equation}
 \rho_{+}(x,y)=\frac{\rt{V(x)V(y)}}{2\cosh \frac{x}{k}+2\cosh \frac{y}{k}},\qquad
V(x)=\hf \left(2\cosh\frac{x}{2k}\right)^2E(x).
\end{equation}
This is exactly the form to which we can apply the Tracy-Widom lemma \cite{TW} and we can easily compute the exact values of
spectral traces.
Once we know the trace $\Tr\rho_{+}^\ell$ from $\ell=1$ to $\ell=N$,
we can compute the canonical partition function $Z_{+}(N,k,M)$ at fixed $N$. 
Using the lemma in \cite{TW},
the $\ell^\text{th}$ power of $\rho_{+}$ can be systematically computed
by constructing a sequence of functions $\phi_\ell(x)~(\ell=0,1,2,\cdots)$
\begin{equation}
 \begin{aligned}
 \rho^\ell_{+}(x,y)&=\frac{\rt{V(x)V(y)}}{2\cosh \frac{x}{k}+(-1)^{\ell-1}2\cosh\frac{y}{k}}
\sum_{j=0}^{\ell-1}(-1)^j \phi_j(x)\phi_{\ell-1-j}(y),\\
\phi_\ell(x)&=\frac{1}{\rt{V(x)}}\int_{-\infty}^\infty dy\, \rho_{+}(x,y)\rt{V(y)}\phi_{\ell-1}(y),\quad\phi_0(x)=1.
\end{aligned}
\label{TW-rho}
\end{equation}
Then $\Tr\rho_{+}^\ell$ is given by
\begin{equation}
 \begin{aligned}
  \Tr\rho^{2n}_{+}&=\int_{-\infty}^\infty  dx 
\frac{kV(x)}{2\sinh \frac{x}{k}}
\sum_{j=0}^{2n-1}(-1)^j \frac{d\phi_j(x)}{dx}\phi_{2n-1-j}(x),\\
\Tr\rho^{2n+1}_{+}&=\int_{-\infty}^\infty  dx 
\frac{V(x)}{4\cosh \frac{x}{k}}
\sum_{j=0}^{2n}(-1)^j \phi_j(x)\phi_{2n-j}(x).
 \end{aligned}
\label{TW-tr}
\end{equation}
The integrals  in \eqref{TW-rho} and \eqref{TW-tr}
can be evaluated by
rewriting them as contour integrals as in  \cite{PY,HMO2,HO}.
On the other hand, the traces of $\rho_{-}$ can be computed by using the following relation found in \cite{HMO2}
\begin{equation}
 \sum_{\ell=0}^\infty \phi_\ell(0)z^\ell=\det\left(\frac{1+z\rho_{-}}{1-z\rho_{+}}\right).
\end{equation}
Using the above algorithm, we have computed the exact values of the partition functions
$Z_\pm(N,k,M)$ for various $(k,M)$ up to $N=N_\text{max}$, where $N_\text{max}$ is about 10-30
\footnote{The data of the exact values of $Z_\pm(N,k,M)$ are attached as ancillary files to the {\tt arXiv} submission of this paper.}.
For the case of $k=8,12$, we have computed $Z_\pm(N,k,M)$ only 
for odd $M$. 

As in the case of ABJM theory \cite{HMO2},
by matching the exact values of $Z_\pm(N,k,M)$
with the expansion in terms of Airy function and its derivatives \eqref{Zexpand}
and \eqref{Zpert}, we can fix the coefficients in the modified grand potential order by order for the first few instanton numbers.
In this way, we find that the modified grand potential for $\rho_\pm$
can be written in a similar form as that in the ABJ theory \eqref{J-decomp}
\begin{equation}
 J_\pm(\mu,k,M)=J^\text{pert}_\pm(\mu,k,M)+J^\text{np}_\pm(\mu,k,M).
\end{equation}
Again, the perturbative part is a cubic polynomial in $\mu$
\begin{equation}
 J^\text{pert}_\pm(\mu,k,M)=\frac{C_\pm(k)}{3}\mu^3+B_\pm(k,M)\mu+A_\pm(k,M),
  \label{Jpm-pert}
\end{equation}
where the coefficients $C_\pm(k)$
and $B_\pm(k,M)$ in \eqref{Jpm-pert}
are related to those in the ABJ theory by
\begin{equation}
C_\pm(k)=\hf C(k),\qquad B_\pm(k,M)=\hf B(k,M)\pm \qu. 
\label{CBpm}
\end{equation}
By matching the exact values of $Z_\pm(N,k,M)$,
 we conjecture that the constant term $A_\pm(k,M)$
 in \eqref{Jpm-pert} is given by
\begin{equation}
 A_\pm(k,M)=\hf A_c(k)+\hf F_\text{coni}(k,M)\pm
  \left(F_\text{non-ori}(k,M)-\hf\log2\right).
\label{Apm}
\end{equation}
Here $A_c(k)$ and $F_\text{coni}(k,M)$ are
given by \eqref{Acon} and \eqref{Fconi}, respectively.
Very interestingly, $A_\pm(k,M)$ contains the contributions from
non-orientable worldsheet instantons $F_\text{non-ori}(k,M)$
on the orientifold of resolved conifold, which in turn  is related to the
$SO(M+1)_k$ pure Chern-Simons theory on $S^3$ via large $N$ duality \cite{Sinha:2000ap}
\begin{equation}
 -\log Z_\text{CS}(SO(M+1)_k)=\hf F_\text{coni}(k,M)
  -F_\text{non-ori}(k,M).
\end{equation}
A useful integral representation of $F_\text{non-ori}(k,M)$ is found in
\cite{HO2}
\begin{equation}
 F_\text{non-ori}(k,M)=\frac{k}{4\pi}\text{Im}\Big[\text{Li}_2(e^{-i \pi b})-\text{Li}_2(-e^{-i \pi b})\Big]
+\int_0^\infty \frac{dx}{e^{2\pi x}+1}\arctan\left(\frac{\sin\pi b}{\sinh \frac{2\pi x}{k}}\right).
\label{Fnon-ori}
\end{equation}
The appearance of non-orientable contributions in $A_\pm(k,M)$
is the first indication
that the projection of ABJ Fermi gas into $R=\pm 1$
is closely related to the orientifolding.

In the next section, we will directly derive
\eqref{CBpm} and \eqref{Apm}  from the small $k$ expansion (WKB expansion)
of the grand potential.

\begin{figure}[tbh]
\begin{center}
\vskip3mm
\includegraphics[width=10cm]{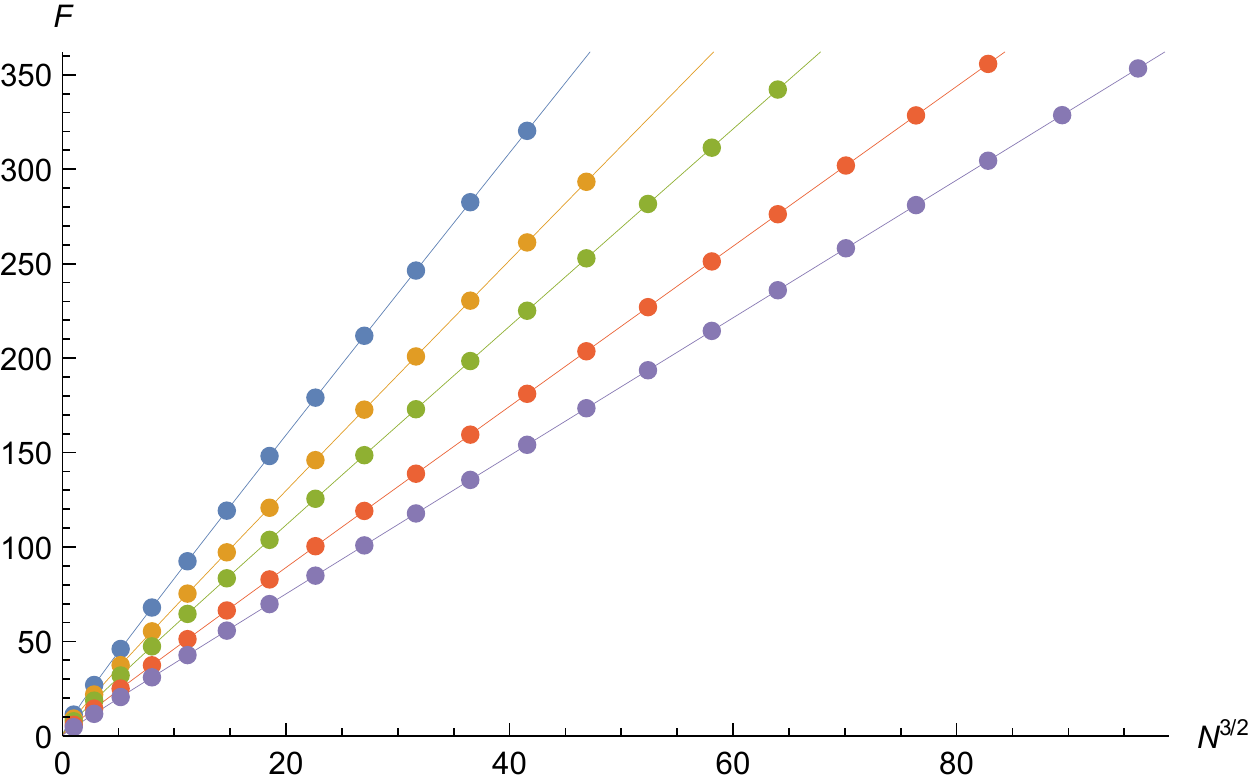}
\end{center}
  \caption{We show the  plot of free energy $F=-\log Z_{-}(N,k,M)$ with $M=3$,
for $k=3,4,6,8,12$.  $k$ increases from the right curve ($k=3$) 
to the left curve ($k=12$).
Note that the horizontal axis is $N^{3/2}$.
The dots are the exact values of free energy at integer $N$,
while
the solid curves represent 
the perturbative free energy
given by the Airy function \eqref{Zpert}
with $C,B$ in \eqref{CBpm} and $A$ in \eqref{Apm}.
}
\label{fig:Zpert}
\end{figure}

\begin{figure}[tbh]
\begin{center}
\vskip3mm
\includegraphics[width=7cm]{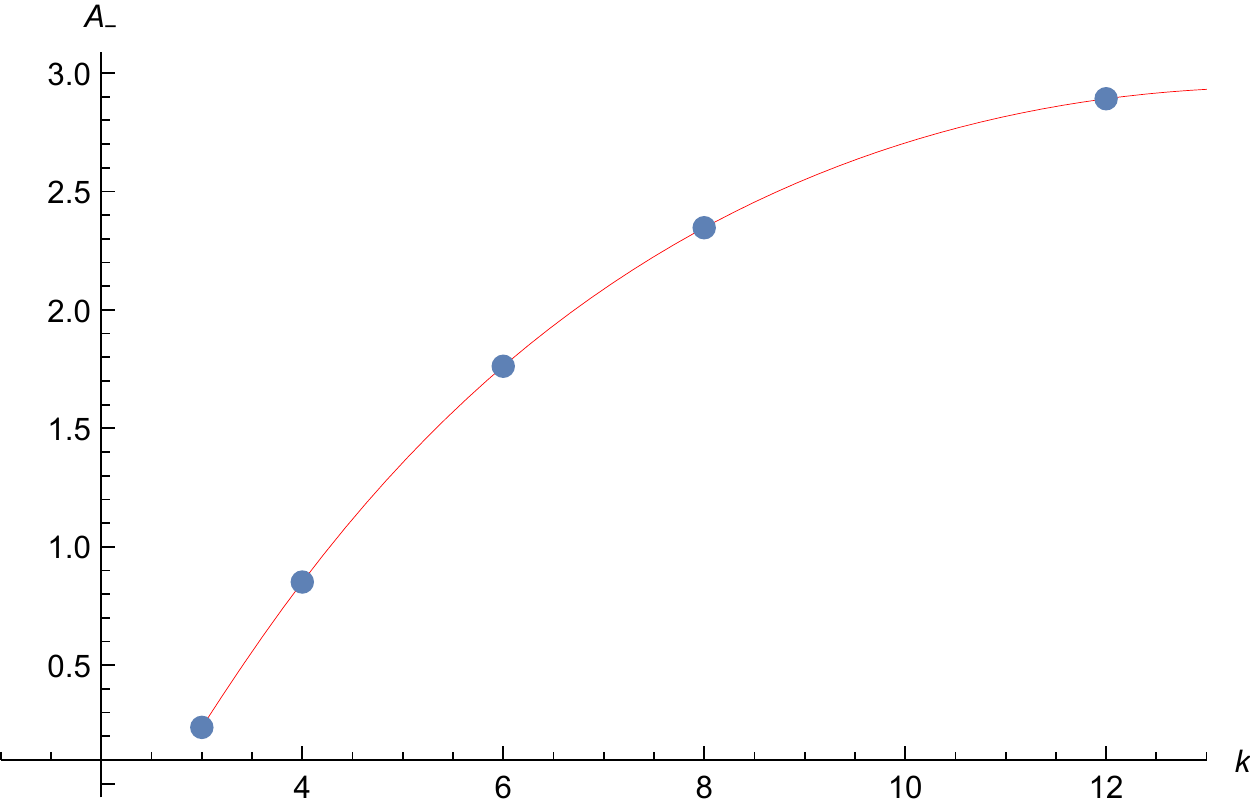}
\end{center}
  \caption{This is the plot of $A_{-}(k,M=3)$  as a function of $k$.
The dots are the numerical
values of $A_{-}(k,3)$ for $k=3,4,6,8,12$ computed from \eqref{Aestimate}, 
while the solid curve is the plot of 
our conjecture of $A_{-}(k,3)$ in \eqref{Apm}.
}
\label{fig:Aminus}
\end{figure}

In Figure \ref{fig:Zpert}, we show the plot of free energy 
$F=-\log Z_{-}(N,k,M)$ as a function of $N$ for
$M=3$ with $k=3,4,6,8,12$, as an example.
As we can see, the exact values of the free energy exhibit a nice
agreement with the perturbative one given by the Airy function
\eqref{Zpert} if we use the correct coefficients $C,B$ in \eqref{CBpm} 
and $A$ in  \eqref{Apm}.
We find the similar agreement for all other cases.

As discussed in \cite{HO1},
we can estimate
the numerical values of  $A_\pm(k,M)$ by
\begin{equation}
 A_\pm(k,M)\approx \log\left[
\frac{Z_\pm(N,k,M)}{C_\pm(k)^{-\frac{1}{3}}\text{Ai}\Bigl[C_\pm(k)^{-\frac{1}{3}}(N-B_\pm(k,M))\Bigr]}\right],\quad(N\gg1).
\label{Aestimate}
\end{equation}
In practice, we compute the numerical value of $A_\pm(k,M)$
from the exact value of $Z_\pm(N_\text{max},k,M)$
by setting $N=N_\text{max}$ in \eqref{Aestimate}.
In Figure \ref{fig:Aminus} we show the plot of  $A_{-}(k,M)$ for $M=3$,
as an example.
One can clearly see a nice agreement between the numerical value of $A_{-}(k,3)$
estimated by using \eqref{Aestimate} and our conjecture of $A_{-}(k,3)$ in \eqref{Apm}.
Let us take a closer look at the case of $A_{-}(8,3)$. 
We have computed the exact values of $Z_{-}(N,8,3)$ up to 
$N_\text{max}=14$, and the numerical estimation \eqref{Aestimate}
with $N=14$
gives
\begin{equation}
 A_{-}(8,3)\approx 2.34669920... .
\label{Anum}
\end{equation}
On the other hand, our proposal of the exact value \eqref{Apm} is
\begin{equation}
A_{-}(8,3)= -\frac{\zeta (3)}{16 \pi ^2}+\frac{5 \log (2)}{8}-2 \log
   \left(\sin \frac{\pi }{8}\right)= 2.34669923144...,
\label{Aex}
\end{equation}
which is in good agreement with the numerical estimation \eqref{Anum}.
The difference between \eqref{Anum} and \eqref{Aex} can be attributed to the instanton corrections.
The first instanton correction
is of order $e^{-\frac{4\mu}{k}}=e^{-\frac{\mu}{2}}$ for $k=8$
in the grand canonical picture, which in the canonical picture corresponds
to the correction of the order
\begin{equation}
\exp\left(-\frac{1}{2}\rt{\frac{N_\text{max}}{C_{-}(8)}}\right)\approx 6.0\times 10^{-8}.
\end{equation}
This is indeed the same order as the difference between \eqref{Anum} and \eqref{Aex}.

Once we know the perturbative grand potential,
we can continue
the above procedure to fix the coefficients in the non-perturbative part 
$J^\text{np}_\pm(\mu,k,M)$ using our exact data of partition functions
$Z_\pm(N,k,M)$.
The results are summarized in Appendix \ref{app:Jpm-inst}.
We find that there are three types of instanton corrections
with the weight:
\begin{equation}
 \mathcal{O}(e^{-\frac{4\mu}{k}}),\qquad
\mathcal{O}(e^{-2\mu}),\qquad
\mathcal{O}(e^{-\mu}).
\label{3-type}
\end{equation}
The first two types have direct analogue in the ABJ theory, namely, worldsheet instantons
and membrane instantons. On the other hand,
the last type in \eqref{3-type} is a new contribution coming from the
 effect of orientifold projection,
which we call ``half-instantons'', following  \cite{Okuyama:2015auc}. 
It is curious to observe that there is no contribution with the half-weight of worldsheet instanton 
of the order $\mathcal{O}(e^{-2\mu/k})$.
Only the half-weight of membrane instantons of order
$\mathcal{O}(e^{-\mu})$ appear.

For the cases of $k=2,4,8$, by looking at the coefficients
of expansion in Appendix \ref{app:Jpm-inst}, one can easily guess
the all order resummation of instanton expansion,
and find the closed form expression of grand partition function
$\Xi_\pm(\mu,k,M)$, which we will consider in section
\ref{sec:k=2}, \ref{sec:k=4}, and \ref{sec:k=8}.
By expanding the closed form
expression
of $\Xi_\pm(\mu,k,M)$ around $e^\mu=0$, one can read off
the exact values of canonical partition functions
$Z_\pm(N,k,M)$ up to arbitrarily high $N$, in principle.
Therefore, one can ``bootstrap'' the computation of
partition functions for $k=2,4,8$:
\begin{equation}
 \begin{aligned}
 &\text{exact~values~of}~Z_\pm(N,k,M)~~~(N\leq N_\text{max})\\
  \Rightarrow~~
  &\text{closed~form~of}~~\Xi_\pm(\mu,k,M)\\
  \Rightarrow~~
   &\text{exact~values~of}~Z_\pm(N,k,M)~~~(N> N_\text{max}).  
 \end{aligned}
\end{equation}
For example, we have computed the exact values of $Z_{-}(N,8,3)$
up to $N=20$ from the expansion of the closed form expression of
the grand partition function
$\Xi_{-}(\mu,8,3)$ in \eqref{Xipm83orb}. Using the exact value of $Z_{-}(N=20,8,3)$,
the numerical estimation of $A_{-}(8,3)$ in \eqref{Aestimate} gets
improved
\begin{equation}
A_{-}(8,3)\approx2.3466992303...,
\end{equation}
which is closer to the exact value \eqref{Aex} than the numerical value in \eqref{Anum}
obtained from  $Z_{-}(N=14,8,3)$, as expected.
\subsection{Relation to $\mathcal{N}=5$ $O(n)\times USp(n')$ theory}
As discussed in \cite{ABJ},
by introducing orientifold planes into the brane construction of ABJ theory,
we can obtain $\mathcal{N}=5$
Chern-Simons-matter theories 
with gauge group $O(n)\times USp(n')$  \cite{ABJ,Hosomichi:2008jb}.
Now it is natural to ask
whether such $\mathcal{N}=5$
theories are related to the projection of ABJ Fermi gas into even/odd sectors.
In a recent paper \cite{Honda2},
it is observed that the grand partition function constructed from
$\rho_{-}$ of ABJM theory ($M=0$) is equivalent to
the grand partition function
of $\mathcal{N}=5$ $O(2N+1)_{2k}\times USp(2N)_{-k}$
Chern-Simons-matter theory\footnote{We are informed by Sanefumi Moriyama that
the grand partition function of $\mathcal{N}=5$ $O(2N+m)_{2k}\times USp(2N)_{-k}$
theory with odd $m$ can be also written as a Fermi gas system \cite{Mo}.} 
\begin{equation}
 \Xi_{-}(\mu,k,0)=\Xi_{\mathcal{N}=5~O(2N+1)_{2k}\times USp(2N)_{-k}}(\mu).
\end{equation}
Also, a Fermi gas formalism of $\mathcal{N}=5$
theory with different gauge group $O(2N)_{2k}\times USp(2N)_{-k}$
is studied in \cite{MS}.
We find that the exact values of the canonical partition functions 
in Table.1 of \cite{MS} all coincide with
the canonical partition functions $Z_{+}(N,k,1)$
computed from the density matrix
$\rho_{+}$ of orientifold ABJ theory with $M=1$.
From this non-trivial agreement of the canonical partition functions at small $N$,
we conjecture that the grand partition functions of these two theories are actually
equivalent
\begin{equation}
 \Xi_{+}(\mu,k,1)=\Xi_{\mathcal{N}=5~O(2N)_{2k}\times USp(2N)_{-k}}(\mu).
\end{equation}
It would be interesting to find a direct proof of this equality.

\section{WKB expansion}
\label{sec:WKB}
In this section, we will consider the small $k$ expansion (WKB expansion)
of the grand potential $J_\pm(\mu,k,M)$ of orientifold ABJ theory.
As discussed in \cite{Okuyama:2015auc},
the \emph{total grand potential} can be written as
\begin{equation}
 J_\pm(\mu,k,M)=\frac{J(\mu,k,M)\pm J^R(\mu,k,M)}{2},
\label{totalJ}
\end{equation}
where
\begin{equation}
  J(\mu,k,M)=\sum_{\ell=1}^\infty \frac{(-1)^{\ell-1}e^{\ell\mu}}{\ell}\Tr(\rho^\ell),\quad
J^R(\mu,k,M)=\sum_{\ell=1}^\infty \frac{(-1)^{\ell-1}e^{\ell\mu}}{\ell}\Tr(\rho^\ell R),
\label{Jellsum}
\end{equation}
with $\rho$ being the density matrix of ABJ theory.
As we will see in the next section, at the level of modified grand potential
\eqref{totalJ} does not hold in general.
In particular, the sum of $J_{+}$ and $J_{-}$
does not agree with the modified grand potential
of ABJ theory\footnote{After the submission of this paper to {\tt arXiv},
the paper \cite{Mo} appeared. We realized that  Table.2 in \cite{Mo}
for the ABJM case ($M=0$) can be generalized to the orientifold ABJ theory as
\begin{align}
 \begin{aligned}
  J_{+}+J_{-}-J=&~Q_1Q_2
+4\cos^2\frac{2\pi}{k}(Q_1^2Q_2+Q_1Q_2^2)+\left(1+\frac{\sin^2\frac{6\pi}{k}}{\sin^2\frac{2\pi}{k}}\right)(Q_1^3Q_2+Q_1Q_2^3)\\
&+\left(1+
\Bigl(1-\sin\frac{2\pi}{k}\sin\frac{6\pi}{k}\Bigr)\frac{16\cos^2\frac{2\pi}{k}\sin\frac{6\pi}{k}}{\sin\frac{2\pi}{k}}\right)Q_1^2Q_2^2+\mathcal{O}(Q^5),
 \end{aligned}
\end{align}
where $Q_{1,2}=e^{-T_{1,2}}$.
}
\begin{equation}
 J_{+}(\mu,k,M)+J_{-}(\mu,k,M)\not= J(\mu,k,M).
  \label{Jdif}
\end{equation}
On the other hand, we find that the 
difference $J_{+}-J_{-}$ is always equal to $J^R$ in the
examples listed in Appendix \ref{app:Jpm-inst}.
We should stress that the total grand grand partition function is
completely factorized and the total grand potentials satisfy
\begin{equation}
 \log\Xi_{+}(\mu,k,M)+\log\Xi_{-}(\mu,k,M)=\log\Xi(\mu,k,M).
\end{equation}
The difference in \eqref{Jdif}
comes from rewriting a double sum into a single sum
in \eqref{periodic-sum} 
\begin{equation}
 \sum_{n_{+}\in\mathbb{Z}}e^{J_{+}(\mu+2\pi in_+,k,M)}
\sum_{n_{-}\in\mathbb{Z}}e^{J_{-}(\mu+2\pi in_-,k,M)}
=\sum_{n\in \mathbb{Z}}e^{J(\mu+2\pi in,k,M)}.
\end{equation}
For the case of $k=2,4$, we can write down the difference in
\eqref{Jdif} explicitly using the exact form of grand partition functions,
as we will see in section \ref{sec:k=2} and section \ref{sec:k=4}.

To find the large $\mu$ expansion of $J_\pm(\mu,k,M)$,
it is useful to rewrite them in the Mellin-Barnes representation \cite{Hatsuda-zeta}
\begin{equation}
 J(\mu,k,M)=-\int_\ga\frac{ds}{2\pi i}\Ga(s)\Ga(-s)\Tr(\rho^s)e^{s\mu},\quad
 J^R(\mu,k,M)=-\int_\ga\frac{ds}{2\pi i}\Ga(s)\Ga(-s)\Tr(\rho^s R)e^{s\mu}.
\end{equation}
The contour $\ga$ is taken to be parallel
to the imaginary $s$-axis with $0<\Re(s)<1$.
Picking up the poles at $s=\ell~(\ell=1,2,\cdots)$, we recover \eqref{Jellsum}.
On the other hand,
closing the contour in
the direction $\Re(s)\leq0$ and picking up  the poles on the negative real
$s$-axis, we can find the large $\mu$ expansion.
Thus, the behavior of $J_\pm(\mu,k,M)$
is encoded in the analytic properties of the \emph{spectral trace}
$\Tr(\rho^s)$ and  the \emph{twisted spectral trace} $\Tr(\rho^s R)$. 
In this section, we will consider the WKB expansion of $\Tr(\rho^s)$
and $\Tr(\rho^s R)$, following a similar computation in \cite{Okuyama:2015auc}.

\subsection{The density matrix of local $\mathbb{F}_0$}
Since the density matrix of ABJ Fermi gas
\eqref{rho-ABJ}
depends on $k$ explicitly,
it is not so obvious how can we compute the WKB expansion
of (twisted) spectral traces.
Also, the expression of density matrix in \eqref{rho-ABJ} makes sense only for integer $M$.
In \cite{Kashaev:2015wia}
an analytic continuation of $\rho$
to arbitrary $M$ is found using the relation to the mirror curve of local $\mathbb{F}_0$ (note that $\mathbb{F}_0=\mathbb{P}^1\times\mathbb{P}^1$).
It turns out that we can systematically compute the small $k$ expansion with fixed $b=M/k$.

Let us briefly recall the construction in \cite{Kashaev:2015wia}.
The density matrix of ABJ theory is
identified as
\begin{equation}
 \rho=O_{\mathbb{F}_0}^{-1}
\end{equation}
where $O_{\mathbb{F}_0}$ is the quantized mirror curve of local $\mathbb{F}_0$
\begin{equation}
O_{\mathbb{F}_0}=e^X+me^{-X}+e^Y+e^{-Y},
\label{OF0}
 \end{equation}
with 
\begin{equation}
  X=x+ \frac{p}{2},\qquad Y=\frac{p}{2},\qquad [x,p]=i\hbar.
\end{equation}
The Planck constant $\hbar$ is related to the Chern-Simons level
$k$ by
\begin{equation}
 \hbar=2\pi k,
\label{hbar-def}
\end{equation}
and the mass parameter $m$ in \eqref{OF0} is related to  the parameters in the ABJ theory as
\begin{equation}
 m=e^{\bt},\qquad\bt=\pi ik\left(1-2b\right).
\end{equation}

As shown in \cite{Kashaev:2015wia},
the inverse operator of $O_{\mathbb{F}_0}$ can be written in terms of the quantum dilogarithm $\Phi(x)$ defined by
\begin{equation}
 \Phi(x)=\prod_{n=0}^\infty \frac{1+e^x\mathfrak{q}^{n+\hf}}{1+e^{\frac{2\pi x}{\hbar}}
\til{\mathfrak{q}}^{n+\hf}},
\label{q-dilog}
\end{equation}
with
\begin{equation}
 \mathfrak{q}=e^{i\hbar},\qquad\til{\mathfrak{q}}=e^{-\frac{4\pi^2i}{\hbar}}.
\end{equation}
From \eqref{q-dilog}, the quantum dilogarithm exhibits the following
quasi-periodicity
\begin{equation}
 \frac{\Phi(x-\frac{i\hbar}{2})}{\Phi(x+\frac{i\hbar}{2})}=1+e^x,\qquad
\frac{\Phi(x-\pi i)}{\Phi(x+\pi i)}=1+e^{\frac{2\pi x}{\hbar}}.
\end{equation}
From this property, 
for the  operators $x,p$ obeying $[x,p]=i\hbar$,
one can show that
\begin{equation}
  \Phi(p)e^x\Phi(p)^{-1}=e^x+e^{p+x}.
\label{phiconj}
\end{equation}
Using this relation \eqref{phiconj}
repeatedly, one finds
\begin{equation}
  e^{\frac{X}{2}}O_{\mathbb{F}_0}e^{\frac{X}{2}}-m
=\Phi(x)^{-1}\Phi(p)e^x\Phi(p)^{-1}\Phi(x)
\end{equation}
and\begin{equation}
 \begin{aligned}
  \Phi(p)^{-1}\Phi(x)e^{\frac{X}{2}}O_{\mathbb{F}_0}e^{\frac{X}{2}}\Phi(x)^{-1}\Phi(p)&=
m(1+e^{x-\bt})=m\frac{\Phi\lf(x-\bt-\frac{i\hbar}{2}\ri)}{\Phi\lf(x-\bt+\frac{i\hbar}{2}\ri)}.
 \end{aligned}
\end{equation}
Now, using the quantum pentagon identity
\begin{equation}
 \Phi(p)\Phi(x)=\Phi(x)\Phi(x+p)\Phi(p),
\end{equation}
the inverse of $O_{\mathbb{F}_0}$ is further rewritten as
\begin{equation}
\begin{aligned}
  mO_{\mathbb{F}_0}^{-1}
&=e^{\frac{X}{2}}\Phi(x)^{-1}\Phi(p)\frac{\Phi\lf(x-\bt+\frac{i\hbar}{2}\ri)}{\Phi\lf(x-\bt-\frac{i\hbar}{2}\ri)}\Phi(p)^{-1}\Phi(x)e^{\frac{X}{2}}\\
&=e^{\frac{X}{2}}\Phi(x)^{-1}\Phi\lf(x-\bt+\frac{i\hbar}{2}\ri)
\frac{\Phi\lf(x+p-\bt+\frac{i\hbar}{2}\ri)}{\Phi\lf(x+p-\bt-\frac{i\hbar}{2}\ri)}\Phi\lf(x-\bt-\frac{i\hbar}{2}\ri)^{-1}\Phi(x)e^{\frac{X}{2}}\\
&=e^{\frac{X}{2}}\Phi(x)^{-1}\Phi\lf(x-\bt+\frac{i\hbar}{2}\ri)\frac{1}{1+e^{x+p-\bt}}\Phi\lf(x-\bt-\frac{i\hbar}{2}\ri)^{-1}\Phi(x)e^{\frac{X}{2}}.
\end{aligned}
\end{equation}
Finally, after redefining the variables
\begin{equation}
 \h{p}=x+p-\bt,\quad \h{x}=x-\frac{\bt}{2},
\end{equation}
we arrive at the following expression of the density matrix $\rho=O^{-1}_{\mathbb{F}_0}$
(up to an overall constant and similarity transformation)
\begin{equation}
 \rho= e^{\frac{\h{x}}{2}}
\frac{\Phi(\h{x}-\frac{\bt}{2}+\frac{i\hbar}{4})\Phi(\h{x}+\frac{\bt}{2}+\frac{i\hbar}{4})}{\Phi(\h{x}-\frac{\bt}{2}-\frac{i\hbar}{4})\Phi(\h{x}+\frac{\bt}{2}-\frac{i\hbar}{4})}\frac{1}{2\cosh\frac{\h{p}}{2}}.
\label{rho-phi}
\end{equation}

We are interested in the small $\hbar$
expansion of $\rho$. In this case the denominator
of \eqref{q-dilog} can be ignored,
since it is non-perturbative in $\hbar$. Then taking the log of \eqref{q-dilog}
we find\footnote{As discussed in \cite{HO2}, the small $\hbar$ expansion of
the first term in \eqref{logPhi} is Borel summable, and the
Borel sum correctly includes the non-perturbative contribution coming from
the denominator
of \eqref{q-dilog}.}
\begin{equation}
 \log\Phi(x)= \sum_{\ell=1}^\infty \frac{(-e^x)^\ell}{\ell(\mathfrak{q}^{\frac{\ell}{2}}-\mathfrak{q}^{-\frac{\ell}{2}})}+\mathcal{O}(e^{-1/\hbar}).
\label{logPhi}
\end{equation}
Plugging this into \eqref{rho-phi}
and expanding around $\hbar=0$, the density matrix is rewritten as
\begin{equation}
 \rho=\exp\left(\sum_{n=1}^\infty \frac{(-1)^n(\hbar/2)^{2n}}{(2n)!}E_{2n}(b)\text{Li}_{1-2n}(-e^{\h{x}})\right)\frac{1}{2\cosh\frac{\h{x}}{2}}\frac{1}{2\cosh\frac{\h{p}}{2}},
\label{rho-smallh}
\end{equation}
where
 $E_n(b)$ is the Euler polynomial defined by
\begin{equation}
 \frac{2e^{xb}}{e^x+1}=\sum_{n=0}^\infty \frac{E_n(b)}{n!}x^n.
\end{equation}
When $b=0$, \eqref{rho-smallh} reduces to the density matrix of ABJM theory
\begin{equation}
 \rho_\text{ABJM}=\frac{1}{2\cosh\frac{\h{x}}{2}}\frac{1}{2\cosh\frac{\h{p}}{2}},
\end{equation}
since $E_{2n}(0)=0~(n\geq1)$.

Now we are ready to perform the WKB expansion (with  $b$ fixed)
of the spectral trace
$\Tr(\rho^s)$ and the twisted spectral trace $\Tr(\rho^s R)$.

\subsection{WKB expansion of spectral trace $\Tr\rho^s$}
\label{sub:wkb-bulk}
Let us first consider the WKB expansion of the spectral trace
$\Tr(\rho^s)$.
As discussed in \cite{Okuyama:2015auc},
this is easily done by using the Wigner transform $\rho_W$
of the density matrix
$\rho$.
In general, the Wigner transform $A_W$ of the operator $A$ is defined by
\begin{align}
 A_W=\int dy e^{\frac{iPy}{\hbar}}\Big\bra X-\frac{y}{2}\Big|A\Big|X+\frac{y}{2}\Big\ket.
 \label{Wignerdef}
\end{align} 
Using the property of Wigner transformation
\begin{equation}
\begin{aligned}
& (AB)_W=A_W\star B_W \equiv A_W e^{\frac{i\hbar}{2}(\overleftarrow{\del_X}\overrightarrow{\del_P}
-\overleftarrow{\del_P}\overrightarrow{\del_X})} B_W,\\
&f(\h{x})_W=f(X),\qquad
g(\h{p})_W=g(P),
  \label{Wprod} 
\end{aligned}
\end{equation}
we find that the Wigner transform of the density matrix $\rho$ \eqref{rho-smallh} is given by
\begin{equation}
 \rho_W=\exp\left(\sum_{n=1}^\infty \frac{(-1)^n(\hbar/2)^{2n}}{(2n)!}E_{2n}(b)\text{Li}_{1-2n}(-e^{X})\right)\frac{1}{2\cosh\frac{X}{2}}\star\frac{1}{2\cosh\frac{P}{2}}.
\end{equation}
From \eqref{Wprod},
the Wigner transform of the $\ell^\text{th}$
power of $\rho$ is easily obtained by the star-product of
$\rho_W$'s
\begin{equation}
 (\rho^\ell)_W=\underbrace{ \rho_W \star\cdots\star \rho_W }_\ell.
\end{equation}
The WKB expansion of $(\rho^\ell)_W$ can be computed recursively
from the obvious relation
$(\rho^\ell)_W=\rho_W \star (\rho^{\ell-1})_W$.
Finally, the WKB expansion of  trace
$\Tr(\rho^\ell)$ can be found
by integrating $(\rho^\ell)_W$ on the classical phase space
\begin{equation}
 \Tr(\rho^\ell)=\int\frac{dXdP}{2\pi\hbar}(\rho^\ell)_W.
\end{equation}

As discussed in \cite{Hatsuda-zeta},
the WKB expansion of the spectral trace
$\Tr(\rho^s)$ takes the following form
\begin{equation}
 \Tr(\rho^s)=Z_0(s)\left(1+\sum_{n=1}^\infty D_n(s)\hbar^{2n}\right).
\end{equation}
The leading term is simply given by
\begin{equation}
 Z_0(s)=\int \frac{dXdP}{2\pi\hbar}\left(\frac{1}{2\cosh\frac{X}{2}}\frac{1}{2\cosh\frac{P}{2}}\right)^s=\frac{1}{2\pi\hbar}\frac{\Ga(s/2)^4}{\Ga(s)^2}.
\end{equation}
The correction terms $D_n(s)$ can be systematically computed
by making an ansatz that $D_n(s)$
is a rational function of $s$ and fixing the coefficients
in the ansatz by matching the values at integer $s$ \cite{Hatsuda-zeta,Okuyama:2015auc}.
In this way, we have computed $D_n(s)$ up to $n=10$.
The first three terms are
\begin{equation}
 \begin{aligned}
D_1(s)&=\frac{-s^3-12 s^2 \left(b-b^2\right)+s^2}{384 (s+1)},\\
D_2(s)&=\frac{s^3 }{1474560
   (s+1) (s+3)}\Big[7 s^3-20 s^2 (6 b^2-6 b-1),\\
&\qquad +s
   (720 b^4-1440 b^3+360 b^2+360 b-11)+16
   (120 b^4-240 b^3+120 b^2-1)\Big]\\
D_3(s)&=-\frac{s^3}{3963617280 (s+1) (s+3) (s+5)}
\Big[31 s^6-3 \left(196 b^2-196 b-115\right) s^5\\
&\quad+\left(5040 b^4-10080
   b^3-1344 b^2+6384 b+1273\right) s^4\\
&\quad+\left(-20160 b^6+60480 b^5-11760 b^4-77280
   b^3+24780 b^2+23940 b+1567\right) s^3\\
&\quad -16 \left(10080
   b^6-30240 b^5+21000 b^4+8400 b^3-6804 b^2-2436
   b+41\right) s^2\\
&\quad-768 \left(448
   b^6-1344 b^5+1120 b^4-189 b^2-35 b+3\right) s\\
&\quad -256
   \left(672 b^6-2016 b^5+1680 b^4-336
   b^2+1\right)\Big].
 \end{aligned}
\end{equation}
Then, the WKB expansion of grand potential can be found
by acting the differential operator on the leading term
\begin{equation}
 J(\mu,k,M)=\left(1+\sum_{n=1}^\infty D_n(\del_\mu)\hbar^{2n}\right)J_0(\mu,k,M),
\label{Ddelmu}
\end{equation}
where the leading term $J_0(\mu,k,M)$ is given by
\begin{equation}
 J_0(\mu,k,M)=-\int_\ga\frac{ds}{2\pi i}\Ga(s)\Ga(-s)Z_0(s)e^{s\mu}.
\end{equation}
The perturbative part comes from the pole at $s=0$
\begin{equation}
 J_0^\text{pert}(\mu,k,M)=-\oint_{s=0}\frac{ds}{2\pi i}\Ga(s)\Ga(-s)Z_0(s)e^{s\mu}
=\frac{2}{3\pi^2k}\mu^3+\frac{1}{3k}\mu+\frac{2\zeta(3)}{\pi^2k}.
\label{J0pert}
\end{equation}
To find the $\hbar$ corrections to the perturbative grand potential,
it is sufficient to expand $D_n(s)$ up to
$s^3$, since $J_0^\text{pert}(\mu,k,M)$ is a cubic polynomial in $\mu$.
In the small $s$ expansion, we find
that $D_n(s)$ behaves as
\begin{equation}
\begin{aligned}
 D_n(s)&=\frac{1-12b+12b^2}{384}s^2\cob_{n,1}\\
&+\frac{(-1)^ns^3}{16(2n)!}
\Big[4^{1-n}B_{2n-2}B_{2n}+ B_{2n-2}(B_{2n}(b)-B_{2n})\Big]+\mathcal{O}(s^4). 
\end{aligned}
\end{equation}
Here $B_{2n}(b)$ and $B_{2n}=B_{2n}(0)$ denote the Bernoulli polynomial and
the Bernoulli number, respectively.
By acting the differential operator \eqref{Ddelmu} on \eqref{J0pert},
we find that the constant $C(k)$ and $B(k,M)$ in \eqref{CB-ABJ} are
correctly reproduced. For instance, the $\mu$-linear term in $J^\text{pert}$
becomes
\begin{equation}
\frac{1}{3k}\mu+\hbar^2\frac{1-12b+12b^2}{384}\del_\mu^2\left(\frac{2\mu^3}{3\pi^2k}\right)
=\left[\frac{1}{3k}-\frac{k}{12}+\frac{k}{2}\left(\hf-b\right)^2\right]\mu,
\end{equation}
which correctly reproduces  $B(k,M)$ in \eqref{CB-ABJ}.
Also, we can read off the WKB expansion of constant term $A(k,M)$
\begin{equation}
\begin{aligned}
 A(k,M)&=\frac{2\zeta(3)}{\pi^2k}+\frac{1}{2\pi\hbar}\sum_{n=1}^\infty 
\frac{4^{1-n}(-1)^n B_{2n-2}B_{2n}\hbar^{2n}}{(2n)!}\\
&\qquad +
\frac{1}{2\pi\hbar}\sum_{n=1}^\infty \frac{(-1)^n B_{2n-2}(B_{2n}(b)-B_{2n})\hbar^{2n}}{(2n)!}. 
\end{aligned}
\label{Awkb}
\end{equation}
The first line of \eqref{Awkb} agrees with the constant term in ABJM theory \cite{KEK}
\begin{equation}
 A_c(k)=\frac{2\zeta(3)}{\pi^2k}+\sum_{n=1}^\infty \frac{(-1)^n B_{2n-2}B_{2n}\pi^{2n-2}k^{2n-1}}{(2n)!}.
\end{equation} 
As we will see below,
the second line in \eqref{Awkb} corresponds to the WKB expansion of 
the free energy $F_\text{coni}(k,M)$
of  the $U(M)_k$ pure Chern-Simons theory.
 To see this, let us consider the WKB expansion of $F_\text{coni}(k,M)$.
In terms of $\hbar=2\pi k$ \eqref{hbar-def}, $F_\text{coni}(k,M)$ in \eqref{Fconi-int} is rewritten as
\begin{equation}
  F_\text{coni}(k,M)
=-\left(\frac{\hbar}{4\pi^2}\right)^2F_0(b)+\frac{\hbar}{4\pi^2}\int_0^\infty
dy \frac{\hbar y}{e^{\hbar y}-1}\log\left(\frac{\cosh 2\pi y-\cos2\pi b}{\cosh 2\pi y-1}\right).
\end{equation}
By expanding the integrand in $\hbar$
\begin{equation}
 \frac{\hbar y }{e^{\hbar y}-1}=1-\frac{\hbar y}{2}+
\sum_{n=2}^\infty \frac{B_{2n-2}}{(2n-2)!}(\hbar y)^{2n-2},
\end{equation}
and making use of the identities
\begin{equation}
\begin{aligned}
\int_0^\infty
dy y\log\left(\frac{\cosh 2\pi y-\cos2\pi b}{\cosh 2\pi y-1}\right)
&=-\frac{1}{2\pi^2}F_0(b),\\
 \int_0^\infty\frac{dy}{2\pi}
y^{2n-2}\log\left(\frac{\cosh 2\pi y-\cos2\pi b}{\cosh 2\pi y-1}\right)&=
\frac{(-1)^n( B_{2n}(b)-B_{2n})}{2n(2n-1)},\quad(n\geq2),
\end{aligned}
\end{equation}
one can see that the small $\hbar$ expansion of $F_\text{coni}(k,M)$ indeed agrees with the second line 
of \eqref{Awkb}. To summarize, the perturbative grand potential of ABJ theory with the coefficients
$C(k), B(k,M)$ in \eqref{CB-ABJ} and $A(k,M)$ in \eqref{A-ABJ}
are correctly reproduced from the WKB expansion of the spectral trace $\Tr(\rho^s)$.

\subsection{WKB expansion of twisted spectral trace $\Tr(\rho^sR)$}
In a similar manner as in the previous subsection
\ref{sub:wkb-bulk}, we can compute the WKB expansion of the twisted spectral trace $\Tr(\rho^sR)$.
This can be done using the fact that $\Tr(AR)$ for some operator $A$
is easily obtained from its Wigner transform $A_W$ by simply setting $X=P=0$ \cite{Okuyama:2015auc} 
\begin{equation}
 \Tr(AR)=\hf A_W\Big|_{X=P=0}.
\end{equation}
We find that the WKB expansion of $\Tr(\rho^sR)$
is written as
\begin{equation}
 \Tr(\rho^sR)=Z_0^R(s)\left(1+\sum_{n=1}^\infty D_n^R(s)\hbar^{2n}\right).
\end{equation}
The leading term is given by
\begin{equation}
 Z_0^R(s)=2^{-1-2s}.
\end{equation}
We have computed the correction terms $D_n^R(s)$ up to $n=12$.
The first three terms are given by
\begin{equation}
 \begin{aligned}
  D_1^R(s)&=-\frac{s^2}{128}+\frac{1}{32} s (b-1) b,\\
D_2^R(s)&=\frac{5 s^4}{98304}+\frac{s^3 \left(-2 b^2+2
   b+1\right)}{8192}+\frac{s^2 \left(6 b^4-12 b^3+6
   b+1\right)}{12288}+\frac{s \left(b^4-2
   b^3+b\right)}{3072},\\
D_3^R(s)&=-\frac{61
   s^6}{188743680}+\frac{\left(5 b^2-5 b-7\right) s^5}{3145728}+\frac{\left(-18
   b^4+36 b^3+30 b^2-48 b-29\right)
   s^4}{4718592}\\
&\quad+\frac{\left(2 b^6-6 b^5-b^4+12
   b^3+4 b^2-11 b-3\right) s^3}{393216}+\frac{\left(6 b^6-18 b^5+30 b^3+3 b^2-21
   b-2\right) s^2}{589824}\\
&\quad+\frac{b
   \left(b^5-3 b^4+5 b^2-3\right) s}{184320}.
 \end{aligned}
\end{equation}

Again, the large $\mu$ expansion of $J^R(\mu,k,M)$
is found by acting differential operators on the leading term 
\begin{equation}
 J^R(\mu,k,M)=\left(1+\sum_{n=1}^\infty D_n^R(\del_\mu)\hbar^{2n}\right)J_0^R(\mu,k,M),
  \label{JR-exp}
\end{equation}
where the leading term $J_0^R(\mu,k,M)$
is given by
\begin{equation}
 J_0^R(\mu,k,M)=-\int_\ga\frac{ds}{2\pi i}\Ga(s)\Ga(-s)Z_0^R(s)e^{s\mu}.
\label{JR0-int}
\end{equation}

\paragraph{Perturbative part.}
Let us first consider the
perturbative part of $J^R(\mu,k,M)$, which comes from the pole at $s=0$.
The leading term is give by
\begin{equation}
 \hf J_0^{R,\text{pert}}(\mu,k,M)=-\hf \oint_{s=0}\frac{ds}{2\pi i}\Ga(s)\Ga(-s)Z_0^R(s)e^{s\mu}
=\frac{\mu}{4}-\hf \log2.
\label{JR-pert}
\end{equation}
This is consistent with the $\pm 1/4$ shift of $B_\pm(k,M)$ in \eqref{CBpm}.
As in the previous subsection \ref{sub:wkb-bulk},
the contribution $F_\text{non-ori}$ in \eqref{Apm} is obtained by summing all order corrections
in $\hbar$. Since \eqref{JR-pert} is a linear function in $\mu$,
in order to compute the term $F_\text{non-ori}$ it is sufficient
to expand $D_n^R(s)$ up to the linear order in $s$.
We find that $D_n^R(s)$ behaves as
\begin{equation}
 D_n^R(s)= \hf\frac{(-1)^n E_{2n-1}(0)E_{2n}(b)}{(2n)! 2^{2n}}s+\mathcal{O}(s^2).
\label{DnR-s}
\end{equation}
As we will see below, this
 indeed reproduces the WKB expansion of $F_\text{non-ori}$ in \eqref{Fnon-ori}.
 By integration by parts and rescaling of integral variable,
 \eqref{Fnon-ori} is rewritten as
\begin{equation}
 F_\text{non-ori}=\frac{\hbar}{8\pi^2}\text{Im}\Big[\text{Li}_2(e^{-i \pi b})-\text{Li}_2(-e^{-i \pi b})\Big]
+\qu\log2
-\int_0^\infty dx \log(1+e^{-\frac{\hbar x}{2}})
\frac{\sin\pi b\cosh\pi x}{\cosh 2\pi x-\cos2\pi b}.
\end{equation}
By expanding the integrand $\log(1+e^{-\frac{\hbar x}{2}})$ in small $\hbar$
\begin{equation}
 \log(1+e^{-\frac{\hbar x}{2}})=\log2-\frac{\hbar x}{4}
-\hf\sum_{n=1}^\infty\frac{E_{2n-1}(0)}{(2n)!}\Bigl(\frac{\hbar x}{2}\Bigr)^{2n},
\end{equation}
and using the formula
\begin{equation}
 \begin{aligned}
\int_0^\infty dx
\frac{x^{2n}\sin\pi b\cosh\pi x}{\cosh 2\pi x-\cos2\pi b}&=
\frac{(-1)^n}{4}E_{2n}(b),\quad(n\geq0),\\
\int_0^\infty dx\frac{x\sin\pi b\cosh\pi x}{\cosh 2\pi x-\cos2\pi b}&=-\frac{1}{2\pi^2}\text{Im}\Big[\text{Li}_2(e^{-i \pi b})-\text{Li}_2(-e^{-i \pi b})\Big],
 \end{aligned}
\end{equation}
the small $\hbar$ expansion of $F_\text{non-ori}$
with fixed $b$ is found to be
\begin{equation}
 F_\text{non-ori}= \frac{1}{8}\sum_{n=1}^\infty \frac{(-1)^n E_{2n-1}(0)E_{2n}(b)}{(2n)!2^{2n}}\hbar^{2n}.
  \label{Fnon-WKB}
\end{equation}
We can compare this with the WKB expansion of twisted spectral trace
in \eqref{DnR-s}.
By acting the differential operator $D_n(\del_\mu)$
\eqref{DnR-s} on the leading term $J_0^{R,\text{pert}}$ in \eqref{JR-pert},
one can easily show that
the constant term of the perturbative
part of $\hf J^R$ agrees with
 the WKB expansion of $F_\text{non-ori}$ in \eqref{Fnon-WKB}
\begin{equation}
  \hf J^{R,\text{pert}}= \hf\left(1+\sum_{n=1}^\infty D_n^R(\del_\mu)\hbar^{2n}\right)J_0^{R,\text{pert}}
=\frac{\mu}{4}-\hf\log2 +F_\text{non-ori}.
\end{equation}
This correctly reproduces the shift in $A_\pm(k,M)$ in \eqref{Apm}.

\paragraph{Instanton corrections.}
$J^R(\mu,k,M)$ can be written as a sum of the
perturbative part and
instanton corrections
\begin{equation}
 J^R(\mu,k,M)= J^{R,\text{pert}}(\mu,k,M)+J^{R,\text{np}}(\mu,k,M).
\end{equation}
By picking up the poles at negative integers $s=-\ell~(\ell=1,2,\cdots)$ in \eqref{JR0-int},
we can compute the non-perturbative correction $J^{R,\text{np}}(\mu,k,M)$, which we identify as the
``half-instanton'' corrections in \eqref{3-type}.
From \eqref{JR-exp} and \eqref{JR0-int},
one can show that the instanton corrections are given by
\begin{equation}
 \begin{aligned}
  J^{R,\text{np}}(\mu,k,M)&=\sum_{\ell=1}^\infty r_\ell(\hbar,b)e^{-\ell\mu},\\
  r_\ell(\hbar,b)&=\left(1+\sum_{n=1}^\infty D_n^R(-\ell)\hbar^{2n}\right)
\frac{(-1)^{\ell-1}2^{-1+2\ell}}{\ell}.
 \end{aligned}
\end{equation}
By matching the WKB expansion, we find the first few instanton coefficients $r_\ell(\hbar,b)$
in closed forms
\begin{equation}
 \begin{aligned}
  r_1(\hbar,b)&=2\cos\th_b,\\
r_2(\hbar,b)&=-2(\cos2\th_b+\cos2\th_0),\\
r_3(\hbar,b)&=\frac{4}{3}(\cos3\th_b+3\cos\th_b\cos^22\th_0),\\
r_4(\hbar,b)&=-4(\cos4\th_b+4\cos2\th_b\cos^32\th_0+\cos^24\th_0+2\cos^22\th_0),\\
r_5(\hbar,b)&=\frac{32}{5}(\cos5\th_b+5\cos3\th_b\cos^42\th_0+5\cos\th_b\cos^22\th_0(1+\cos^24\th_0)),\\
r_6(\hbar,b)&=-\frac{32}{3}\Bigl(\cos6\th_b+6\cos4\th_b\cos^52\th_0+3\cos2\th_b\frac{\cos^54\th_0-1}{\cos4\th_0-1}
\\
&\qquad+\cos^32\th_0(6\cos^34\th_0-3\cos^24\th_0+3\cos4\th_0+4)\Bigr),\\
r_7(\hbar,b)&=\frac{128}{7}\Bigl(\cos7\th_b+7\cos5\th_b\cos^62\th_0+7\cos3\th_b\cos^22\th_0(\cos^24\th_0+2\cos^22\th_0)(\cos^24\th_0+2\sin^22\th_0)\\
&\qquad+7\cos\th_b\cos^42\th_0(4\cos^44\th_0-4\cos^34\th_0+3\cos^24\th_0+2)\Bigr).
 \end{aligned}
\label{rell}
\end{equation}
Here we have introduced the notation $\th_b$ and $\th_0$ by
\begin{equation}
 \th_b=\frac{\hbar(1-2b)}{8},\qquad \th_0=\frac{\hbar}{8}.
\label{theta-b}
\end{equation}
When $k$ and $M$ are both integers,
$r_\ell(\hbar,b)$ becomes simpler and 
we can easily guess the instanton coefficients $r_\ell(\hbar,b)$ for all order in $\ell$.
From \eqref{theta-b} and the periodicity of trigonometric functions,
 the result depends on the value of $k$ modulo 8.
For odd $M$ we find
\begin{equation}
 \begin{aligned}
  J^{R,\text{np}}(k\equiv1,3~\text{mod}8)=-J^{R,\text{np}}(k\equiv5,7~\text{mod}8)
&=\qu\log\left(\frac{1+2\rt{2}(-1)^{\frac{M-1}{2}}e^{-\mu}+4e^{-2\mu}}{1-2\rt{2}(-1)^{\frac{M-1}{2}}e^{-\mu}+4e^{-2\mu}}\right),\\
J^{R,\text{np}}(k\equiv2~\text{mod}8)=-J^{R,\text{np}}(k\equiv6~\text{mod}8)
&=\qu\log\frac{1+4(-1)^{\frac{M-1}{2}}e^{-\mu}}{1-4(-1)^{\frac{M-1}{2}}e^{-\mu}},\\
J^{R,\text{np}}(k\equiv0,4~\text{mod}8)
&=0,
\label{JR-Modd}
 \end{aligned}
\end{equation}
while for even $M$ we find
\begin{equation}
 \begin{aligned}
  J^{R,\text{np}}(k\equiv1,7~\text{mod}8)=-J^{R,\text{np}}(k\equiv3,5~\text{mod}8)
&=\qu\log\left(\frac{1+2\rt{2}(-1)^{\frac{M}{2}}e^{-\mu}+4e^{-2\mu}}{1-2\rt{2}(-1)^{\frac{M}{2}}e^{-\mu}+4e^{-2\mu}}\right),\\
J^{R,\text{np}}(k\equiv2,6~\text{mod}8)
&=\qu\log(1+16e^{-2\mu}),\\
J^{R,\text{np}}(k\equiv0~\text{mod}8)
&=\hf\log(1+4(-1)^{\frac{M}{2}}e^{-\mu}),\\
J^{R,\text{np}}(k\equiv4~\text{mod}8)
&=\hf\log(1-4(-1)^{\frac{M}{2}}e^{-\mu}).
 \end{aligned}
\label{JR-Meven}
\end{equation}
For $M=1$, \eqref{JR-Modd} reproduces the results of
half-instantons in $\mathcal{N}=5$
$O(2N)_{2k}\times USp(2N)_{-k}$ theory found
in \cite{MS}.
Also, for all cases in Appendix \ref{app:Jpm-inst}
with various integral $k$ and $M$,
the difference between the modified grand potentials
$J_{+}$ and $J_{-}$
agrees with the results in \eqref{JR-Modd} and \eqref{JR-Meven}
\begin{equation}
 J_{+}^\text{np}(\mu,k,M)-J_{-}^\text{np}(\mu,k,M)=J^{R,\text{np}}(\mu,k,M).
\end{equation}

\section{Exact grand partition functions for $k=2$}
\label{sec:k=2}
In this section, we will write down the exact grand partition functions of orientifold ABJ theory $\Xi_\pm(\mu,k,M)$ for $k=2$
with $M=0,1$. 
In general, in order to find the exact grand partition function,
first we need to find the closed form expression of the
modified grand potential.
Then, by performing the periodic sum \eqref{periodic-sum},
we can construct the grand partition function.
It turns out that for $k=2$ the
modified grand potential is determined by the genus-zero and genus-one
free energies of topological string,
and the exact grand partition function
is written in terms of Jacobi theta functions,
as in \cite{CGM,GHM}.
The genus-zero free energy of ABJ(M) theory is encoded in the
classical periods of local $\mathbb{P}^1\times \mathbb{P}^1$,
which we review first in the next subsection.
Then we proceed to write down the exact grand partition functions
of orientifold ABJ theory for $k=2$.

\subsection{Periods of diagonal local $\mathbb{P}^1\times \mathbb{P}^1$}\label{sec:period}
In this subsection, we summarize the known results of classical periods
of the spectral curve of ABJM matrix model, or the mirror curve of  diagonal local $\mathbb{P}^1\times \mathbb{P}^1$ \cite{DMP,GHM}.
The spectral curve of ABJM matrix model has genus one, hence there are two independent periods:
A-period and B-period.
These two periods  are
characterized by the Picard-Fuchs equation
\begin{equation}
[\th^3-4z\th(2\th+1)^2]\Pi=0,
\label{PF}
\end{equation}
where
\begin{equation}
 \th=z\del_z,\quad z=e^{-2\mu}.
\end{equation}
To study the small $z$ expansion and the large $z$ expansion of periods
systematically,
it is useful to write the solution of \eqref{PF} in a Mellin-Barnes type integral representation.
By plugging the ansatz
\begin{equation}
 \Pi=\int\frac{ds}{2\pi i}f(s)z^s,
\end{equation}
into \eqref{PF},
we find that $f(s)$ should satisfy
\begin{equation}
 s^3f(s)-2f(s-1)\frac{\Ga(2s)^2}{\Ga(2s-2)\Ga(2s-1)}=0.
\end{equation}
One can easily find two independent solutions:
\begin{equation}
  f_A(s)=-\frac{4s\Ga(-s)\Ga(2s)^2}{\Ga(s+1)^3},\quad
f_B(s)=-\frac{\Ga(-s)^3\Ga(2s)}{\Ga(s+1)\Ga(-2s)}.
\end{equation}
In the large radius frame (small $z$ region),
these two solutions correspond to the flat coordinate $t$
(A-period)
and the derivative of the genus-zero free energy (B-period)
\begin{equation}
  \int_{\ga'}\frac{ds}{2\pi i}f_A(s)z^s=t,\qquad
\int_{\ga'}\frac{ds}{2\pi i}f_B(s)z^s=\del_t F_0+\frac{\pi^2}{6},
\label{defperi}
\end{equation}
where the integration contour $\ga'$ is taken to be
parallel to the imaginary $s$-axis with
$-1/2<\Re(s)<0$.
The small $z$ expansion of those periods are easily obtained by
picking up the poles at $s=0,1,2,\cdots$, 
\begin{equation}
 \begin{aligned}
t&=-\sum_{n=0}^\infty \text{Res}_{s=n}\Big[f_A(s)z^s\Big]=  -\log z+w_a,\\
\del_t F_0+\frac{\pi^2}{6}&=-\sum_{n=0}^\infty \text{Res}_{s=n}\Big[f_B(s)z^s\Big]
=\hf(t^2-w_a^2+w_b).
\label{tdeltF}
 \end{aligned}
\end{equation}
Note that the constant $\pi^2/6$ in the B-period comes from the pole at $s=0$
\begin{equation}
 -\text{Res}_{s=0}\Big[f_B(s)z^s\Big]=\hf (\log z)^2+\frac{\pi^2}{6}.
\end{equation}
$w_a$ and $w_b$ in \eqref{tdeltF} are given by
\begin{equation}
 \begin{aligned}
  w_a&=-\sum_{n=1}^\infty \frac{1}{n}\left(\frac{\Ga(n+\hf)}{\Ga(\hf)n!}\right)^2(-16z)^n,\\
w_b&=\sum_{n=1}^\infty \frac{8}{n}\left(\frac{\Ga(n+\hf)}{\Ga(\hf)n!}\right)^2\left[\psi(2n)-\psi(n)
-\frac{3}{4n}\right](-16z)^n,
 \end{aligned}
\end{equation}
where $\psi(n)$ denotes the digamma function.
The A-period and the B-period in \eqref{tdeltF} can be written in closed forms
in terms of a hypergeometric function and a Meijer $G$-function, respectively
\begin{equation}
 \begin{aligned}
  t&=-\log z+4z{}_4F_3\left(1,1,\frac{3}{2},\frac{3}{2};2,2,2;-16z\right),\\
\del_t F_0+\frac{\pi^2}{6}&=\frac{1}{\pi}G^{3,2}_{3,3}\lf(\begin{matrix}\hf,&\hf,&1\\
					       0,&0,&0
					      \end{matrix}\Big|-16z\ri)
+\pi it.
 \end{aligned}
\label{tFinhyper}
\end{equation}
From these relations, one can find the genus-zero free energy $F_0(t)$ as a function of $t$.
The integration constant in  $F_0(t)$ is found to be $-2\zeta(3)$, and the large $t$
expansion of $F_0(t)$ is given by\footnote{
One could absorb the constant $\pi^2/6$ into the definition of $F_0(t)$
\[
 \del_t\h{F}_0=\del_t F_0+\frac{\pi^2}{6},\qquad
\h{F}_0=\frac{t^3}{6}+\frac{\pi^2}{6}t-2\zeta(3)+\mathcal{O}(e^{-t}).
\]
However, we will not do so and we will stick to the convention in \cite{GHM}.
Note that the $t$-linear term in $F_0(t)$ drops out in the combination
of $f_0$ in \eqref{f0}. On the other hand,
this constant $\pi^2/6$ plays an important role
in $\xi_L$ \eqref{xiLdef} and the modular property of the grand partition function.
}
\begin{equation}
 F_0(t)=\frac{t^3}{6}-2\zeta(3)+4e^{-t}-\frac{9}{2}e^{-2t}+\frac{328}{27}e^{-3t}
-\frac{777}{16}e^{-4t}+\frac{30004}{125}e^{-5t}+\cdots.
\end{equation}
It turns out that the logarithmic derivatives of $t$ and $\del_tF_0$ with respect to $z$
are simpler than $t$ and $\del_tF_0$ themselves,
and they are given by the complete elliptic integral
of the first kind\footnote{We follow the convention of elliptic integral in {\tt Mathematica}
\[
 K(k^2)=\int_0^1 \frac{dt}{\rt{(1-t^2)(1-k^2t^2)}}.
\]
}
\begin{equation}
 \begin{aligned}
-z\del_z t&=\frac{2}{\pi}K(-16z),\\
-z\del_z\del_t F_0(t)&=2K(1+16z)+2iK(-16z).
\label{logdel}
 \end{aligned}
\end{equation}
Taking the ratio of them, we find the modulus of the spectral curve
\begin{equation}
 \tau_L=\frac{i}{\pi}\del_t^2 F_0(t)=
i\frac{K(1+16z)}{K(-16z)}-1.
\label{tauL}
\end{equation}
Here the subscript $L$ stands for the ``large radius frame''.

On the other hand, by deforming the contour $\ga'$ of \eqref{defperi}
in the direction
$\Re(s)<0$ and
picking up the poles at $s=-(n+1/2),~(n=0,1,2,\cdots)$,
we find the small $\ka$
expansion of the periods, where $\ka$ is defined by
\begin{equation}
 \ka=e^\mu,\quad z=\ka^{-2}.
\end{equation}
Then the two periods in \eqref{defperi} become
\begin{equation}
 \begin{aligned}
  \int\frac{ds}{2\pi i}f_A(s)z^s&= \sum_{n=0}^\infty
\text{Res}_{s=-(n+1/2)}\Big[f_A(s)\ka^{-2s}\Big]=
\frac{1}{\pi^2}\del_\la\mathcal{F}_0,\\
\int\frac{ds}{2\pi i}f_B(s)z^s&=\sum_{n=0}^\infty
\text{Res}_{s=-(n+1/2)}\Big[f_B(s)\ka^{-2s}\Big]=
4\pi^2\la.
\label{ladella}
 \end{aligned}
\end{equation}
By performing the sum of residues in \eqref{ladella},
we find that the flat coordinate $\la$ and the genus-zero free energy
$\mathcal{F}_0(\la)$ in the ``orbifold frame'' are given by
\begin{equation}
 \begin{aligned}
  \la&=\frac{\ka}{8\pi}{}_3F_2\lf(\hf,\hf,\hf;1,\frac{3}{2};-\frac{\ka^2}{16}\ri),\\
\del_\la\mathcal{F}_0&=\frac{\ka}{4}G^{2,3}_{3,3}\lf(\begin{matrix}\hf,&\hf,&\hf\\
					       0,&0,&-\hf
					      \end{matrix}\Big|-\frac{\ka^2}{16}\ri)
+4\pi^3i\la.
 \end{aligned}
\end{equation}
The small $\la$ expansion of the free energy $\mathcal{F}_0(\la)$ reads
\begin{equation}
 \mathcal{F}_0(\la)=2\pi^2\la^2\lf(3-2\log\frac{\pi\la}{2}\ri)+\frac{4\pi^4\la^4}{9}+\cdots.
\end{equation}
In this case, the ordinary derivatives of $\la$ and
$\del_\la\mathcal{F}_0$ with respect to $\ka$
become the complete elliptic integral
\begin{equation}
\begin{aligned}
 8\pi\del_\ka\la&=\frac{2}{\pi}K\Bigl(-\frac{\ka^2}{16}\Bigr),\\
\del_\ka\del_\la \mathcal{F}_0&=\pi K\Bigl(1+\frac{\ka^2}{16}\Bigr)+ \pi i K\Bigl(-\frac{\ka^2}{16}\Bigr),
\end{aligned}
\label{odel}
\end{equation}
and the torus modulus in
the orbifold frame is given by
\begin{equation}
 \tau_\text{orb}=\frac{i}{4\pi^3}\del_\la^2 \mathcal{F}_0=i\frac{K\Bigl(1+\frac{\ka^2}{16}\Bigr)}{K\Bigl(-\frac{\ka^2}{16}\Bigr)}-1.
\label{tauorb}
\end{equation}
As we will see below, the difference of the derivatives between \eqref{logdel} and \eqref{odel}
plays an important role in the modular covariance of the
grand partition functions.

Since \eqref{defperi} and \eqref{ladella} are just the different expansions of the same
integrals, they are related by\footnote{The second relation in \eqref{Pirel} can be rewritten as
\[
 \del_t F_0=4\pi^2\h{\la},\qquad
\h{\la}=\la-\frac{1}{24}.
\]
As observed in \cite{DMP}, this shifted 't Hooft parameter $\h{\la}$
naturally appears in the large $\la$ expansion of the free energy.
However, we will not use $\h{\la}$ in this paper.  
}
\begin{equation}
\begin{aligned}
 &t=\frac{\del_\la\mathcal{F}_0}{\pi^2},\qquad
\del_t F_0+\frac{\pi^2}{6}=4\pi^2\la.
\label{Pirel}
\end{aligned}
\end{equation}
In other words, the role of A-cycle and B-cycle are essentially exchanged
when going from the large radius frame to the orbifold frame\footnote{We should emphasize that the $s$-plane
in the Mellin-Barnes representation is an auxiliary object, and
the contour $\ga'$ on the $s$-plane has nothing to do with the A-cycle and B-cycle on the mirror curve of local 
$\mathbb{P}^1\times\mathbb{P}^1$.}.
In fact, $\tau_L$ \eqref{tauL} and $\tau_\text{orb}$ \eqref{tauorb} are related by the $S$-transformation
\begin{equation}
 \tau_\text{orb}=-\frac{1}{\tau_L}.
\end{equation}
Finally, the genus-zero free energy of the large radius frame $F_0$ and
the orbifold frame $\mathcal{F}_0$ are related by
\begin{equation}
 F_0+\frac{\pi^2 t}{6}=\int d\ka \del_\ka t \left(\del_t F_0+\frac{\pi^2}{6}\right)=
\int d\ka \frac{\del_\ka \del_\la\mathcal{F}_0}{\pi^2}
\cdot 4\pi^2\la=-4(\mathcal{F}_0-\la\del_\la \mathcal{F}_0).
\end{equation}

\subsection{The case of $(k,M)=(2,0)$}
In this subsection, we write down the closed form expression of the grand partition functions
$\Xi_\pm(\mu,2,0)$ of orientifold ABJ theory. 
As in \cite{CGM, GHM}, we find that $\Xi_\pm(\mu,2,0)$ are written in terms of the
Jacobi theta functions. Also, they have a nice property under modular transformation, which
enables us to find the expansion around the orbifold point $\ka\to0$ and extract the 
canonical partition function at fixed $N$.

\subsubsection{$\Xi(\mu,2,0)$}
\label{sub:Xi20}
Let us recall  the grand partition function $\Xi(\mu,2,0)$ of ABJM theory \cite{CGM} before projection to $R=\pm1$.
In \cite{CGM}, it was found that the modified grand potential is essentially determined by
the genus-zero and genus-one data of the (refined) topological string on local $\mathbb{P}^1\times\mathbb{P}^1$.
The large $\mu$ expansion of the modified grand potential is given by \cite{HMO2}
\begin{equation}
 \begin{aligned}
 J(\mu,2,0)&=\frac{\mu^3}{3\pi^2}+\frac{\mu}{4}-\frac{\zeta(3)}{2\pi^2}\\
&+\left[\frac{4\mu^2+2\mu+1}{\pi^2}\ri]e^{-2\mu}
+\left[-\frac{52\mu^2+\mu+9/4}{2\pi^2}+2\right]e^{-4\mu}\\
&+\left[\frac{736\mu^2-304/3\mu+154/9}{3\pi^2}-32\right]e^{-6\mu}\\
&+\left[-\frac{2701\mu^2-13949/24\mu+11291/192}{\pi^2}+466\right]e^{-8\mu}+\cdots,
\label{J20}
\end{aligned}
\end{equation}
where the first line corresponds to the perturbative part,
while the remaining terms are instanton corrections.
As observed in \cite{CGM},
the terms with and without $1/\pi^2$ coefficient, 
which we call $f_0$ and $f_1$, respectively, have different origins.
Accordingly, it is natural to decompose the modified grand potential as
\begin{equation}
 J(\mu,2,0)=f_0+f_1,
\end{equation}
where
\begin{equation}
 \begin{aligned}
  f_0&=\frac{\mu^3}{3\pi^2}-\frac{\zeta(3)}{2\pi^2}+\frac{4\mu^2+2\mu+1}{\pi^2}e^{-2\mu}
-\frac{52\mu^2+\mu+9/4}{2\pi^2}e^{-4\mu}+\cdots,\\
f_1&=\frac{\mu}{4}+2e^{-4\mu}-32e^{-6\mu}+466e^{-8\mu}+\cdots.
\label{f0f1expand}
 \end{aligned}
\end{equation}
It turns out that $f_0$ can be written in a closed form
in terms of the genus-zero free energy $F_0(t)$
\begin{equation}
 f_0=\frac{1}{4\pi^2}\left(F_0-t\del_t F_0 +\frac{t^2}{2}\del_t^2 F_0\right).
\label{f0}
\end{equation}
On the other hand, $f_1$ is given by the sum of the genus-one free energies
of the standard topological string
and the refined topological string in the Nekrasov-Shatashvili limit 
\begin{equation}
 f_1=\frac{\mu}{4}+F_1+F_1^\text{NS}
\label{f1}
\end{equation}
where the first term comes from  the perturbative
grand potential\footnote{Our definition of $f_1$ includes the perturbative term  $\mu/4$, which is different from the definition of $f_1$ in \cite{GHM}.}.
Explicit forms of $F_1$ and $F_1^\text{NS}$ are given by \cite{Nakajima:2003uh,Huang:2010kf,Aganagic:2006wq}
\begin{equation}
\begin{aligned}
 F_1&=-\frac{1}{12}\log[z(1+16z)]-\hf\log(-z\del_z t),\\
F_1^\text{NS}&=\frac{1}{12}\log z-\frac{1}{24}\log(1+16z),
\label{F1inz}
\end{aligned}
\end{equation}
and the sum of them becomes
\begin{equation}
 F_1+F_1^\text{NS}=
-\frac{1}{8}\log(1+16z)-\hf\log(-z\del_z t).
\label{F1+F1NS}
\end{equation}
Note that $-z\del_z t$ is given by the complete elliptic integral \eqref{logdel}.
One can easily check 
that \eqref{f0} and \eqref{F1+F1NS} correctly reproduce the large $\mu$ expansion 
\eqref{f0f1expand}.

\paragraph{Modular expression of genus-one part.}
The genus-one free energies $F_1$ and $F_1^\text{NS}$
have a nice expression as modular forms.
To see this,
it is convenient to introduce $\tau$ by
\begin{equation}
 \tau=\tau_L+1=i\frac{K(1+16z)}{K(-16z)}.
\end{equation}
Using the formula in Appendix \ref{app:theta},
various
quantities appearing in the genus-one free energies \eqref{F1inz} can be written in terms of $\tau$ 
\begin{equation}
 -16z=\frac{\vartheta_2(0,\tau)^4}{\vartheta_3(0,\tau)^4},
\quad
1+16z=\frac{\vartheta_4(0,\tau)^4}{\vartheta_3(0,\tau)^4},\quad
-z\del_z t=\vartheta_3(0,\tau)^2.
\label{zintheta}
\end{equation}
Plugging \eqref{zintheta} into \eqref{F1inz}, and using the
expression of Dedekind eta function $\eta(\tau)$ in
\eqref{id-eta},
we recover the well known expression
of the genus-one free energy of the standard topological string
\begin{equation}
 F_1=-\log\eta(\tau).
  \label{F1eta}
\end{equation}
In a similar manner, one can also show that\footnote{
\eqref{F1+F1NS} and \eqref{F1eta} imply
\[
 F_1^\text{NS}=-\log\left(\frac{\vartheta_4(0,2\tau)}{\eta(\tau)}\right)=-\log\left(q^{-\frac{1}{24}}\prod_{n=1}^\infty(1-q^{2n-1})\right),\quad(q=e^{2\pi i\tau}).
\]
This agrees with the result in \cite{Nakajima:2003uh} (note that $G$ in \cite{Nakajima:2003uh} is 
equal to our $-F_1^\text{NS}$).
}
\begin{equation}
F_1+ F_1^\text{NS}=-\hf\log\Big[\vartheta_3(0,\tau)\vartheta_4(0,\tau)\Big]=
-\log\vartheta_4(0,2\tau).
\label{F1FNS-theta}
\end{equation}

For later purposes, it is also convenient to rewrite 
the genus-one free energy in terms of the original variable $\tau_L$. We find
\begin{equation}
 F_1+ F_1^\text{NS}=-\log\vartheta_4(0,2\tau_L).
\label{F1FNSL}
\end{equation}
The first term $\mu/4$ in $f_1$ \eqref{f1} can also be written 
in terms of $\tau_L$ using \eqref{zintheta} 
\begin{equation}
 \frac{\mu}{4}=\hf\log\left(\frac{2\vartheta_4(0,\tau_L)}{\vartheta_2(0,\tau_L)}\right).
\label{mu/4L}
\end{equation}
From \eqref{mu/4L} and \eqref{F1FNSL}, we find
that $f_1$ is written as
\begin{equation}
 f_1=-\hf \log\Big[\vartheta_2(0,\tau_L)\vartheta_3(0,\tau_L)\Big]
+\hf\log2=-\log \vartheta_2\Bigl(0,\frac{\tau_L}{2}\Bigr)+\log2.
\label{f1theta}
\end{equation}

\paragraph{Exact grand partition function.}
Once we find the modified grand potential, we can
construct the grand partition function by summing over the
periodic shift $\mu\to\mu+2\pi in$
\eqref{periodic-sum}.
As noticed in \cite{CGM}, using the relation 
\begin{equation}
 \frac{n^3-n}{3}\in 2\mathbb{Z}\qquad(n\in\mathbb{Z}),
\label{n3n1mod}
\end{equation}
the $n^3$ term coming from the cubic term $\frac{\mu^3}{3\pi^2}$ in
$J^\text{pert}(\mu,2,0)$ can be rewritten as
a liner term in $n$
\begin{equation}
 \exp\left(\frac{(2\pi in)^3}{3\pi^2}\right)
  =\exp\left(-\frac{8\pi in}{3}\right)=
  \exp\left(-\frac{2\pi in}{3}\right).
  \label{n3exp}
\end{equation}
Then, the summand in \eqref{periodic-sum}
is written as
\begin{equation}
 e^{J(\mu+2\pi in,2,0)}=e^{J(\mu,2,0)+
  2\pi i n^2\tau_L+4\pi in\xi_L},
\end{equation}
and the periodic sum \eqref{periodic-sum}
becomes the Jacobi theta function
\begin{equation}
 \Xi(\mu,2,0)=e^{J(\mu,2,0)}\sum_{n\in\mathbb{Z}}e^{
  2\pi i n^2\tau_L+4\pi in\xi_L}=e^{J(\mu,2,0)}\vartheta_3(2\xi_L,2\tau_L),
\label{Xi20}
\end{equation}
where $\tau_L$ is defined in \eqref{tauL} and
$\xi_L$ is given by\footnote{$\tau$ and $\xi$ in \cite{CGM} and our $\tau_L$ and $\xi_L$ are related by
\[
 \tau=2\tau_L,\quad
\xi-\frac{1}{12}=2\xi_L.
\]
}
\begin{equation}
 \xi_L=\frac{t\del_t^2F_0-\del_t F_0}{4\pi^2}-\frac{1}{24}.
  \label{xiLdef}
\end{equation}
Note that the constant term $-1/24$ in \eqref{xiLdef}
comes from the $2\pi in$-shift of
$\mu/4$ in
$J^\text{pert}(\mu,2,0)$ and \eqref{n3exp}
\begin{equation}
 \hf\left(\qu-\frac{1}{3}\right)=-\frac{1}{24}.
\end{equation}
Using the expression of $f_1$ in \eqref{f1theta}, $\Xi(\mu,2,0)$ can also be written as
\begin{equation}
 \Xi(\mu,2,0)=e^{f_0+f_1}\vartheta_3(2\xi_L,2\tau_L)=
e^{f_0}\frac{2\vartheta_3(2\xi_L,2\tau_L)}{\vartheta_2\lf(0,\frac{\tau_L}{2}\ri)}.
\end{equation}

As we will see below, the constant $-1/24$ in 
\eqref{xiLdef} is necessary for the grand partition function to
behave properly under the modular transformation.
In fact, $\xi_L$ can be written as
\begin{equation}
 \xi_L=\frac{1}{4\pi^2}\left(t\del_t^2F_0-\del_t F_0-\frac{\pi^2}{6}\right),
\label{xiLinF0}
\end{equation}
and the last two terms are exactly the B-period in \eqref{defperi}.
One can rewrite $\xi_L$ in a form that the modular property
is more transparent. As in \cite{Eynard:2008he},
by introducing the
appropriately normalized A-period $\Pi_A$
and B-period $\Pi_B$ 
\begin{equation}
 \Pi_A=-\frac{t}{2},\qquad
  \Pi_B=\frac{1}{2\pi i}\left(\del_t F_0+\frac{\pi^2}{6}\right),
  \label{PiAB}
\end{equation}
$\xi_L$ in \eqref{xiLinF0} is written as
\begin{equation}
 \xi_L=\frac{1}{2\pi i}(\Pi_B-\tau_L\Pi_A),
\end{equation}
with
\begin{equation}
 \tau_L=\frac{d\Pi_B}{d\Pi_A}.
\end{equation}
Also, one can show that
$f_0$ can be written in terms of $\Pi_{A}$ and
 $\Pi_B$ in \eqref{PiAB} as
\begin{equation}
 f_0=\frac{1}{2\pi i}\Bigl(2\int d\Pi_A \Pi_B
  -2\Pi_A\Pi_B+\tau_L\Pi_A^2\Bigr).
\end{equation}
After integration by parts, we find a surprisingly simple expression for $f_0$
\begin{equation}
 f_0=\int \frac{d\tau_L}{2\pi i} \Pi_A^2.
  \label{f0inA}
\end{equation}
In a similar manner, we can also rewrite $\xi_L$ as
\begin{equation}
 \xi_L=-\int \frac{d\tau_L}{2\pi i} \Pi_A.
  \label{xiLinA}
\end{equation}
The integration constants in \eqref{f0inA} and \eqref{xiLinA} 
should be fixed so that the large $t$ expansions
of $f_0$ and $\xi_L$ are correctly
reproduced
\begin{equation}
 \begin{aligned}
  f_0&=\frac{t^3}{24\pi^2}-\frac{\zeta(3)}{2\pi^2}
+\frac{t^2+2t+2}{2\pi^2}e^{-t}-\frac{9(2t^2+2t+1)}{8\pi^2}e^{-2t}+\cdots,\\
\xi_L&=\frac{t^2}{8\pi^2}-\frac{1}{24}+\frac{t+1}{\pi^2}e^{-t}-\frac{9(2t+1)}{4\pi^2}e^{-2t}
+\cdots.
 \end{aligned}
\end{equation}

To understand the modular property of
grand partition function better,
it is desirable to write $f_0$ and $\xi_L$ explicitly as modular forms.
We find that the derivative of A-period with respect to $\tau_L$
can be written in a closed form
\begin{equation}
 \frac{\del\Pi_A}{\del\tau_L} =\frac{\pi i}{2}\vartheta_3(0,\tau_L)^4\vartheta_4(0,\tau_L)^2,
\label{dPiA}
\end{equation}
but we could not find a closed form expression of $\Pi_A$ itself.
We leave this as an interesting future problem.

\paragraph{$S$-transformation and orbifold expansion.}
As discussed in \cite{CGM,GHM},
in order to find the small $\ka$ expansion (orbifold expansion) of the
grand partition function,
we have to perform the modular $S$-transformation.
Indeed,
 the combination
 $e^{f_0+f_1}\vartheta_3(2\xi_L,2\tau_L)$
 is exactly the form of ``non-perturbative partition function''
 proposed in \cite{Eynard:2008he} which has a nice modular property.
 Using the $S$-transformation of theta function in \eqref{thetaS},
 we find
\begin{equation}
 \Xi(\mu,2,0)=e^{\b{f}_0+\b{f}_1}\vartheta_3\lf(\xi_\text{orb},\frac{\tau_\text{orb}}{2}\ri),
  \label{Xi20orb}
\end{equation}
where $\xi_\text{orb}$ and $\tau_\text{orb}$ are given by
\begin{equation}
 \xi_\text{orb}=\frac{\xi_L}{\tau_L},\qquad
  \tau_\text{orb}=-\frac{1}{\tau_L}
  = -\frac{d\Pi_A}{d\Pi_B},
  \label{xiorb}
\end{equation}
and $\b{f}_0$ and $\b{f}_1$ in \eqref{Xi20orb} are
given by
\begin{equation}
 \b{f}_0=f_0-\frac{2\pi i\xi_L^2}{\tau_L},\qquad
  \b{f}_1=f_1-\hf\log(-2i\tau_L).
  \label{barf0f1}
\end{equation}

One can easily show that $\xi_\text{orb}$ is written as
\begin{equation}
 \xi_\text{orb}=-\frac{1}{2\pi i}\Bigl(\Pi_A+\tau_\text{orb}\Pi_B\Bigr),
\end{equation}
where the two periods in the orbifold frame are given by
\begin{equation}
 \Pi_B=-2\pi i\la,\qquad
  \Pi_A=-\frac{\del_\la\mathcal{F}_0}{2\pi^2}.
\end{equation}
Also, one can show that $\b{f}_0$ and $\xi_\text{orb}$
can be recast in the same form as $f_0$ in \eqref{f0inA} and $\xi_L$
in \eqref{xiLinA}, as expected
\begin{equation}
 \b{f}_0=\int \frac{d\tau_\text{orb}}{2\pi i} \Pi_B^2,\qquad
  \xi_\text{orb}=-\int \frac{d\tau_\text{orb}}{2\pi i} \Pi_B,
\end{equation}
and the derivative of $\Pi_B$ with respect to $\tau_\text{orb}$
is given by the $S$-transformation of 
\eqref{dPiA}
\begin{equation}
 \frac{\del \Pi_B}{\del\tau_\text{orb}} =-\frac{\pi }{2}\vartheta_2(0,\tau_\text{orb})^2\vartheta_3(0,\tau_\text{orb})^4.
\end{equation}
In terms of the orbifold free energy $\mathcal{F}_0(\la)$, $\b{f}_0$ and $\xi_\text{orb}$ are written as
\begin{equation}
 \b{f}_0=-\frac{1}{\pi^2}\left(\mathcal{F}_0-\la\del_\la\mathcal{F}_0+\frac{\la^2}{2}
			  \del_\la^2\mathcal{F}_0\right),\qquad
 \xi_\text{orb}
 =\frac{i}{4\pi^3}\Bigl(\la\del_\la^2\mathcal{F}_0-\del_\la \mathcal{F}_0\Bigr).
\end{equation}

For the genus-one part, we find
\begin{equation}
 \b{f}_1=-\frac{1}{8}\log\Bigl(1+\frac{\ka^2}{16}\Bigr)-\hf\log(8\pi\del_\ka\la)
=-\log\vartheta_4(0,2\tau_\text{orb}).
\label{barf1}
\end{equation}
Note that $\del_\ka\la$ is given by the complete elliptic integral
\eqref{odel}. 
Interestingly, the difference of the logarithmic derivative  $-z\del_zt$ in \eqref{F1+F1NS}
and the ordinary derivative
$\del_\ka\la$ in \eqref{barf1}
is compensated by the perturbative term $\mu/4$ and the term $-\hf\log(-2i\tau_L)$
coming from the $S$-transformation of theta function
in \eqref{barf0f1}.
Also, it is interesting to observe that
$\b{f}_1$ in \eqref{barf1} has a similar form
as $F_1+F_1^\text{NS}$ in \eqref{F1+F1NS} and \eqref{F1FNSL}. 
Using the expression of $\b{f}_1$ as a theta function \eqref{barf1}, the grand partition function can also be written as
\begin{equation}
 \Xi(\mu,2,0)=
e^{\b{f}_0}\frac{\vartheta_3\lf(\xi_\text{orb},\frac{\tau_\text{orb}}{2}\ri)}{\vartheta_4(0,2\tau_\text{orb})}.
\end{equation}

The small $\ka$
expansion of the grand partition function
can be found from the expansion of $\xi_\text{orb}$,
$\tau_\text{orb}$ and $\b{f}_0$
\begin{equation}
 \begin{aligned}
 2\pi i\xi_\text{orb}&=\frac{\ka}{2\pi}-\frac{7\ka^3}{1152\pi}+\cdots,\\
\pi i\tau_\text{orb}&=2\log\frac{\ka}{16}-\frac{\ka^2}{32}+\frac{13\ka^4}{16384}+\cdots,\\
\b{f}_0&=\frac{\ka^2}{32\pi^2}-\frac{\ka^4}{1536\pi^2}+\cdots.
 \end{aligned}
\end{equation}
Putting all together, finally we find
\begin{equation}
 \Xi(\mu,2,0)=1+\frac{\ka}{8}+\frac{\ka^2}{32\pi^2}+\frac{10-\pi^2}{512\pi^2}\ka^3+\cdots.
\end{equation}
The coefficient of $\ka^N~(N=1,2,\cdots)$ correctly reproduces the canonical
partition function $Z(N,2,0)$ of
ABJM theory computed in \cite{HMO2}.

\subsubsection{$\Xi_\pm(\mu,2,0)$}
\label{sub:Xipm20}
In this subsection, we consider the closed form expression of $\Xi_\pm(\mu,2,0)$. 
For even $k$,
it is convenient to redefine the constants
in the perturbative part of grand potential as
\begin{equation}
 \til{A}(k,M)=A(k,M)+\frac{\zeta(3)}{\pi^2k},\qquad
  \til{A}_\pm(k,M)=A_\pm(k,M)+\frac{\zeta(3)}{2\pi^2k},
\label{tilA}
\end{equation}
so that $\til{A}(k,M)$ and $\til{A}_\pm(k,M)$ do not contain the term
proportional to $\zeta(3)$.

From the results of $J^\text{np}_\pm(\mu,2,0)$ \eqref{Jpm20} in Appendix \ref{app:Jpm-inst},
and the perturbative coefficients
\begin{equation}
\til{A}_\pm(2,0)=\mp\hf\log2,\qquad B_\pm(2,0)=\frac{1}{8}\pm\qu,
\end{equation}
we find the modified grand potentials  in closed forms
\begin{equation}
\begin{aligned}
 J_{\pm}(\mu,2,0)&=\hf f_0+F_1+F_1^\text{NS}+\frac{\mu}{8}\pm
  \left(\frac{\mu}{4}+\frac{1}{8}\log(1+16e^{-2\mu})-\hf\log2\right)\\
&=\hf f_0+f_1-\frac{\mu}{8}\pm
  \frac{1}{8}\log\lf(1+\frac{e^{2\mu}}{16}\ri),
\label{Jpm20closed}
\end{aligned}
\end{equation}
where $f_0$ and $f_1$ are defined in \eqref{f0} and \eqref{f1}, respectively.
Note that the constant proportional to $\zeta(3)$ is contained in $f_0/2$.
After finding the modified grand potential in a closed form,
we are ready to perform the periodic sum
$\eqref{periodic-sum}$ to construct the exact grand partition function.
As in the previous subsection \ref{sub:Xi20}, the $n^3$ term coming from the $\frac{\mu^3}{6\pi^2}$
term
in the perturbative part can be rewritten as a linear term in $n$ using \eqref{n3n1mod}
\begin{equation}
 \exp\lf(\frac{(2\pi in)^3}{6\pi^2}\ri)=\exp\lf(-\frac{4\pi in}{3}\ri)
=\exp\lf(\frac{2\pi in}{3}\ri).
\label{n3inn}
\end{equation}
Again, we find that $\Xi_\pm(\mu,k,M)$ is proportional to the
theta function $\vartheta_3(v,\tau_L)$, where the parameter $v$ is given by
\begin{equation}
 v=\lf(\xi_L+\frac{1}{24}\ri)+\frac{1}{3}+B_\pm(2,0)=\xi_L+\hf\pm\qu.
\label{vfor20pm}
\end{equation}
The first term $\xi_L+1/24$ in \eqref{vfor20pm} comes from the periodic shift of $f_0/2$,
while the second term
$1/3$ and the third term $B_\pm(2,0)$ come from \eqref{n3inn} and
the $B_\pm\mu$ term in $J^\text{pert}_\pm$, respectively.
Finally, we find the grand partition function in a closed form
\begin{equation}
 \Xi_\pm(\mu,2,0)=e^{J_\pm(\mu,2,0)}\vartheta_3\Bigl(
  \xi_L+\hf\pm\qu,\tau_L\Bigr),
\label{Xipm20}
\end{equation} 
which can also be written as
\begin{equation}
 \Xi_+(\mu,2,0)=e^{J_+(\mu,2,0)}\vartheta_4\Bigl(
  \xi_L+\qu,\tau_L\Bigr),
\quad
\Xi_-(\mu,2,0)=e^{J_-(\mu,2)}\vartheta_3\Bigl(
\xi_L+\qu,\tau_L\Bigr).
\end{equation}
We note in passing that all terms in $J_\pm(\mu,2,0)$ \eqref{Jpm20closed},
except for the genus-zero term $f_0/2$, can be written in terms of
Jacobi theta functions,
but the resulting expression is not so illuminating.
This is true for all other cases considered in this paper.

Using the identity  \eqref{id-landen} of theta functions 
in Appendix \ref{app:theta}, 
the product of $\Xi_{\pm}(\mu,2,0)$ becomes
\begin{equation}
 \Xi_{+}(\mu,2,0)\Xi_{-}(\mu,2,0)=e^{J_+(\mu,2,0)+J_-(\mu,2,0)}
\vartheta_4(0,2\tau_L)
\vartheta_3(2\xi_L,2\tau_L).
\label{20prod}
\end{equation}
This should be equal to the grand partition function $\Xi(\mu,2,0)$
of ABJM theory.
Comparing \eqref{20prod} and \eqref{Xi20},
we find the difference between $J_{+}+J_{-}$ and
$J$ comes from $\vartheta_4(0,2\tau_L)$
\begin{equation}
J_{+}(\mu,2,0)+J_{-}(\mu,2,0)-J(\mu,2,0)=-\log \vartheta_4(0,2\tau_L)=F_1+F_1^\text{NS}.
\end{equation}
From \eqref{J20} and \eqref{Jpm20},
one can easily check
the small $z$
expansion of $F_1+F_1^\text{NS}$ indeed agrees with
the difference of the modified grand potential.

\paragraph{Orbifold expansion.}
The small $\ka$ expansion of $\Xi_\pm(\mu,2,0)$
can be found by performing the $S$-transformation of 
theta function, as in the previous subsection \ref{sub:Xi20}.
This is easily done by noticing that
$\Xi_\pm(\mu,2,0)$ in \eqref{Xipm20}
can be written as a theta function with characteristic $(a,b)=(0,\mp\qu)$\footnote{As did in \cite{CGM,GHM}, one could find the small $\ka$ expansion of $\Xi_\pm(\mu,2,0)$ using the identity \eqref{id-4tau} of Jacobi theta functions, without using the theta function with characteristic.}
\begin{equation}
 \Xi_\pm(\mu,2,0)=e^{J_\pm(\mu,2,0)}\vartheta\Bigl[\begin{matrix} 0\\ \mp\qu 
						   \end{matrix}\Bigr](\xi_L,\tau_L).
\end{equation}
By the $S$-transformation, this becomes
\begin{equation}
 \Xi_\pm(\mu,2,0)
=e^{\b{J}_\pm(\mu,2,0)}\vartheta\Bigl[\begin{matrix}\pm\qu\\ 0 
						   \end{matrix}\Bigr](\xi_\text{orb},\tau_\text{orb})
\label{Xipm20orb}
\end{equation}
where $\b{J}_\pm(\mu,2,0)$ is given by
\begin{equation}
\begin{aligned}
\b{J}_\pm(\mu,2,0)&=J(\mu,2,0)-\frac{\pi i\xi_L^2}{\tau_L}-\hf\log(-i\tau_L)\\
&=\hf\b{f}_0+\b{f}_1+\hf\log2-\frac{1}{8}\log\ka\pm \frac{1}{8}\log\left(1+\frac{\ka^2}{16}\ri). 
\end{aligned}
\label{J1pm}
\end{equation}
Note that the $\hf\log2$ term in \eqref{J1pm} appears from our definition of $\b{f}_1$ in \eqref{barf0f1}
\begin{equation}
 f_1-\hf\log(-i\tau_L)=\b{f}_1+\hf\log2.
\end{equation}
One can easily show that the theta function with characteristic $(a,b)=(\pm\qu,0)$ in \eqref{Xipm20orb}
can also be written in terms of the Jacobi theta functions as
\begin{equation}
 \vartheta\Bigl[\begin{matrix}\pm\qu\\ 0 
						   \end{matrix}\Bigr](\xi_\text{orb},\tau_\text{orb})
=\hf\lf(\vartheta_2\left(\frac{\xi_\text{orb}}{2},\frac{\tau_\text{orb}}{4}\right)
\pm i \vartheta_1\left(\frac{\xi_\text{orb}}{2},\frac{\tau_\text{orb}}{4}\right)\ri).
\end{equation}
The $-\frac{1}{8}\log\ka$ term in \eqref{J1pm}
is precisely canceled by the factor $e^{\pi i\tau_\text{orb}/16}$ coming from
the theta function, and hence the fractional
power of $\ka$ does not appear in the expansion of $\Xi_\pm(\mu,2,0)$, as expected.
Finally, the small $\ka$ expansion of $\Xi_\pm(\mu,2,0)$ is found to be
\begin{equation}
 \begin{aligned}
  \Xi_{+}(\ka,2,0)&=1+\frac{\pi+2}{16\pi}\ka+\frac{3(\pi-2)^2}{512\pi^2}\ka^2+
\frac{168+396 \pi -58 \pi ^2-27 \pi ^3}{73728 \pi ^3}\ka^3+\cdots,\\
  \Xi_{-}(\ka,2,0)&=1+\frac{\pi-2}{16\pi}\ka+
  \frac{12+12\pi-5\pi^2}{512\pi^2}\ka^2+
\frac{-168+396 \pi +202 \pi ^2-99 \pi ^3}{73728 \pi ^3}\ka^3+\cdots.
 \end{aligned}
\end{equation}
They correctly reproduce the exact values of $Z_\pm(N,2,0)$ computed in 
section \ref{subsec:Zpm}.

\subsection{The case of $(k,M)=(2,1)$}
\subsubsection{$\Xi(\mu,2,1)$}\label{sub:Xi21}
For even $k$ with integral $M$, the effective chemical potential 
is given by \cite{MaMo,HO}
\begin{equation}
 2\mu_\text{eff}=2\mu-4\varepsilon e^{-2\mu}{}_4F_3\left(1,1,\frac{3}{2},\frac{3}{2};2,2,2;16
\varepsilon e^{-2\mu}\right),
\end{equation}
where $\varepsilon$ is a sign
\begin{equation}
 \varepsilon=(-1)^{\frac{k}{2}-M}.
\label{ep-sign}
\end{equation}
In the case of $(k,M)=(2,1)$,
this sign is $\varepsilon=+1$ and the modulus in the large radius frame is identified with $z=-e^{-2\mu}$.
In this case, $2\mu_\text{eff}$
is related to the A-period $t$ in \eqref{tFinhyper} by the shift
\begin{equation}
 \mu\to\mu+\frac{\pi i}{2},
\label{shift}
\end{equation}
but we have to subtract the imaginary part of 
$-\log z=2\mu+\pi i$
in \eqref{tFinhyper}.
Therefore, $2\mu_\text{eff}$ and the A-period $t'$ 
are related by
\begin{equation}
 2\mu_\text{eff}=t'-\pi i.
\label{t-prime}
\end{equation}
Here and in what follows, we put prime to indicate the shift \eqref{shift}, except for the subtlety in the genus-zero part
as we will mention below.

Again, the modified grand potential can be written 
in a closed form.
For the perturbative part, we find
\begin{equation}
 \til{A}(2,1)=\hf\log2,\qquad B(2,1)=0,
\end{equation}
where $\til{A}(k,M)$ is defined in \eqref{tilA}.
From the results in \cite{MaMo,HO},
the modified grand potential is given by
\begin{equation}
J(\mu,2,1)=\til{A}(2,1)+f_0'+f_1'-\frac{\mu'}{4}, 
\end{equation}
where $\mu'=\mu+\pi i/2$.

The genus-zero part $f_0'$ is obtained from
\eqref{f0inA} by replacing
\begin{equation}
 \Pi_A\to -\hf(t'-\pi i)=\Pi_A'+\frac{\pi i}{2},
\end{equation}
where $\Pi_A'=-t'/2$.
Thus we find\footnote{Note that $f_0'\not=f_0(t')$, so our notation
of ``prime'' is not completely consistent for $f_0'$.
We should emphasize that this
inconsistency of notation occurs only in $f_0'$.}
\begin{equation}
  f_0'=\int\frac{d\tau_L'}{2\pi i}\left(\Pi_A'+\frac{\pi i}{2}\right)^2
   =f_0(t')-\pi i\xi_L'+\frac{\pi i}{8}\tau_L',
\label{f'01}   
\end{equation}
where $\xi_L'$ and $\tau_L'$ are given by
\begin{equation}
  \xi_L'=\frac{1}{4\pi^2}\left(t'\del_tF_0(t')-\del_t^2 F_0(t')\right)-\frac{1}{24},\qquad
  \tau_L'=\frac{i}{\pi}\del_t^2 F_0(t').
\end{equation}
On the other hand, the genus-one part
 $f_1'$ is literally related to
 $f_1$ \eqref{f1}
 by the shift $\mu\to\mu+\pi i/2$
\begin{equation}
f_1'=f_1(t')=-\frac{1}{8}\log(1-16e^{-2\mu})-\hf\log\left(\frac{2K(16e^{-2\mu})}{\pi}\right)+\frac{\mu'}{4}.
\end{equation}
By performing the periodic sum \eqref{periodic-sum},
the grand partition function becomes a theta function
$\vartheta_3(v,\tau)$ with
\begin{equation}
v=2\lf(\xi_L'+\frac{1}{24}\ri)-\frac{\tau_L'}{2}-\qu-\frac{1}{3}
=2\xi_L'-\frac{\tau_L'+1}{2}.
\end{equation}
Note that the first three terms correspond to the three terms
in $f_0'$ \eqref{f'01}, and the last term $-1/3$ comes from \eqref{n3exp}.
The $\tau$-parameter in the theta function $\vartheta_3(v,\tau)$ is given by
\begin{equation}
 \tau=2(\tau_L'+1),
\end{equation}
where the additional constant $+2$ comes from the second term
$-i\pi \xi_L'$ in $f_0'$ \eqref{f'01}.
Then the grand partition function becomes
\begin{equation}
 \Xi(\mu,2,1)=e^{J(\mu,2,1)}\vartheta_3\left(2\xi_L'-\frac{\tau_L'+1}{2},2\tau_L'\right)
=e^{J(\mu,2,1)}\vartheta_4\left(2\xi_L'-\frac{\tau_L'}{2},2\tau_L'\right).
\label{Xi21}
\end{equation}
As we will see below, the combination $\xi_L'-\tau_L'/4$
appearing in \eqref{Xi21}
\begin{equation}
 \xi_L'-\frac{\tau_L'}{4}=-
  \int \frac{d\tau_L'}{2\pi i}\left(\Pi_A'+\frac{\pi i}{2}\right),
\end{equation}
behave nicely under the $S$-transformation when
combined with $f_0'$ in \eqref{f'01}.

The small $\ka$ expansion can be found by performing the $S$-transformation
\begin{equation}
 \Xi(\mu,2,1)=e^{\til{A}(2,1)+\b{f}_0'+\b{f}_1'-\frac{\mu'}{4}}\vartheta_2\left(\xi'_\text{orb}-\frac{1}{4},\frac{\tau'_\text{orb}}{2}\right),
\end{equation}
with
\begin{equation}
 \b{f}_0'=f_0'-\frac{2\pi i}{\tau_L'}\left(\xi_L'-\frac{\tau_L'}{4}\right)^2=
f_0(t')-\frac{2\pi i\xi_L'^2}{\tau_L'}=
\int\frac{d\tau_\text{orb}'}{2\pi i}(\Pi_B')^2,
\end{equation}
\begin{equation}
 \b{f}_1'-\frac{\mu'}{4}=-\frac{1}{8}\log\lf(1-\frac{\ka^2}{16}\ri)
-\hf\log\left(\frac{2K(\ka^2/16)}{\pi}\right)-\frac{1}{4}\log(i\ka).
\label{f'1}
\end{equation}
Again, the small $\ka$ 
expansion of $\Xi(\mu,2,1)$ does not contain the fractional power of $\ka$,
since the $-\qu\log(i\ka)$ term in \eqref{f'1} is 
precisely canceled by the factor $e^{\pi i\tau_\text{orb}'/8}$ coming from
the theta function $\vartheta_2$. 
Finally, the small $\ka$ 
expansion of $\Xi(\mu,2,1)$ is found to be
\begin{equation}
 \Xi(\mu,2,1)=1+\frac{\ka}{4\pi}+\frac{\pi^2-8}{128\pi^2}\ka^2+\frac{5\pi^2-48}{4608\pi^3}\ka^3+\cdots
\cdots,
\end{equation}
which correctly reproduces the results of canonical partition function
$Z(N,2,1)$ of ABJ theory computed in \cite{MaMo,HO}.

\subsubsection{$\Xi_\pm(\mu,2,1)$}
Let us consider the grand partition function $\Xi_\pm(\mu,2,1)$
of orientifold ABJ theory.
From \eqref{Jpm21} in Appendix 
\ref{app:Jpm-inst},
we find that the modified grand potential $J_\pm(\mu,2,1)$
can be written in a closed form:
\begin{equation}
 \begin{aligned}
  J_\pm(\mu,2,1)=\til{A}_\pm(2,1)+\hf f_0'+f_1'-\frac{\mu'}{4}\pm 
\left(\frac{\mu}{4}+\frac{1}{8}\log\left(\frac{1+4e^{-\mu}}{1-4e^{-\mu}}\right)\right),
 \end{aligned}
 \label{Jpm21closed}
\end{equation}
where the constant $\til{A}_\pm(2,1)$ is given by
\begin{equation}
 \til{A}_\pm(2,1)=\qu\log2\mp \frac{3}{4}\log2.
\end{equation}
Then, performing the periodic sum \eqref{periodic-sum}
we find that the grand partition function
is proportional to a theta function $\vartheta_3(v,\tau)$.
The $\tau$-parameter in $\vartheta_3(v,\tau)$ is given by
\begin{equation}
 \tau=\tau_L'+1,
\end{equation}
where the constant $+1$ comes from the second term in $f_0'$ \eqref{f'01},
and the $v$-parameter is given by
\begin{equation}
 v=\lf[\lf(\xi_L'+\frac{1}{24}\ri)-\frac{\tau_L'}{4}-\frac{1}{8}\ri]+\frac{1}{3}\pm\qu=\xi_L'-\frac{\tau_L'}{4}+\qu\pm\qu.
\end{equation}
Note that the first three terms in $v$ correspond to
the three terms in $f_0'$ \eqref{f'01},
while the contribution $1/3$ comes from
\eqref{n3inn},
and the last term $\pm\qu$ is the contribution of $B_\pm(2,1)=\pm\qu$.
Finally, we find
the closed form expressions of $\Xi_{\pm}(\mu,2,1)$
\begin{equation}
\begin{aligned}
 \Xi_{+}(\mu,2,1)&=
e^{J_{+}(\mu,2,1)}\vartheta_3\left(\xi_L'-\frac{\tau_L'}{4}+\hf,\tau_L'+1\right)
=e^{J_{+}(\mu,2,1)}\vartheta_3\left(\xi_L'-\frac{\tau_L'}{4},\tau_L'\right),\\
\Xi_{-}(\mu,2,1)&=
e^{J_{-}(\mu,2,1)}\vartheta_3\left(\xi_L'-\frac{\tau_L'}{4},\tau_L'+1\right)
=e^{J_{-}(\mu,2,1)}\vartheta_4\left(\xi_L'-\frac{\tau_L'}{4},\tau_L'\right).
\end{aligned}
\end{equation}

Now, using the identity \eqref{id-landen} of theta functions,
the product of $\Xi_{\pm}(\mu,2,1)$
becomes 
\begin{equation}
\Xi_{+}(\mu,2,1)\Xi_{-}(\mu,2,1)=e^{J_{+}(\mu,2,1)+J_{-}(\mu,2,1)}
\vartheta_4\left(0,2\tau_L'\right)
\vartheta_4\left(2\xi_L'-\frac{\tau_L'}{2},2\tau_L'\right).
\label{Xipm21}
\end{equation}
As in the case of $(k,M)=(2,0)$,
comparing \eqref{Xipm21} and $\Xi(\mu,2,1)$ in \eqref{Xi21}, 
the difference between $J_{+}+J_{-}$ and $J$ 
can be attributed to the contribution of
the factor $\vartheta_4\left(0,2\tau_L'\right)$, 
\begin{equation}
J_{+}(\mu,2,1)+J_{-}(\mu,2,1)- J(\mu,2,1)=-\log \vartheta_4\left(0,2\tau_L'\right).
\end{equation}

\paragraph{Orbifold expansion.}
As in the previous subsection \ref{sub:Xi21},
the small $\ka$ expansion
of $\Xi_\pm(\mu,2,1)$ can be found by performing the $S$-transformation:
\begin{equation}
 \begin{aligned}
  \Xi_{+}(\mu,2,1)&=
e^{\b{J}_{+}(\mu,2,1)}\vartheta_3\left(\xi_\text{orb}'-\qu,\tau_\text{orb}'\right),\\
\Xi_{-}(\mu,2,1)&=e^{\b{J}_{-}(\mu,2,1)}\vartheta_2\left(\xi_\text{orb}'-\qu,\tau_\text{orb}'\right),
 \end{aligned}
\end{equation}
where
\begin{equation}
\begin{aligned}
 \b{J}_{\pm}(\mu,2,1)=\til{A}_\pm(2,1)+\hf \b{f}_0'+\b{f}_1'+\hf\log2
-\frac{1}{4}\log(i\ka)\pm\left(\qu\log(i\ka)+\frac{1}{8}\log\frac{1+\frac{\ka}{4}}{1-\frac{\ka}{4}}\right). 
\end{aligned}
\end{equation}
Note that $\hf\log2$ term comes from the $S$-transformation and the definition of $\b{f}_1$
in \eqref{barf0f1}
\begin{equation}
 f_1'-\hf\log(-i\tau_L')=\b{f}_1'+\hf\log2.
\end{equation}
Finally, the small $\ka$ expansion of $\Xi_\pm(\mu,2,1)$
is found to be
\begin{equation}
 \begin{aligned}
  \Xi_{+}(\mu,2,1)&=1+\frac{1}{16}\ka+\frac{\pi^2-8}{512\pi^2}\ka^2+
\frac{-8-32 \pi +11 \pi ^2}{8192 \pi ^2}\ka^3+\cdots,\\
 \Xi_{-}(\mu,2,1)&=1+\frac{4-\pi}{16\pi}\ka+\frac{5\pi^2-8\pi-24}{512\pi^2}\ka^2+
\frac{-480+216 \pi +404 \pi ^2-135 \pi ^3}{73728 \pi ^3}\ka^3+\cdots,
 \end{aligned}
\end{equation}
which correctly reproduces the exact computation of $Z_\pm(N,2,1)$ in section \ref{subsec:Zpm}.

\section{Exact grand partition functions for $k=4$}
\label{sec:k=4}
For $k=4$, thanks to the Seiberg-like duality \eqref{sei-dual}, it is sufficient to consider
the cases with $M=0,1,2$. 
As observed in \cite{GHM}, the grand partition functions of 
$k=4$ ABJ theories are related to $\Xi_\pm(\mu,2,M)$ obtained in the previous section \ref{sec:k=2}
\begin{equation}
 \begin{aligned}
  \Xi(\mu,4,0)&=\Xi_{+}(\mu,2,1)=
e^{J_{+}(\mu,2,1)}\vartheta_3\lf(\xi_L'-\frac{\tau_L'}{4},\tau_L'\ri),\\
\Xi(\mu,4,1)&=\Xi_{-}(\mu,2,0)=e^{J_{-}(\mu,2,0)}\vartheta_3\lf(\xi_L+\frac{1}{4},\tau_L\ri),\\
\Xi(\mu,4,2)&=\Xi_{-}(\mu,2,1)=e^{J_{-}(\mu,2,1)}\vartheta_4\lf(\xi_L'-\frac{\tau_L'}{4},\tau_L'\ri).
\label{Xik4ABJ}
 \end{aligned}
\end{equation}
We have checked that the small $\ka$ expansion of the right hand side
of \eqref{Xik4ABJ} correctly reproduces the known canonical partition functions $Z(N,k,M)$ of ABJ(M) theory at $k=4$ computed in
\cite{HMO2,MaMo,HO}.
Note that \eqref{Xik4ABJ} implies that the modified grand potential
on the both sides of  \eqref{Xik4ABJ} are equal:
\begin{equation}
  J(\mu,4,0)=J_{+}(\mu,2,1),\quad
  J(\mu,4,1)=J_{-}(\mu,2,0),\quad
  J(\mu,4,2)=J_{-}(\mu,2,1).
  \label{J4toJ2}
\end{equation}

In this section we will find the closed form expressions
of the grand partition function $\Xi_\pm(\mu,4,M)$ 
of orientifold ABJ theory with $M=0,1,2$.

\subsection{$\Xi_\pm(\mu,4,0)$}\label{sub:Xipm40}
In this subsection, we will write down the closed form expression of
$\Xi_\pm(\mu,4,0)$.
In this case the sign in \eqref{ep-sign} is $\varepsilon=+1$,
and the effective chemical potential is $2\mu_\text{eff}=t'-\pi i$ as in the case of $(k,M)=(2,1)$.
The constants in the perturbative part are given by
\begin{equation}
 \til{A}_\pm(4,0)=-\qu\log2\mp\hf\log2,\qquad
B_\pm(4,0)=\frac{1}{8}\pm\qu.
\end{equation}
From the results \eqref{Jpm40} in Appendix \ref{app:Jpm-inst},
the modified grand potential can be found in a closed form
\begin{equation}
 J_\pm(\mu,4,0)=\til{A}_\pm(4,0)+\qu f_0'+f_1'-\frac{1}{8}\log(1-4e^{-\mu})-\frac{\mu'}{4}+\frac{\mu}{8}
\pm\lf(\frac{\mu}{4}+\qu\log(1-4e^{-\mu})\ri).
\end{equation}
By performing the periodic sum, the grand partition function 
becomes the Jacobi theta function $\vartheta_3(v,\tau)$
with the $\tau$-parameter
\begin{equation}
 \tau=\frac{\tau_L'+1}{2}.
\end{equation}
The constant $1/2$ comes from the second term
of $f_0'$ in \eqref{f'01}. 
Using \eqref{n3n1mod},
the $n^3$ term coming from the perturbative part $\frac{\mu^3}{12\pi^2}$ is rewritten
as a linear term in $n$
\begin{equation}
 \exp\left(\frac{(2\pi in)^3}{12\pi^2}\right)=\exp\left(-\frac{2\pi in}{3}\right).
\end{equation}
Then, the parameter $v$ in the theta function is
found to be
\begin{equation}
 v=\hf\lf[\left(\xi_L'+\frac{1}{24}\right)-\frac{\tau_L'}{4}-\frac{1}{8}\ri]-\frac{1}{3}+B_\pm(4,0)
=\frac{\xi_L'}{2}-\frac{\tau_L'}{8}-\qu\pm\qu,
\end{equation}
and we arrive at
the following form of the exact grand partition functions
\begin{equation}
 \Xi_{\pm}(\mu,4,0)=e^{J_\pm(\mu,4,0)}\vartheta_3\left(\frac{\xi_L'}{2}-\frac{\tau_L'}{8}-\qu\pm\qu,\frac{\tau_L'+1}{2}\right).
\end{equation}
From the property of theta functions in Appendix \ref{app:theta},
they can be also written as
\begin{equation}
 \Xi_{+}(\mu,4,0)=e^{J_{+}(\mu,4,0)}\vartheta_3\left(\frac{\xi_L'}{2}-\frac{\tau_L'}{8},\frac{\tau_L'+1}{2}\right),\quad
\Xi_{-}(\mu,4,0)=e^{J_{-}(\mu,4,0)}\vartheta_4\left(\frac{\xi_L'}{2}-\frac{\tau_L'}{8},\frac{\tau_L'+1}{2}\right).
\end{equation}
Then using the identity in \eqref{id-landen},
the product of $\Xi_{\pm}(\mu,4,0)$ becomes
\begin{equation}
\Xi_{+}(\mu,4,0)\Xi_{-}(\mu,4,0)=e^{J_{+}(\mu,4,0)+J_{-}(\mu,4,0)} \vartheta_3(0,\tau_L')
\vartheta_3\left(\xi_L'-\frac{\tau_L'}{4},\tau_L'\right)
\end{equation}
which is indeed proportional to $\Xi(\mu,4,0)$ in \eqref{Xik4ABJ}.
The difference of modified grand potential $J_{+}+J_{-}$ and $J(\mu,4,0)$ comes from the factor
$\vartheta_3(0,\tau_L')$
\begin{equation}
\begin{aligned}
 J_{+}(\mu,4,0)+J_{-}(\mu,4,0)-J(\mu,4,0)&=-\log\vartheta_3(0,\tau_L') \\
&=-\qu\log(1-16e^{-2\mu})-\hf \log\left(\frac{2}{\pi}K(16e^{-2\mu})\right).
\end{aligned}
\end{equation}
From the results in \eqref{Jpm40} and \cite{HMO2,HMO-bound},
one can check that this indeed reproduces the difference
between $J_{+}+J_{-}$ and $J$.

\paragraph{Orbifold expansion.}
The small $\ka$ expansion of $\Xi_{\pm}(\mu,4,0)$
can be found, again, by performing the $S$-transformation.
However, before doing the $S$-transformation, we have to
rewrite the theta function $\vartheta_a(v,\tau)$
in such a way that the $\tau$ parameter is proportional to
$\tau_L'$.
Using the identity  \eqref{id-4tau} in Appendix \ref{app:theta}, we find
\begin{equation}
\begin{aligned}
\Xi_{\pm}(\mu,4,0)&=e^{J_\pm(\mu,4,0)}
\lf[\vartheta_3\left(\xi_L'-\frac{\tau_L'}{4},2(\tau_L'+1)\right)
\pm \vartheta_2\left(\xi_L'-\frac{\tau_L'}{4},2(\tau_L'+1)\right)\ri] \\
&=e^{J_\pm(\mu,4,0)}
\lf[\vartheta_3\left(\xi_L'-\frac{\tau_L'}{4},2\tau_L'\right)\pm i\vartheta_2\left(\xi_L'-\frac{\tau_L'}{4},2\tau_L'\right)\ri].
\end{aligned}
\end{equation}
Now we are ready to perform the $S$-transformation.  We get
\begin{equation}
 \Xi_{\pm}(\mu,4,0)=e^{\b{J}_\pm(\mu,4,0)}
  \lf[\vartheta_3\left(\frac{\xi_\text{orb}'}{2}-\frac{1}{8},\frac{\tau_\text{orb}'}{2}\right)\pm i\vartheta_4\left(\frac{\xi_\text{orb}'}{2}-\frac{1}{8},\frac{\tau_\text{orb}'}{2}\right)\ri],
  \label{Xipm40orb}
\end{equation}
where $\b{J}_\pm(\mu,4,0)$ is given by
\begin{equation}
 \b{J}_\pm(\mu,4,0)=\til{A}_\pm(4,0)+\qu \b{f}_0'+\b{f}_1'
-\frac{1}{8}\log\left(4-\ka\right)
\pm\left(\frac{1}{4}\log\left(4-\ka\right)+\frac{\pi i}{4}\right).
\end{equation}
One can easily find the small $\ka$ expansion of
 \eqref{Xipm40orb}
\begin{equation}
 \begin{aligned}
  \Xi_{+}(\mu,4,0)&=1+\frac{-1+2\rt{2}}{32}\ka+\frac{-16+32\rt{2}\pi-(7+4\rt{2})\pi^2}
{2048\pi^2}\ka^2+\cdots,\\
\Xi_{-}(\mu,4,0)&=1+\frac{3-2\rt{2}}{32}\ka+\frac{-16-32\rt{2}\pi+(33-12\rt{2})\pi^2}
{2048\pi^2}\ka^2+\cdots,
 \end{aligned}
\end{equation}
which correctly reproduces the exact computation
 of $Z_\pm(N,4,0)$ in section \ref{subsec:Zpm}.

 \subsection{$\Xi_\pm(\mu,4,2)$}
 Next we consider $\Xi_\pm(\mu,4,2)$.
In this case, the sign in \eqref{ep-sign} is $\varepsilon=+1$, hence the 
expression of modified grand potential is similar to the $(k,M)=(4,0)$ case in the previous subsection \ref{sub:Xipm40}. 
The constants in the perturbative part are given by
\begin{equation}
 \til{A}_\pm(4,2)=\hf\log2\pm \lf(\qu \log2+\log\Bigl(\sin\frac{\pi}{8}\Bigr)\ri),\qquad
B_\pm(4,2)=-\frac{1}{8}\pm\qu.
\end{equation}
From the results \eqref{Jpm42} in Appendix \ref{app:Jpm-inst},
the modified grand potential can be found in a closed form
\begin{equation}
 J_\pm(\mu,4,2)=\til{A}_\pm(4,2)+\qu f_0'+f_1'-\frac{1}{8}\log(1+4e^{-\mu})-\frac{\mu'}{4}-\frac{\mu}{8}
\pm\lf(\frac{\mu}{4}+\qu\log(1+4e^{-\mu})\ri).
\end{equation}
By performing the periodic sum \eqref{periodic-sum},
the grand partition function 
becomes the Jacobi theta function $\vartheta_3(v,(\tau_L'+1)/2)$ with
\begin{equation}
 v=\hf\lf[\left(\xi_L'+\frac{1}{24}\right)-\frac{\tau_L'}{4}-\frac{1}{8}\ri]-\frac{1}{3}+B_\pm(4,2)
=\frac{\xi_L'}{2}-\frac{\tau_L'}{8}-\hf\pm\qu.
\end{equation}
Then, we arrive at the
following closed form expression of the grand partition functions
\begin{equation}
 \Xi_{\pm}(\mu,4,2)=e^{J_\pm(\mu,4,2)}\vartheta_4\left(\frac{\xi_L'}{2}-\frac{\tau_L'}{8}\pm\qu,\frac{\tau_L'+1}{2}\right),
\end{equation}
which can be also written as
\begin{equation}
\begin{aligned}
 \Xi_{+}(\mu,4,2)&=e^{J_{+}(\mu,4,2)}\vartheta_3\left(\frac{\xi_L'}{2}-\frac{\tau_L'}{8}-\qu,\frac{\tau_L'+1}{2}\right),\\
 \Xi_{-}(\mu,4,2)&=e^{J_{-}(\mu,4,2)}\vartheta_4\left(\frac{\xi_L'}{2}-\frac{\tau_L'}{8}-\qu,\frac{\tau_L'+1}{2}\right). 
\end{aligned}
\label{Xipm42}
\end{equation}
Using the identity \eqref{id-landen},
the product of $\Xi_{\pm}(\mu,4,2)$ becomes
\begin{equation}
 \Xi_{+}(\mu,4,2)\Xi_{-}(\mu,4,2)=e^{J_{+}(\mu,4,2)+J_{-}(\mu,4,2)}
\vartheta_3(0,\tau_L')\vartheta_4\left(\xi_L'-\frac{\tau_L'}{4},\tau_L'\right),
\end{equation}
which should be equal to $\Xi(\mu,4,2)$ in
\eqref{Xik4ABJ}.
Again, the difference between $J_{+}+J_{-}$ and $J(\mu,4,2)$
comes from the factor $\vartheta_3(0,\tau_L')$
\begin{equation}
 J_{+}(\mu,4,2)+J_{-}(\mu,4,2)-J(\mu,4,2)=-\log \vartheta_3(0,\tau_L'). 
\end{equation}

\paragraph{Orbifold expansion.}
The small $\ka$ expansion of $\Xi_{\pm}(\mu,4,2)$
can be found
by rewriting them in
a similar manner as in the previous subsection \ref{sub:Xipm40}:
\begin{equation}
\begin{aligned}
\Xi_{\pm}(\mu,4,2)&=e^{J_\pm(\mu,4,2)}
\lf[\vartheta_3\left(\xi_L'-\frac{\tau_L'}{4}\pm\hf,2(\tau_L'+1)\right)
- \vartheta_2\left(\xi_L'-\frac{\tau_L'}{4}\pm\hf,2(\tau_L'+1)\right)\ri] \\
&=e^{J_\pm(\mu,4,2)}
\lf[\vartheta_4\left(\xi_L'-\frac{\tau_L'}{4},2\tau_L'\right)\pm i\vartheta_1\left(\xi_L'-\frac{\tau_L'}{4},2\tau_L'\right)\ri].
\end{aligned}
\end{equation}
Then, performing the $S$-transformation, we find
\begin{equation}
 \Xi_{\pm}(\mu,4,2)=e^{\b{J}_\pm(\mu,4,2)}
\lf[\vartheta_2\left(\frac{\xi_\text{orb}'}{2}-\frac{1}{8},\frac{\tau_\text{orb}'}{2}\right)
\mp\vartheta_1\left(\frac{\xi_\text{orb}'}{2}-\frac{1}{8},\frac{\tau_\text{orb}'}{2}\right)\ri],
\label{Xipm42orb}
\end{equation}
where $\b{J}_\pm(\mu,4,2)$ is given by
\begin{equation}
 \b{J}_\pm(\mu,4,2)=\til{A}_\pm(4,2)+\qu \b{f}_0'+\b{f}_1'
-\frac{1}{8}\log\left(4+\ka\right)-\qu\log(i\ka)
\pm\frac{1}{4}\log\left(4+\ka\right).
\end{equation}
The small $\ka$ expansion of $\Xi_\pm(\mu,4,2)$
 in \eqref{Xipm42orb} reads
\begin{equation}
 \begin{aligned}
  \Xi_{+}(\mu,4,2)&=1+\frac{4-4\rt{2}+\pi}{32\pi}\ka+\frac{-32-8(-1+\rt{2})\pi+(-7+8\rt{2})\pi^2}
{2048\pi^2}\ka^2+\cdots,\\
\Xi_{-}(\mu,4,2)&=1+\frac{4+4\rt{2}-3\pi}{32\pi}\ka+\frac{-32-24(1+\rt{2})\pi+(33-8\rt{2})\pi^2}
{2048\pi^2}\ka^2+\cdots,
 \end{aligned}
\label{Xipm42exp}
\end{equation}
which correctly reproduces the exact computation
 of $Z_\pm(N,4,2)$ in section \ref{subsec:Zpm}.

 \subsection{$\Xi_\pm(\mu,4,1)$}
 Here we consider $\Xi_\pm(\mu,4,1)$.
 This case is similar to the $(k,M)=(2,0)$ case
 in the previous section \ref{sec:k=2} with sign $\varepsilon=-1$.
The constants in the perturbative part are given by
\begin{equation}
 \til{A}_\pm(4,1)=\qu\log2\mp\log2,\qquad
B_\pm(4,1)=-\frac{1}{16}\pm\qu.
\end{equation}
From the results \eqref{Jpm41} in Appendix \ref{app:Jpm-inst},
the modified grand potential can be found in a closed form
\begin{equation}
 J_\pm(\mu,4,1)=\til{A}_\pm(4,1)+\qu f_0+f_1-\frac{\mu}{4}-\frac{\mu}{16}
\pm\frac{\mu}{4}.
\end{equation}
One can easily show that
the grand partition function is proportional to $\vartheta_3(v,\tau_L)$ with
\begin{equation}
 v=\hf\left(\xi_L+\frac{1}{24}\right)-\frac{1}{3}+B_\pm(4,1)=\frac{\xi_L}{2}-\frac{3}{8}\pm\qu.
\end{equation}
Finally, we find the following form of the exact grand partition functions
\begin{equation}
 \begin{aligned}
  \Xi_{+}(\mu,4,1)&=e^{J_{+}(\mu,4,1)}\vartheta_3\left(\frac{\xi_L}{2}-\frac{1}{8},\frac{\tau_L}{2}\right),\\
 \Xi_{-}(\mu,4,1)&=e^{J_{-}(\mu,4,1)}\vartheta_4\left(\frac{\xi_L}{2}-\frac{1}{8},\frac{\tau_L}{2}\right).
 \end{aligned}
 \label{Xipm41}
\end{equation}
Again, the product of $\Xi_\pm(\mu,4,1)$ can be simplified
by using \eqref{id-landen} in Appendix \ref{app:theta} as
\begin{equation}
  \Xi_{+}(\mu,4,1) \Xi_{-}(\mu,4,1)=e^{J_{+}(\mu,4,1)+J_{-}(\mu,4,1)}
   \vartheta_4(0,\tau_L)\vartheta_3\left(\xi_L+\frac{1}{4},\tau_L\right),
 \label{prod41}
\end{equation}
which is indeed proportional to $\Xi(\mu,4,1)$ in \eqref{Xik4ABJ}.
The difference between $J_{+}+J_{-}$ and $J(\mu,4,1)$ comes from
$\vartheta_4(0,\tau_L)$ in \eqref{prod41}
\begin{equation}
 J_{+}(\mu,4,1)+J_{-}(\mu,4,1)-J(\mu,4,1)=-\log \vartheta_4(0,\tau_L)=
-\hf\log\left(\frac{2}{\pi}K(-16e^{-2\mu})\right).
\end{equation}

\paragraph{Orbifold expansion.}
The small $\ka$ expansion of $\Xi_\pm(\mu,4,1)$
can be easily found by rewriting \eqref{Xipm41}
in terms of the theta function with characteristic
\begin{equation}
 \Xi_\pm(\mu,4,1)=e^{J_\pm(\mu,4,1)}
\vartheta\Bigl[\begin{matrix} 0\\ \frac{1}{8}\mp\qu 
						   \end{matrix}\Bigr]
\Bigl(\frac{\xi_L}{2},\frac{\tau_L}{2}\Bigr)
=e^{\b{J}_\pm(\mu,4,1)}\vartheta\Bigl[\begin{matrix}-\frac{1}{8}\pm\qu\\ 0 
 \end{matrix}\Bigr](\xi_\text{orb},2\tau_\text{orb}).
 \label{Xipm41orb}
\end{equation}
where $\b{J}_\pm(\mu,4,1)$ are given by
\begin{equation}
 \b{J}_\pm(\mu,4,1)=\til{A}_\pm(4,1)+\qu \b{f}_0+\b{f}_1+\log2 -\frac{5}{16}\log\ka\pm \qu\log\ka.
\end{equation}
Note that $\log2$ term comes from 
\begin{equation}
 f_1-\hf\log\Bigl(-i\frac{\tau_L}{2}\Bigr)=\b{f}_1+\log2.
\end{equation}
The small $\ka$ expansion of $\Xi_{\pm}(\mu,4,1)$
is easily found from \eqref{Xipm41orb} 
\begin{equation}
 \begin{aligned}
  \Xi_{+}(\mu,4,1)&=1+\frac{1}{16\pi}\ka+\frac{10-\pi^2}
{1024\pi^2}\ka^2+\frac{78-121 \pi ^2+36 \pi ^3}{147456 \pi ^3}\ka^3+\cdots,\\
\Xi_{-}(\mu,4,1)&=1+\frac{\pi-3}{16\pi}\ka+
\frac{26+20 \pi -9 \pi ^2}{1024 \pi ^2}\ka^2+
\frac{-126+174 \pi +193 \pi ^2-75 \pi ^3}{49152 \pi ^3}\ka^3+\cdots,
 \end{aligned}
\end{equation}
which correctly reproduces the exact computation
of $Z_\pm(N,4,1)$ in section \ref{subsec:Zpm}.
\section{Exact grand partition functions for $k=8$}
\label{sec:k=8}
For $k=8$, up to Seiberg-like duality \eqref{sei-dual}
the independent grand partition functions 
are
$\Xi_\pm(\mu,8,M)$ with $M=0,1,\cdots,4$.
We find that the grand partition functions at $k=8$
are qualitatively different from the $k=2,4$ cases studied in the previous sections \ref{sec:k=2} and \ref{sec:k=4}.
In the case of $k=8$, the worldsheet instanton part of
the modified grand potential receives
all order corrections in the genus expansion.
In general, it is hopeless to sum over all order corrections,
but somewhat
miraculously 
for the $k=8$ case the all-order corrections in
the genus expansion can be resummed in a closed form.
\subsection{$\Xi_\pm(\mu,8,1)$}\label{sub:Xipm81}
Let us first consider $\Xi_\pm(\mu,8,1)$.
The constants in the perturbative part are given by
\begin{equation}
 \til{A}_\pm(8,1)=-\frac{1}{8}\log2\mp\frac{5}{4}\log2,\qquad
B_\pm(8,1)=-\frac{1}{32}\pm\qu.
\end{equation}
From the results in \eqref{Jpm81},
we find the modified grand potential in a closed form
\begin{equation}
 J_\pm(\mu,8,1)=j_\pm(\mu,8,1)+\log G_o(\mu).
  \label{Jpm81closed}
\end{equation}
Here $j_\pm(\mu,8,1)$ represents the perturbative part and the membrane
instanton corrections
\begin{equation}
 j_\pm(\mu,8,1)=\til{A}_\pm(8,1)+\frac{1}{8}f_0+f_1-\frac{\mu}{4}-\frac{\mu}{32}\pm \frac{\mu}{4}.
\end{equation}
On the other hand, $\log G_o(\mu)$ term
in \eqref{Jpm81closed} represents the worldsheet instanton
corrections.
Note that the worldsheet 1-instanton factor for $k=8$ is
\begin{equation}
 e^{-\frac{4\mu}{k}}=e^{-\frac{\mu}{2}}.
\end{equation}
By looking at the worldsheet instanton coefficients in \eqref{Jpm81}
for the first few instanton numbers,
one can easily guess the all order
resummation of such worldsheet instanton corrections
\begin{equation}
  G_o(\mu)=\left(\frac{1+2\rt{2}e^{-\frac{\mu}{2}}+4e^{-\mu}}{1-2\rt{2}e^{-\frac{\mu}{2}}+4e^{-\mu}}\right)^{\frac{1}{8}}.
\label{defG}
\end{equation}

In the periodic sum \eqref{periodic-sum},
we find that the summand can be written as
\begin{equation}
e^{J_\pm(\mu+2\pi in)}=G_o(\mu+2\pi in)e^{j_\pm(\mu,8,1)+\frac{\pi in^2}{4}\tau_L+2\pi inv},
\end{equation}
where $v$ is given by
\begin{equation}
 v=\qu\left(\xi_L+\frac{1}{24}\right)-\frac{1}{6}+B_\pm(8,1)=\frac{\xi_L}{4}-\frac{3}{16}\pm\qu.
\end{equation}
Above,
the second term $-1/6$ comes from the $2\pi in$-shift of
perturbative part $\frac{\mu^3}{24\pi^2}$
\begin{equation}
 \exp\left(\frac{(2\pi in)^3}{24\pi^2}\right)=\exp\left(-\frac{2\pi in}{6}\right).
\end{equation}
Since $G_o(\mu)$ is only invariant under 
the $4\pi i$-shift
\begin{equation}
 G_o(\mu+2\pi i)=G_o(\mu)^{-1},\quad
 G_o(\mu+4\pi i)=G_o(\mu),
\end{equation}
we have to consider the  periodic sum
 \eqref{periodic-sum}
 with even $n$ and odd $n$ separately.
Finally, we find the closed from expression of
$\Xi_\pm(\mu,8,1)$
\begin{equation}
 \Xi_\pm(\mu,8,1)=e^{j_\pm(\mu,8,1)}\left(G_o(\mu)\vartheta\Bigl[\begin{matrix} 0\\ -\frac{3}{8}\pm\hf 
						   \end{matrix}\Bigr]\Bigl(\frac{\xi_L}{2},\tau_L\Bigr)+G_o(\mu)^{-1}\vartheta\Bigl[\begin{matrix} \hf\\ -\frac{3}{8}\pm\hf 
						   \end{matrix}\Bigr]\Bigl(\frac{\xi_L}{2},\tau_L\Bigr)\right). 
\end{equation}

Let us consider the small $\ka$ expansion of $\Xi_\pm(\mu,8,1)$.
By performing the $S$-transformation,
we find
\begin{equation}
 \Xi_\pm(\mu,8,1)
=e^{\b{j}_\pm(\mu,8,1)}\left(G_o(\mu)\vartheta\Bigl[\begin{matrix} \frac{3}{8}\mp\hf \\0
						   \end{matrix}\Bigr]\Bigl(\frac{\xi_\text{orb}}{2},\tau_\text{orb}\Bigr)+e^{\pi i(-\frac{3}{8}\pm\hf)}G_o(\mu)^{-1}\vartheta\Bigl[\begin{matrix}  \frac{3}{8}\mp\hf\\ \hf
						   \end{matrix}\Bigr]\Bigl(\frac{\xi_\text{orb}}{2},\tau_\text{orb}\Bigr)\right),
\label{Xipm81orb}
\end{equation}
where $\b{j}_\pm(\mu,8,1)$ is given by
\begin{equation}
 \b{j}_\pm(\mu,8,1)=\til{A}_\pm(8,1)+\frac{1}{8}\b{f}_0+\b{f}_1+\hf\log2
-\frac{9}{32}\log\ka\pm\qu\log\ka.
\end{equation}
Now it is straightforward to find the small $\ka$ expansion
of \eqref{Xipm81orb}
\begin{equation}
 \begin{aligned}
  \Xi_{+}(\mu,8,1)&=1+\frac{-2+\pi}{64\pi}\ka+\frac{36-4 \pi +\left(16 \sqrt{2}-25\right) \pi ^2}{8192 \pi ^2}\ka^2+\cdots,\\
\Xi_{-}(\mu,8,1)&=1+\frac{-2+(4\rt{2}-5)\pi}{64\pi}\ka+
\frac{36+4 \left(5+28 \sqrt{2}\right) \pi +\left(40 \sqrt{2}-117\right) \pi ^2}{8192
   \pi ^2}\ka^2+\cdots,
 \end{aligned}
\label{Xipm81exp}
\end{equation}
which indeed reproduces the exact values of $Z_\pm(N,8,1)$ computed in section \ref{subsec:Zpm}.

A few comments are in order here: 
\begin{itemize}
 \item[1.] $\Xi_\pm(\mu,8,1)$ is a sum of two theta functions
multiplied by some factor $G_o^{\pm1}$,
and $\Xi_\pm(\mu,8,1)$ cannot be simply written as a single theta function.
This is qualitatively different from
the expression of the grand partition functions at $k=2,4$.
This also implies that the zeros of the grand partition function
$\Xi_\pm(\mu,8,1)=0$, which determines the exact quantization condition
for the spectrum of the kernel $\rho_\pm$,
cannot be simply expressed as the zeros of the Jacobi theta function $\vartheta_3(v,\tau)$
as in the case of $k=2,4$ studied in \cite{GHM}.
As we will see below, this is also the case for $(k,M)=(8,3)$, and we believe this is a general
phenomenon for $k\not=1,2,4$.

\item[2.] The grand partition function $\Xi(\mu,8,1)$ of ABJ theory 
is given by the product of $\Xi_\pm(\mu,8,1)$
\begin{equation}
 \Xi(\mu,8,1)=\Xi_{+}(\mu,8,1)\Xi_{-}(\mu,8,1),
\end{equation}
but it is difficult to rewrite this product
in a simple form and extract the modified grand potential $J(\mu,8,1)$ in a closed form.
This suggests that for general $k$, $\Xi_\pm(\mu,k,M)$ and $J_\pm(\mu,k,M)$ are more simpler than
$\Xi(\mu,k,M)$ and $J(\mu,k,M)$, and we should regard $\Xi_\pm(\mu,k,M)$ 
as the fundamental building block.
 \item[3.]
	  Although the small $\ka$ expansion of worldsheet instanton factor $G_o(\mu)$
	  includes 
a square root of $\ka$
\begin{equation}
 G_o(\mu)=\left(\frac{\ka+2\rt{2\ka}+4}{\ka-2\rt{2\ka}+4}\right)^{\frac{1}{8}}=
1+\frac{\rt{\ka}}{4\rt{2}}+\frac{\ka}{64}-\frac{5\ka\rt{\ka}}{256\rt{2}}+\cdots,
\end{equation}
such terms $\ka^{n+1/2}~(n=0,1,2,\cdots)$ are canceled by adding the two terms in \eqref{Xipm81orb},
and the small $\ka$ expansion of grand partition function  $\Xi_\pm(\mu,8,1)$ only includes
the integer powers of $\ka$, as it should be.
\item[4.] This factor $G_o(\mu)$ can be written as a modular form
\begin{equation}
 G_o(\mu)=\left[\frac{\vartheta_2(0,\tau_L)^2+\rt{2}\vartheta_2(0,\tau_L)\vartheta_4(0,\tau_L)+\vartheta_4(0,\tau_L)^2}{\vartheta_2(0,\tau_L)^2-\rt{2}\vartheta_2(0,\tau_L)\vartheta_4(0,\tau_L)+\vartheta_4(0,\tau_L)^2}\right]^{\frac{1}{8}}=
\left[\frac{\vartheta_3\left(\frac{\tau_L+1}{4}\right)\vartheta_3\left(\frac{\tau_L-1}{4}\right)}{\vartheta_4\left(\frac{\tau_L+1}{4}\right)\vartheta_4\left(\frac{\tau_L-1}{4}\right)}\right]^{\frac{1}{8}}. 
\end{equation} 
Note that $G_o(\mu)$ is invariant under the $S$-transformation $\tau_L\to\tau_\text{orb}=-1/\tau_L$
and $T^8:\tau_L\to\tau_L+8$. It would be interesting to understand
the modular property of $\Xi_\pm(\mu,8,1)$ better. 

\end{itemize}
\subsection{$\Xi_\pm(\mu,8,3)$}
Next let us consider $\Xi_\pm(\mu,8,3)$. The computation is almost
parallel to the $\Xi_\pm(\mu,8,1)$ case studied in the previous subsection \ref{sub:Xipm81}.
For $(k,M)=(8,3)$,
the constants in the perturbative part
are given by
\begin{equation}
 \til{A}_\pm(8,3)=\frac{1}{8}\log2-\log\Bigl(\sin\frac{\pi}{8}\Bigr)
\pm\lf(-\hf\log2+\log\Bigl(\sin\frac{\pi}{8}\Bigr)\ri),~~
B_\pm(8,3)=-\frac{9}{32}\pm\qu.
\end{equation}
Again, from the results in \eqref{Jpm83}
we can write down the modified grand potential in a closed form
\begin{equation}
 J_\pm(\mu,8,3)=j_\pm(\mu,8,3)-\log G_o(\mu),
\end{equation}
where $G_o(\mu)$ is the same function \eqref{defG} appeared in $J_\pm(\mu,8,1)$,
and $j_\pm(\mu,8,3)$ is given by
\begin{equation}
 j_\pm(\mu,8,3)=\til{A}_\pm(8,3)+\frac{1}{8}f_0+f_1-\frac{\mu}{4}-\frac{9\mu}{32}\pm \frac{\mu}{4}.
\end{equation}
Under the shift $\mu\to\mu+2\pi in$,
 the summand in the periodic
 sum in \eqref{periodic-sum} becomes
\begin{equation}
 e^{J_\pm(\mu+2\pi in,8,3)}=G_o(\mu+2\pi in)^{-1}e^{j_\pm(\mu,8,3)+\frac{\pi in^2}{4}\tau_L+2\pi inv}
\end{equation}
where $v$ is given by
\begin{equation}
 v=\qu\left(\xi_L+\frac{1}{24}\right)-\frac{1}{6}+B_\pm(8,3)=\frac{\xi_L}{4}-\frac{7}{16}\pm\qu.
\end{equation}
Finally, by performing the periodic sum \eqref{periodic-sum} for even $n$ and odd $n$ separately,
we arrive at the closed form expression of grand partition functions
\begin{equation}
 \Xi_\pm(\mu,8,3)=e^{j_\pm(\mu,8,3)}\left(G_o(\mu)^{-1}\vartheta\Bigl[\begin{matrix} 0\\ -\frac{7}{8}\pm\hf 
						   \end{matrix}\Bigr]\Bigl(\frac{\xi_L}{2},\tau_L\Bigr)+G_o(\mu)\vartheta\Bigl[\begin{matrix} \hf\\ -\frac{7}{8}\pm\hf 
						   \end{matrix}\Bigr]\Bigl(\frac{\xi_L}{2},\tau_L\Bigr)\right).
\end{equation}

The small $\ka$ expansion can be found by doing the $S$-transformation
\begin{equation}
 \Xi_\pm(\mu,8,3)=e^{\b{j}_\pm(\mu,8,3)}
\left(G_o(\mu)^{-1}\vartheta\Bigl[\begin{matrix} \frac{7}{8}\mp\hf \\0
						   \end{matrix}\Bigr]\Bigl(\frac{\xi_\text{orb}}{2},\tau_\text{orb}\Bigr)+e^{\pi i(-\frac{7}{8}\pm\hf)}G_o(\mu)\vartheta\Bigl[\begin{matrix}  \frac{7}{8}\mp\hf\\ \hf
						   \end{matrix}\Bigr]\Bigl(\frac{\xi_\text{orb}}{2},\tau_\text{orb}\Bigr)\right),
\label{Xipm83orb}
\end{equation}
where
\begin{equation}
 \b{j}_\pm(\mu,8,3)=\til{A}_\pm(8,1)+\frac{1}{8}\b{f}_0+\b{f}_1+\hf\log2
-\frac{17}{32}\log\ka\pm\qu\log\ka,
\end{equation}
and the small $\ka$ expansion of $\Xi_\pm(\mu,8,3)$ reads
\begin{equation}
 \begin{aligned}
  \Xi_{+}(\mu,8,3)&=1+\frac{6+(1 -2 \sqrt{2}) \pi }{64 \pi }\ka+
\frac{68+4 \left(3+10 \sqrt{2}\right) \pi +\left(20
   \sqrt{2}-57\right) \pi ^2}{8192 \pi ^2}\ka^2+\cdots,\\
\Xi_{-}(\mu,8,3)&=1+\frac{-26-16 \sqrt{2}+(7  +6 \sqrt{2}) \pi }{64 \pi }\ka+
\frac{196+64 \sqrt{2}+20 \left(1+2 \sqrt{2}\right) \pi
   +\left(28 \sqrt{2}-93\right) \pi ^2}{8192 \pi ^2}\ka^2+\cdots.
 \end{aligned}
\label{Xipm83exp}
\end{equation}
One can show that this agrees with the computation
of $Z_\pm(N,8,3)$ in section \ref{subsec:Zpm}.

\subsection{$\Xi_\pm(\mu,8,0)$}\label{sub:Xipm80}
Then let us move on to the even $M$ case.
In this case, the sign $\varepsilon$ in \eqref{ep-sign} is $(-1)^{k/2-M}=1$
and the natural variable is $t'$ in \eqref{t-prime}.

In this subsection we consider the $(k,M)=(8,0)$ case.
From the results in \eqref{Jpm80}, we find the closed form
of modified grand potential
\begin{align}
 J_\pm(8,0)=j_\pm(8,0)+\log G_e(\mu),
\end{align}
where $\log G_e(\mu)$ denotes the worldsheet instanton corrections
\begin{align}
 \log G_e(\mu)=\hf\text{arcsinh}(2e^{-\frac{\mu}{2}}),
\label{defGe}
\end{align}
and $j_\pm(8,0)$ is given by
\begin{align}
\begin{aligned}
 j_\pm(8,0)&=\til{A}_\pm(8,0)+\frac{f_0'}{8}+f_1'-\frac{\mu'}{4}+\frac{3\mu}{16} 
\\
&-\frac{1}{8}\log(1-4e^{-\mu})-\frac{3}{16}\log(1+4e^{-\mu})
\pm\left(\frac{\mu}{4}+\qu\log(1+4e^{-\mu})\right).
\end{aligned}
\end{align}
From this exact form of modified grand potential,
we can construct the exact grand partition function
by performing the periodic sum \eqref{periodic-sum}.
As in the case of $(k,M)=(8,1),(8,3)$ in the previous subsections,
the worldsheet instanton factor $G_e(\mu)$ in \eqref{defGe}
is not invariant under the $2\pi i$-shift of $\mu$
but invariant under the $4\pi i$-shift. Therefore, we have to devide the summation
into two parts.
In a similar manner as in the previous subsections, we arrive at the
exact form of $\Xi_\pm(8,0)$
\begin{align}
 \begin{aligned}
 \Xi_\pm(8,0)&=e^{j_\pm(8,0)}\left[G_e(\mu)
\vartheta\Bigl[\begin{matrix} 0\\ \pm\hf
						   \end{matrix}\Bigr]\Bigl(\frac{\xi_L'}{2}-\frac{\tau_L'}{8},\tau_L'+1\Bigr)
+G_e(\mu)^{-1}
\vartheta\Bigl[\begin{matrix} \hf\\ \pm\hf
						   \end{matrix}\Bigr]\Bigl(\frac{\xi_L'}{2}-\frac{\tau_L'}{8},\tau_L'+1\Bigr)
\right]\\
 &=e^{j_\pm(8,0)}\left[
G_e(\mu)\vartheta_3\Bigl(\frac{\xi_L'}{2}-\frac{\tau_L'}{8},\tau_L'\Bigr)
\mp e^{\frac{\pi i}{4}}G_e(\mu)^{-1}\vartheta_1\Bigl(\frac{\xi_L'}{2}-\frac{\tau_L'}{8},\tau_L'\Bigr)
\right].
\end{aligned}
\end{align}

The small $\ka$ expansion can be found by performing the $S$-transformation
\begin{align}
 \Xi_\pm(8,0)=e^{\b{j}_\pm(8,0)}\left[
\b{G_e}\vartheta_3\Bigl(\frac{\xi_\text{orb}'}{2}-\frac{1}{8},\tau_\text{orb}'\Bigr)
\mp ie^{\frac{\pi i}{4}}\b{G_e}^{-1}\ka^\hf\vartheta_1\Bigl(\frac{\xi_\text{orb}'}{2}-\frac{1}{8},\tau_\text{orb}'\Bigr)
\right],
\end{align}
where
\begin{align}
\b{G_e}
=\exp\left(\hf\log(2+\rt{\ka+4})\right),
\label{barGe}
\end{align}
and
\begin{align}
 \begin{aligned}
 \b{j}_\pm(8,0)&= \til{A}_\pm(8,0)+\frac{\b{f}_0'}{8}+\b{f}_1'+\hf\log2
-\frac{1}{8}\log(\ka-4)-\frac{3}{16}\log(\ka+4)\pm\qu\log(\ka+4).
 \end{aligned}
\end{align}
Finally we find
\begin{align}
 \begin{aligned}
  \Xi_{+}(8,0)&=1+\frac{1}{64} \left(5-4 \sqrt{2-\sqrt{2}}\right)\ka
+\lf(\frac{17-32 \sqrt{2}-8 \sqrt{2-\sqrt{2}}}{8192}-\frac{1}{256
   \pi ^2}+\frac{\sqrt{2+\sqrt{2}}}{128 \pi }\ri)\ka^2+\cdots\\
\Xi_{-}(8,0)&=1+\frac{1}{64} \left(4 \sqrt{2-\sqrt{2}}-3\right)\ka
+\left(\frac{129-32 \sqrt{2}-56
   \sqrt{2-\sqrt{2}}}{8192}-\frac{1}{256 \pi
   ^2}-\frac{\sqrt{2+\sqrt{2}}}{128 \pi }\right)\ka^2+\cdots.
 \end{aligned}
\end{align}

\subsection{$\Xi_{\pm}(\mu,8,4)$}
Let us consider $\Xi_{\pm}(8,4)$.
From the result in \eqref{Jpm84},
we find the closed form of modified grand potential
\begin{align}
 J_\pm(8,4)=j_\pm(8,4)-\log G_e(\mu)
\end{align}
where $G_e(\mu)$ is the same function in \eqref{defGe}
appeared in the $(k,M)=(8,0)$ case, and
$j_\pm(8,4)$ is given by
\begin{align}
 \begin{aligned}
   j_\pm(8,4)&=\til{A}_\pm(8,4)+\frac{f_0'}{8}+f_1'-\frac{\mu'}{4}-\frac{5\mu}{16} 
\\
&-\frac{1}{8}\log(1-4e^{-\mu})-\frac{3}{16}\log(1+4e^{-\mu})
\pm\left(\frac{\mu}{4}+\qu\log(1+4e^{-\mu})\right).
 \end{aligned}
\end{align}
By performing
the periodic sum \eqref{periodic-sum},
we find the exact form of grand partition function as
\begin{align}
 \begin{aligned}
  \Xi_\pm(8,4)&=e^{j_\pm(8,4)}\left[G_e^{-1}
\vartheta\Bigl[\begin{matrix} 0\\ -1\pm\hf
						   \end{matrix}\Bigr]\Bigl(\frac{\xi_L'}{2}-\frac{\tau_L'}{8},\tau_L'+1\Bigr)
+G_e
\vartheta\Bigl[\begin{matrix} \hf\\ -1\pm\hf
						   \end{matrix}\Bigr]\Bigl(\frac{\xi_L'}{2}-\frac{\tau_L'}{8},\tau_L'+1\Bigr)
\right]\\
&=e^{j_\pm(8,4)}\left[
G_e^{-1}\vartheta_3\Bigl(\frac{\xi_L'}{2}-\frac{\tau_L'}{8},\tau_L'\Bigr)
\pm e^{\frac{\pi i}{4}}G_e\vartheta_1\Bigl(\frac{\xi_L'}{2}-\frac{\tau_L'}{8},\tau_L'\Bigr)\right].
 \end{aligned}
\end{align}

Again, the small $\ka$ expansion can be found by the $S$-transformation
\begin{align}
 \Xi_\pm(8,4)=e^{\b{j}_\pm(8,4)}\left[
\b{G_e}^{-1}\vartheta_3\Bigl(\frac{\xi_\text{orb}'}{2}-\frac{1}{8},\tau_\text{orb}'\Bigr)
\pm ie^{\frac{\pi i}{4}}\b{G_e}\ka^{-\hf}\vartheta_1\Bigl(\frac{\xi_\text{orb}'}{2}-\frac{1}{8},\tau_\text{orb}'\Bigr)
\right],
\end{align}
where $\b{G}_e$ is defined in \eqref{barGe}
and $\b{j}_\pm(8,4)$ is given by
\begin{align}
 \begin{aligned}
 \b{j}_\pm(8,4)&= \til{A}_\pm(8,4)+\frac{\b{f}_0'}{8}+\b{f}_1'+\hf\log2
-\frac{1}{8}\log(\ka-4)-\frac{3}{16}\log(\ka+4)\pm\qu\log(\ka+4).
 \end{aligned}
\end{align}
Finally, we find the small $\ka$ expansion of $\Xi_\pm(8,4)$
\begin{align}
 \begin{aligned}
  &\Xi_{+}(8,4)=1+\frac{\sin\frac{\pi}{16}\sin\frac{3\pi}{16}}{\sin^2\frac{\pi}{8}}
\Biggl[\frac{1}{64} \left(10-5 \sqrt{2}+2 \sin \left(\frac{\pi
   }{8}\right)\right)-\frac{1}{4 \sqrt{2} \pi }\Biggr]\ka\\
&+\frac{\sin\frac{\pi}{16}\sin\frac{3\pi}{16}}{\sin^2\frac{\pi}{8}}
\Biggl[-\frac{5}{256 \sqrt{2} \pi }+\frac{-6+3 \sqrt{2}-2 \sin
   \left(\frac{\pi }{8}\right)}{256 \pi
   ^2}+\frac{\left(82-64 \sqrt{2}\right) \sin
   \left(\frac{\pi }{8}\right)-81
   \left(\sqrt{2}-2\right)}{8192}\Biggr]\ka^2+\cdots,\\
 &\Xi_{-}(8,4)=1+\frac{1}{\sin^2\frac{\pi}{8}\sin\frac{\pi}{16}\sin\frac{3\pi}{16}}
\Biggl[\frac{1}{64 \pi }+\frac{1}{512} \left(-3+3 \sqrt{2}-7
   \sqrt{2} \sin \left(\frac{\pi }{8}\right)\right)\Biggr]\ka\\
&+\frac{1}{\sin^2\frac{\pi}{8}\sin\frac{\pi}{16}\sin\frac{3\pi}{16}}
\Biggl[-\frac{3}{4096 \pi }+\frac{-3+3 \sqrt{2}-\sqrt{2} \sin
   \left(\frac{\pi }{8}\right)}{2048 \pi ^2}+\frac{\left(217
   \sqrt{2}-64\right) \sin \left(\frac{\pi }{8}\right)-193
   \left(\sqrt{2}-1\right)}{65536}\Biggr]\ka^2+\cdots.
 \end{aligned}
\end{align}

\subsection{$\Xi_{\pm}(\mu,8,2)$}
Let us consider $\Xi_{\pm}(8,2)$.
From the result in \eqref{Jpm82},
we find the closed form of modified grand potential
\begin{align}
\begin{aligned}
J_\pm(8,2)&=\til{A}_\pm(8,2)+\frac{f_0'}{8}+f_1'-\frac{\pi i}{8}-\frac{7\mu}{16}\\
  &-\frac{1}{8}\log(1+4e^{-\mu})-\frac{3}{16}\log(1-4e^{-\mu})\pm \left(\frac{\mu}{4}+\qu\log(1-4e^{-\mu})\right).
 \end{aligned}
\end{align}
Interestingly, in this case the modified grand potential $J_\pm(8,2)$
is invariant under the $2\pi i$-shift of $\mu$, and the 
periodic sum is written as a single theta function
\begin{align}
\begin{aligned}
 \Xi_{+}(8,2)&=e^{J_{+}(8,2)}
\vartheta_3\left(\frac{\xi_L'}{4}-\frac{\tau_L'}{16}-\frac{1}{8},\frac{\tau_L'+1}{4}\right),\\
\Xi_{-}(8,2)&=e^{J_{-}(8,2)}
\vartheta_4\left(\frac{\xi_L'}{4}-\frac{\tau_L'}{16}-\frac{1}{8},\frac{\tau_L'+1}{4}\right). 
\end{aligned}
\end{align}
Using the formula of theta function in Appendix \ref{app:theta},
the product of $\Xi_{+}(8,2)$ and $\Xi_{-}(8,2)$
can be rewritten as
\begin{align}
 \Xi(8,2)=e^{J_{+}(8,2)+J_{-}(8,2)}\vartheta_4\left(0,\frac{\tau_L'+1}{2}\right)
\vartheta_4\left(\frac{\xi_L'}{2}-\frac{\tau_L'}{8}-\frac{1}{4},\frac{\tau_L'+1}{2}\right).
\end{align}
One can show that this is equal to $\Xi_{-}(4,2)$ in \eqref{Xipm42},
as conjectured in \cite{GHM}
\begin{align}
\Xi(8,2)= \Xi_{-}(4,2)=e^{J_{-}(4,2)}\vartheta_4\left(\frac{\xi_L'}{2}-\frac{\tau_L'}{8}-\frac{1}{4},\frac{\tau_L'+1}{2}\right).
\label{82-vs-42}
\end{align}

The small $\ka$ expansion of $\Xi_\pm(8,2)$ can be found by performing the
$S$-transformation
\begin{align}
 \begin{aligned}
  \Xi_\pm(8,2)&=e^{J_\pm(8,2)}\left[\vartheta_3\left(\frac{\xi_L'}{2}-\frac{\tau_L'}{8}-\frac{1}{4},\tau_L'+1\right)\pm \vartheta_2\left(\frac{\xi_L'}{2}-\frac{\tau_L'}{8}-\frac{1}{4},\tau_L'+1\right)\right]\\
&=e^{J_\pm(8,2)}\left(\vartheta\Bigl[\begin{matrix} 0\\ \qu 
						   \end{matrix}\Bigr]\Bigl(\frac{\xi_L'}{2}-\frac{\tau_L'}{8},\tau_L'\Bigr)\pm e^{\frac{\pi i}{4}}
\vartheta\Bigl[\begin{matrix} \hf\\ -\qu 
						   \end{matrix}\Bigr]
\Bigl(\frac{\xi_L'}{2}-\frac{\tau_L'}{8},\tau_L'\Bigr)\right)\\
&=e^{\b{J}_\pm(8,2)}\left(\vartheta\Bigl[\begin{matrix} -\qu\\ 0 
						   \end{matrix}\Bigr]\Bigl(\frac{\xi_\text{orb}'}{2}-\frac{1}{8},\tau_\text{orb}'\Bigr)\pm
\vartheta\Bigl[\begin{matrix} \qu\\ \hf 
						   \end{matrix}\Bigr]\Bigl(\frac{\xi_\text{orb}'}{2}-\frac{1}{8},\tau_\text{orb}'\Bigr)\right),
 \end{aligned}
\end{align}
where $\b{J}_\pm(8,2)$ is given by
\begin{align}
 \begin{aligned}
  \b{J}_\pm(8,2)&=\til{A}_\pm+\frac{\b{f}_0'}{8}+\b{f}_1'+\hf\log2+\frac{\pi i}{16}-\frac{\mu}{8}\\
&-\frac{1}{8}\log(4+\ka)-\frac{3}{16}\log(4-\ka)\pm\left(-\frac{\pi i}{4}+\qu\log(4-\ka)\right).
 \end{aligned}
\end{align}
Finally, we find
\begin{align}
 \begin{aligned}
 &\Xi_{+}(8,2)=1+\left(\frac{1}{64} \left(4
   \sqrt{2+\sqrt{2}}-7\right)+\frac{1+\sqrt{2}-2
   \sqrt{1+\frac{1}{\sqrt{2}}}}{16 \pi }\right)\ka\\
&+\left(\frac{65-24 \sqrt{2+\sqrt{2}}}{8192}-\frac{3}{512 \pi
   ^2}+\frac{3 \left(-5-5 \sqrt{2}+\sqrt{52+34
   \sqrt{2}}\right)}{1024 \pi }\right)\ka^2+\cdots,\\
  &\Xi_{-}(8,2)=1+\left(\frac{1}{64} \left(1-4
   \sqrt{2+\sqrt{2}}\right)+\frac{1+\sqrt{2}+\sqrt{2
   \left(2+\sqrt{2}\right)}}{16 \pi }\right)\ka\\
&+\left(\frac{5 \left(29-8
   \sqrt{2+\sqrt{2}}\right)}{8192}-\frac{3}{512 \pi
   ^2}+\frac{-7-7 \sqrt{2}-2
   \sqrt{37-\frac{23}{\sqrt{2}}}}{1024 \pi }\right)\ka^2+\cdots.
 \end{aligned}
\end{align}

\subsection{Comment on the $k=6$ case}
For $k=3,6,12$, we find the instanton expansion of grand potential from the
exact values of $Z_\pm(N,k,M)$, which is summarized in Appendix \ref{app:Jpm-inst}.
For these cases, however, it is difficult to find the closed form expression of the modified grand potential.
In particular,
it is difficult to resum the series of worldsheet instanton corrections,
since these corrections come from  all genera in the
string coupling expansion with $g_s=2/k$. 
However, for $k=6$,
we find that the membrane instanton corrections of the form $e^{-2n\mu}~(n=1,2,\cdots)$  can be resummed in a closed form.
The terms with $1/\pi^2$ coefficient is given by
the genus-zero free energy, as in the case of $k=2,4$
\begin{equation}
 \frac{1}{6}f_0~~~(\text{even}~M),\qquad\text{or}\qquad \frac{1}{6}f_0' ~~~(\text{odd}~M).
\end{equation}
We find that the terms without  $1/\pi^2$ coefficient can be  also
written in closed forms.
For even $M$ we find
\begin{equation}
 J_{\pm}(6,M)\Big|_{e^{-2n\mu}} =
  -\frac{2}{9}\log(1+16e^{-2\mu})-\frac{2}{3}
  \log\Bigl[\frac{2}{\pi}K(-16e^{-2\mu})\Bigr]
\pm\frac{1}{8}\log(1+16e^{-2\mu}),
\end{equation}
while for odd $M$ we find
\begin{equation}
 J_{\pm}(6,M)\Big|_{e^{-2n\mu}} =
-\frac{2}{9}\log(1-16e^{-2\mu})-\frac{2}{3}\log\Bigl[\frac{2}{\pi}K(16e^{-2\mu})\Bigr]. 
\end{equation}

Suppose we have managed to sum over the worldsheet instanton corrections to
the grand potential, which we denote as $\log W(\mu)$.
For definiteness, let us consider  the $(k,M)=(6,0)$ case.
In this case, the perturbative coefficients are given by
\begin{equation}
 \til{A}_\pm(6,0)=-\frac{1}{3}\log3\pm\hf\log2,\quad
B_\pm(6,0)=\frac{11}{72}\pm\qu
\end{equation}
and the modified grand potential can be written as
\begin{equation}
 J_\pm(\mu,6,0)=j_\pm(6,0)+\log W(\mu),
  \label{j6}
\end{equation}
where $j_\pm(6,0)$ includes the perturbative part and
the membrane instanton corrections
\begin{equation}
 \begin{aligned}
j_\pm(6,0)&=\til{A}_\pm(6,0)+\frac{1}{6}f_0+B_\pm(6,0)\mu\\
  &\quad-\frac{2}{9}\log(1+16e^{-2\mu})-\frac{2}{3}
  \log\Bigl[\frac{2}{\pi}K(-16e^{-2\mu})\Bigr]
  \pm\frac{1}{8}\log(1+16e^{-2\mu}).
 \end{aligned}
\end{equation}
From the results in \eqref{Jpm60},
$\log W(\mu)$ has the following expansion
\begin{equation}
 \log W(\mu)=\frac{2}{3}e^{-\frac{2\mu}{3}}
  -\frac{17}{9}e^{-\frac{8\mu}{3}}
  +\frac{2}{15}e^{-\frac{10\mu}{3}}
  +\frac{2776}{189}e^{-\frac{14\mu}{3}}+\cdots.
\end{equation}
On the other hand, the worldsheet instanton corrections
(and bound states with membrane instantons)
of ABJM theory at $k=6$
is \cite{HMO2}
 \begin{equation}
  J^\text{WS}(\mu,6,0)=\frac{4}{3}e^{-\frac{2\mu}{3}}
   -2e^{-\frac{4\mu}{3}}+\frac{88}{9}e^{-\frac{8\mu}{3}}
   +\frac{108}{5}e^{-\frac{10\mu}{3}}+
   \frac{25208}{189}e^{-\frac{14\mu}{3}}+\cdots,
 \end{equation}
which is different from $2\log W(\mu)$
\begin{equation}
 J^\text{WS}(\mu,6,0)\not=2\log W(\mu).
\end{equation}
 This suggests that the worldsheet instanton corrections
 in the orientifold ABJ theory receive additional contributions
 from the orientifolding.
 
 From \eqref{j6}, we can consider the possible form of the
 exact grand partition function $\Xi_\pm(\mu,6,0)$
 by performing the periodic sum \eqref{periodic-sum}.
Since  $W(\mu)$ 
is not invariant under the $2\pi i$-shift of $\mu$
but invariant under
the $6\pi i$-shift,
we have to decompose the periodic sum \eqref{periodic-sum}
 into three parts 
according to $n$ modulo 3.
Then, we find
\begin{equation}
 \Xi_\pm(\mu,6,0)=e^{j_\pm(\mu,6,0)}\sum_{r=0,1,2}e^{-\frac{4\pi ir^3}{9}}W(\mu+2\pi ir)
\vartheta\Bigl[\begin{matrix} \frac{r}{3}\\3B_\pm(6,0)+\frac{1}{24}
						   \end{matrix}\Bigr]\Bigl(\xi_L,3\tau_L\Bigr).
\end{equation}
This is similar to the structure of $\Xi_\pm(\mu,8,M=0,1,3,4)$ in the previous subsections.
Also, we speculate that the
factor $W(\mu)$ can be written as a modular
form,  as in the case of $G(\mu)$ for $k=8$. For general even $k$,
the worldsheet 1-instanton factor is given by
\begin{equation}
 e^{-\frac{4\mu}{k}}\propto \left[\frac{\vartheta_2(0,\tau_L)}{\vartheta_4(0,\tau_L)}\right]^{\frac{8}{k}},
\end{equation}
and we expect that 
 $W(\mu)$ is invariant under the modular transformation
 $S$ and $T^k:\tau_L\to\tau_L+k$.
 It would be interesting to find the closed form expression of the function $W(\mu)$. Also,
 we have detailed data of instanton coefficients
 for $k=3$ (see Appendix \ref{app:J3}).   It would be nice to find the closed form expression of
 the modified grand potential for $k=3$.

 \section{Functional relations}
 \label{sec:qw-rel}
In \cite{GHM}, it was conjectured that
the grand partition functions of orientifold ABJ theory satisfy certain functional relations.
These relations involve 
grand partition functions at fixed $k$ but $M$ differing by $\pm1$.
The first relation is
\begin{equation}
 \Xi_{+}^{+}(k,1)\Xi_{+}^{-}(k,1)+\frac{e^\mu}{4k} \Xi_{-}^{+}(k,1)\Xi_{-}^{-}(k,1)
= \Xi(k,0).
\label{rel1}
\end{equation}
Here we suppressed the $\mu$-dependence of grand partition functions
and introduced the notation
$f^\pm$ for a function of $\mu$
\begin{equation}
 f^\pm=f\Bigl(\mu\pm\frac{\pi i}{2}\Bigr).
\label{fshift}
\end{equation}
The second set of relations is
\begin{equation}
\begin{aligned}
 &~e^{-\frac{\pi i M}{2k}}\Xi_+^+(k,M-1)\Xi_-^-(k,M+1)
+e^{\frac{\pi i M}{2k}}\Xi_+^-(k,M-1)\Xi_-^+(k,M+1)\\
=&~2\cos\Bigl(\frac{\pi M}{2k}\Bigr)\Xi(k,M),  
\end{aligned}
\label{QWrel1} 
\end{equation}
\begin{equation}
\begin{aligned}
 &~e^{\frac{\pi i M}{2k}}\Xi_+^+(k,M+1)\Xi_-^-(k,M-1)
-e^{-\frac{\pi i M}{2k}}\Xi_+^-(k,M+1)\Xi_-^+(k,M-1)\\
=&~2i\sin\Bigl(\frac{\pi M}{2k}\Bigr)\Xi(k,M). 
\end{aligned}
\label{QWrel2} 
\end{equation}
Following \cite{GHM},
we will call these relations ``quantum Wronskian relations''.
Using our exact data of $Z_\pm(N,k,M)$,
for $k=2,3,4,6$,
we have checked that these relations are indeed satisfied in the small $\ka$
expansion.

We first notice that the perturbative part of grand potential $J_\pm(k,M)$
is consistent with those relations.
For the first relation \eqref{rel1},
the perturbative part $J_{\pm,\text{pert}}(k,1)$ appearing in the left hand side
is
\begin{equation}
 J_{\pm,\text{pert}}=\frac{\mu^3}{3\pi^2k}+\lf(\frac{ B(k,1)}{2}\pm\qu\ri) \mu+
 A_\pm(k,1),
\end{equation}
with
\begin{equation}
 A_\pm(k,1)=\hf A_c(k)+\qu\log k\mp\qu\log(4k).
\end{equation}
One can show that
\begin{equation}
 \begin{aligned}
J_{+,\text{pert}}^{+}(k,1)+
J_{+,\text{pert}}^{-}(k,1)-J_\text{pert}(k,0)&=-\log2,\\
J_{-,\text{pert}}^{+}(k,1)+
J_{-,\text{pert}}^{-}(k,1)-J_\text{pert}(k,0)&=-\log2-\mu+\log(4k),
 \end{aligned}
\label{shiftJpert1}
\end{equation}
which implies that the perturbative part
satisfies the relation \eqref{rel1}
\begin{equation}
 e^{J_{+,\text{pert}}^{+}(k,1)+
J_{+,\text{pert}}^{-}(k,1)}+\frac{e^\mu}{4k}
e^{J_{-,\text{pert}}^{+}(k,1)+
J_{-,\text{pert}}^{-}(k,1)}=e^{J_\text{pert}(k,0)}.
\end{equation}

Next consider the quantum Wronskian relations \eqref{QWrel1}
and \eqref{QWrel2}, at the level of perturbative part. 
We find that the constant term satisfies
\begin{equation}
 \begin{aligned}
  A_+(k,M-1)+A_{-}(k,M+1)-A(k,M)&=-\log\left(2\sin\frac{\pi M}{2k}\right),\\
A_+(k,M+1)+A_{-}(k,M-1)-A(k,M)&=-\log\left(2\cos\frac{\pi M}{2k}\right),
 \end{aligned}
\end{equation}
and the perturbative grand potential satisfies
\begin{equation}
 \begin{aligned}
 J_{+,\text{pert}}^+\left(k,M-1\right)+
J_{-,\text{pert}}^{-}\left(k,M+1\right)- J_\text{pert}(k,M)&=
+\frac{\pi i}{2}\left(1-\frac{M}{k}\right)-\log\left(2\sin\frac{\pi M}{2k}\right),\\
J_{+,\text{pert}}^-\left(k,M-1\right)+
J_{-,\text{pert}}^{+}\left(k,M+1\right)-J_\text{pert}(k,M)&=
-\frac{\pi i}{2}\left(1-\frac{M}{k}\right)-\log\left(2\sin\frac{\pi M}{2k}\right),\\
J_{+,\text{pert}}^+\left(k,M+1\right)+
J_{-,\text{pert}}^{-}\left(k,M-1\right)-J_\text{pert}(k,M)&=
+\frac{\pi i M}{2k}-\log\left(2\cos\frac{\pi M}{2k}\right),\\
J_{+,\text{pert}}^-\left(k,M+1\right)+
J_{-,\text{pert}}^{+}\left(k,M-1\right)- J_\text{pert}(k,M)&=
-\frac{\pi i M}{2k}-\log\left(2\cos\frac{\pi M}{2k}\right).
 \end{aligned}
 \label{qw-pert}
\end{equation}
From these relations \eqref{qw-pert},
one can show that
the perturbative part
satisfies the quantum Wronskian relations \eqref{QWrel1} and \eqref{QWrel2}
\begin{equation}
 \begin{aligned}
e^{-\frac{\pi iM}{2k}+J_{+,\text{pert}}^{+}(k,M-1)+
J_{-,\text{pert}}^{-}(k,M+1)}+
e^{\frac{\pi iM}{2k}+J_{+,\text{pert}}^{-}(k,M-1)+
J_{-,\text{pert}}^{+}(k,M+1)}&=2\cos\Bigl(\frac{\pi M}{2k}\Bigr)e^{J_\text{pert}(k,M)},\\
e^{\frac{\pi iM}{2k}+J_{+,\text{pert}}^{+}(k,M+1)+
J_{-,\text{pert}}^{-}(k,M-1)}-
e^{-\frac{\pi iM}{2k}+J_{+,\text{pert}}^{-}(k,M+1)+
J_{-,\text{pert}}^{+}(k,M-1)}&=2i\sin\Bigl(\frac{\pi M}{2k}\Bigr)e^{J_\text{pert}(k,M)}.
 \end{aligned}
\end{equation}
We would like to show that not only the perturbative part, but also
the exact grand partition functions
including the instanton corrections,
indeed satisfy the functional relations.
In \cite{GHM}, it is found that the the exact grand partition function at $k=1$ 
satisfies the relation \eqref{rel1}.
In this section, we will prove 
the functional
relations \eqref{rel1}, \eqref{QWrel1}, and \eqref{QWrel2}
for $k=2,4$ using the exact from of grand partition functions
obtained in the previous sections.

\subsection{Proof of functional relations for $k=2$}
In this subsection we will show the functional relations
\eqref{rel1}, \eqref{QWrel1}, and \eqref{QWrel2} for $k=2$.
Interestingly, they all follow from an identity
$\mathcal{N}_2(\tau)=1$ proved in Appendix \ref{proof:N2}.

\subsubsection{Functional relation \eqref{rel1} for $k=2$}
\label{sub:k2rel1}
Let us first consider the first relation
\eqref{rel1} for $k=2$.
For readers convenience, let us recall the exact expression of
grand partition functions appearing in the relation \eqref{rel1}
\begin{equation}
\begin{aligned}
\Xi(2,0)&=e^{J(2,0)} \vartheta_3\left(2\xi_L,2\tau_L\right),\\
 \Xi_{+}(2,1)&=e^{J_+(2,1)}\vartheta_3\left(\xi_L'-\frac{\tau_L'}{4},\tau_L'\right),\quad
 \Xi_{-}(2,1)=e^{J_-(2,1)}\vartheta_4\left(\xi_L'-\frac{\tau_L'}{4},\tau_L'\right),
\end{aligned}
 \end{equation}
and the modified grand potentials are given by
\begin{equation}
 \begin{aligned}
  J(2,0)&=f_0+f_1,\\
J_\pm(2,1)&=\til{A}_\pm(2,1)+\hf f_0'+F_1'+F_1^{\text{NS}'}\pm\left(\frac{\mu}{4}
+\frac{1}{8}\log\left(\frac{1+4e^{-\mu}}{1-4e^{-\mu}}\right)\right).
 \end{aligned}
\end{equation}
Note that $F_1'=F_1(t')$ and $F_1^{\text{NS}'}=F_1^\text{NS}(t')$.

First, let us consider the transformation
of the modified grand potential $J_\pm(2,1)$ under the shift $\mu\to\mu\pm\pi i/2$.
As discussed in \cite{GHM},
the genus-zero and genus-one free energies transforms as
\begin{equation}
\begin{aligned}
 &(f_0')^{+}+(f_0')^{-} =2f_0-\qu\del_t^2F_0,\\
&(F_1'+F_1^{\text{NS}'})^{+}+(F_1'+F_1^{\text{NS}'})^{-}=2(F_1+F_1^\text{NS}). 
\end{aligned}
\label{trff0f1}
\end{equation}
Using these relations \eqref{trff0f1} we find
\begin{equation}
\begin{aligned}
 J_{+}^{+}(2,1)+J_{+}^{-}(2,1)-J(2,0)&=-\log2
+F_1+F_1^\text{NS}-\frac{1}{8}(\del_t^2F_0-2\mu) \\
J_{-}^{+}(2,1)+J_{-}^{-}(2,1)-J(2,0)&=-\log2-\mu+\log(8)
+F_1+F_1^\text{NS}-\frac{1}{8}(\del_t^2F_0-2\mu).
\end{aligned}
\label{k2Jshift}
\end{equation}
In the first line of \eqref{k2Jshift}, the first term on the right hand side  
comes from the perturbative part \eqref{shiftJpert1},
and the remaining terms comes from \eqref{trff0f1}.
Note that the
linear term in $\mu$ in the first line of \eqref{k2Jshift}
is canceled 
on the left hand side,
hence on the right hand side
we subtracted the linear term from $\del_t^2F_0$
so that 
\begin{equation}
 \del_t^2F_0-2\mu=\mathcal{O}(e^{-2\mu}).
\end{equation}
The origin of the last term in the second line of \eqref{k2Jshift} is similar.
Also, note that the ``half-instanton''  contribution $J^{R,\text{np}}(2,1)$ in
\eqref{JR-Modd}
\begin{equation}
  J^{R,\text{np}}(2,1)= \qu\log\left(\frac{1+4e^{-\mu}}{1-4e^{-\mu}}\right)
\end{equation}
 vanishes by adding the $\pm\pi i/2$ shift in \eqref{k2Jshift}
\begin{equation}
 J^{R,\text{np}}(2,1)^{+}+J^{R,\text{np}}(2,1)^{-}=0.
\end{equation}

Next, let us consider the transformation of theta functions under the shift $\mu\to\mu\pm\pi i/2$.
Using the relation
\begin{equation}
\begin{aligned}
 \xi_L'^{+}&=\xi_L+\frac{\tau_L}{2}-\hf,\quad &\tau_L'^{+}&=\tau_L-2,\\
\xi_L'^{-}&=\xi_L,\quad
 &\tau_L'^{-}&=\tau_L,
\end{aligned}
\label{xiL'pm}
\end{equation}
we find
\begin{equation}
 \Xi_{+}^{\pm}(2,1)=e^{J^\pm_+(2,1)}\vartheta_3\left(\xi_L\pm \frac{\tau_L}{4},\tau_L\right),\qquad
\Xi_{-}^{\pm}(2,1)=e^{J^\pm_{-}(2,1)}\vartheta_4\left(\xi_L\pm \frac{\tau_L}{4},\tau_L\right).
\end{equation}
Then, as discussed in \cite{GHM}, using the formula
\eqref{id-theta-add}
one can rewrite the the product of $\Xi_\pm^\pm(2,1)$
appearing on the left hand side of the relation
\eqref{rel1} as
\begin{equation}
\begin{aligned}
 \Xi_{+}^{+}(2,1)\Xi_{+}^{-}(2,1)
=&e^{J^+_+(2,1)+J^-_+(2,1)}\left[\vartheta_3\left(2\xi_L,2\tau_L\right)\vartheta_3\left(\frac{\tau_L}{2},2\tau_L\right)+\vartheta_2\left(2\xi_L,2\tau_L\right)\vartheta_2\left(\frac{\tau_L}{2},2\tau_L\right)\right],
\\
\Xi_{-}^{+}(2,1)\Xi_{-}^{-}(2,1)
=&e^{J^+_-(2,1)+J^-_-(2,1)}\left[\vartheta_3\left(2\xi_L,2\tau_L\right)\vartheta_3\left(\frac{\tau_L}{2},2\tau_L\right)-\vartheta_2\left(2\xi_L,2\tau_L\right)\vartheta_2\left(\frac{\tau_L}{2},2\tau_L\right)\right]. 
\end{aligned}
\label{Xiprod2}
\end{equation}
One can easily see that
the first term in \eqref{Xiprod2} is proportional to $\Xi(2,0)$,
and the $\vartheta_2$ terms in \eqref{Xiprod2} are canceled by 
adding the two lines in \eqref{Xiprod2}
with the proper weight in the functional relation \eqref{rel1}.
Finally, we arrive at the desired functional relation \eqref{rel1}
\begin{equation}
 \Xi_{+}^{+}(2,1)\Xi_{+}^{-}(2,1)+\frac{e^\mu}{8}
\Xi_{-}^{+}(2,1)\Xi_{-}^{-}(2,1)=\mathcal{N}_2(\tau_L)\Xi(2,0),
\label{rel1fork2}
\end{equation}
up to a factor $\mathcal{N}_2(\tau_L)$\footnote{The same factor $\mathcal{N}_2(\tau_L)$ appeared
in the proof of the functional relation \eqref{rel1} for $k=1$ \cite{GHM}.}
\begin{equation}
 \mathcal{N}_2(\tau_L)=e^{f_1-\frac{1}{8}\del_t^2F_0}\vartheta_3\left(\frac{\tau_L}{2},2\tau_L\right)
=e^{f_1}q_L^{\frac{1}{16}}\vartheta_3\left(\frac{\tau_L}{2},2\tau_L\right),
\label{Nfac}
\end{equation}
where $q_L=e^{2\pi i\tau_L}$.
Using the expression of $f_1$ in \eqref{f1theta},
we find that $\mathcal{N}_2(\tau_L)$
 can be written as
\begin{equation}
 \mathcal{N}_2(\tau_L)=\left[\frac{2q_L^{\frac{1}{8}}\vartheta_3\left(\frac{\tau_L}{2},2\tau_L\right)^2}{\vartheta_2(0,\tau_L)\vartheta_3(0,\tau_L)}\right]^\hf.
\label{Nid-k2}
\end{equation}
One can show that this factor is actually unity $\mathcal{N}_2(\tau_L)=1$
(see Appendix \ref{proof:N2} for a proof),
which implies that the functional relation \eqref{rel1} is indeed satisfied for $k=2$. 

\subsubsection{Quantum Wronskian relation \eqref{QWrel1} for $k=2$}
Let us consider the quantum Wronskian relation \eqref{QWrel1}
for $k=2,M=1$
\begin{equation}
 e^{-\frac{\pi i}{4}}\Xi_{+}^{+}(2,0)\Xi_{-}^{-}(2,0)+
e^{\frac{\pi i}{4}}\Xi_{+}^{-}(2,0)\Xi_{-}^{+}(2,0)=\rt{2}\Xi(2,1). 
\label{QW-k2}
\end{equation}
In the case of $k=2,M=1$, the second quantum Wronskian relation \eqref{QWrel2} 
is equivalent to the first quantum Wronskian relation
\eqref{QWrel1}, therefore it is sufficient to consider
\eqref{QWrel1} only.

Again, for readers convenience we recall the exact expression of
the grand partition functions
appearing in  \eqref{QW-k2}
\begin{equation}
\begin{aligned}
\Xi(2,1)&=e^{J(2,1)} \vartheta_4\left(2\xi_L'-\frac{\tau_L'}{2},2\tau_L'\right),\\
 \Xi_{+}(2,0)&=e^{J_{+}(2,0)}\vartheta_4\left(\xi_L+\qu,\tau_L\right),\quad
\Xi_{-}(2,0)=e^{J_{-}(2,0)}\vartheta_3\left(\xi_L+\qu,\tau_L\right),
\end{aligned}
\end{equation}
and
the modified grand potentials are given by
\begin{equation}
 \begin{aligned}
  J(2,1)&=\til{A}(2,1)+f_0'+F_1'+F_1^{\text{NS}'},\\
J_\pm(2,0)&=\til{A}_\pm(2,0)+\hf f_0+F_1+F_1^\text{NS}+\frac{\mu}{8}\pm\left(
\frac{\mu}{4}+\frac{1}{8}\log(1+162^{-2\mu})\right).
 \end{aligned}
\end{equation}
As in the previous subsection \ref{sub:k2rel1},
one can show that the modified grand potential satisfy
\begin{equation}
 \begin{aligned}
  J_{+}^{+}(2,0)+J_{-}^{-}(2,0)-J(2,1)&=-\hf\log2+\frac{\pi i}{4}+F_1'+F_1^{\text{NS}'}
-\frac{1}{8}\left(\del_t^2F_0(t')-2\mu'\right),\\
 J_{+}^{-}(2,0)+J_{-}^{+}(2,0)-J(2,1)&=-\hf\log2-\frac{\pi i}{4}
+F_1'+F_1^{\text{NS}'}-\frac{1}{8}\left(\del_t^2F_0(t')-2\mu'\right).
 \end{aligned}
\end{equation}
The first two terms come from
 the perturbative part and remaining terms come from
 the transformation of $f_0$ and $F_1+F_1^\text{NS}$
 as in \eqref{trff0f1}.
Next we consider the transformation of the theta function part. 
From the relation
\begin{equation}
 \begin{aligned}
  \xi_L^{+}&=\xi_L',\quad
&\tau_L^{+}&=\tau_L',\\
\xi_L^{-}&=\xi_L'-\frac{\tau_L'}{2}-\hf,\quad
&\tau_L^{-}&=\tau_L'+2,
 \end{aligned}
\label{xiLpm}
\end{equation}
we find
\begin{equation}
 \begin{aligned}
  \Xi_{+}^{+}(2,0)&=e^{J^{+}_{+}(2,0)}\vartheta_3\lf(\xi_L'-\qu,\tau_L'\ri),~~
&\Xi_{-}^{-}(2,0)&=e^{J^{-}_{-}(2,0)}\vartheta_3\lf(\xi_L'-\frac{\tau_L'}{2}-\qu,\tau_L'\ri),\\
\Xi_{-}^{+}(2,0)&=e^{J^{-}_{+}(2,0)}\vartheta_4\lf(\xi_L'-\qu,\tau_L'\ri),~~
&\Xi_{+}^{-}(2,0)&=e^{J^{-}_{+}(2,0)}\vartheta_4\lf(\xi_L'-\frac{\tau_L'}{2}-\qu,\tau_L'\ri).
 \end{aligned}
\end{equation}
Again, using the formula \eqref{id-theta-add}
one can show that the functional relation \eqref{QW-k2}  holds
up to the same factor
$\mathcal{N}_2(\tau_L')$ in \eqref{Nfac},
with a simple replacement of the argument $\tau_L\to\tau_L'$
in \eqref{Nfac}.
From the identity $\mathcal{N}_2(\tau_L')=1$
proved in Appendix \ref{proof:N2},
we find that the functional relation
\eqref{QW-k2} is indeed satisfied for $k=2$.

\subsection{Proof of functional relations for $k=4$}
In this subsection we will show the functional relations
\eqref{rel1}, \eqref{QWrel1}, and \eqref{QWrel2} for $k=4$.
As in the case of $k=2$, these relations all follow from
a single identity
$\mathcal{N}_4(\tau)=1$ proved in Appendix \ref{proof:N4}.

\subsubsection{Functional relation \eqref{rel1} for $k=4$}
To prove the functional relation \eqref{rel1} for $k=4$,
let us recall
the exact form of the grand partition functions
appearing in \eqref{rel1}
\begin{equation}
\begin{aligned}
\Xi(4,0)&=e^{J(4,0)}\vartheta_3\left(\xi_L'-\frac{\tau_L'}{4},\tau_L'\right),\\ 
 \Xi_{+}(4,1)&=e^{J_{+}(4,1)}\vartheta_3\left(\frac{\xi_L}{2}-\frac{1}{8},\frac{\tau_L}{2}\right),\quad
\Xi_{-}(4,1)=e^{J_{-}(4,1)}\vartheta_4\left(\frac{\xi_L}{2}-\frac{1}{8},\frac{\tau_L}{2}\right),
\end{aligned}
\end{equation}
and the modified grand potentials are given by
\begin{equation}
 \begin{aligned}
 J(4,0)&=\til{A}(4,0)+\hf f_0'+F_1'+F_1^{\text{NS}'}+\frac{\mu}{4}+\frac{1}{8}
\log\left(\frac{1+4e^{-\mu}}{1-4e^{-\mu}}\right),\\
J_\pm(4,1)&= \til{A}_\pm(4,1)+\qu f_0+F_1+F_1^\text{NS}-\frac{\mu}{16}\pm\frac{\mu}{4}.
 \end{aligned}
\end{equation}
First, the transformation of modified grand potentials
under the shift $\mu\to\mu\pm\pi i/2$
is given by
\begin{equation}
\begin{aligned}
 J_{+}^{+}(4,1)+J_{+}^{-}(4,1)-J(4,0)&=-\log2+F_1'+F_1^{\text{NS}'}
-\frac{1}{16}\left(\del_t^2 F_0(t')-2\mu'\right)
-\frac{1}{8}\log\left(\frac{1+4e^{-\mu}}{1-4e^{-\mu}}\right),\\
 J_{-}^{+}(4,1)+J_{-}^{-}(4,1)-J(4,0)&=-\mu-\log(32)+F_1'+F_1^{\text{NS}'}
-\frac{1}{16}\left(\del_t^2 F_0(t')-2\mu'\right)
-\frac{1}{8}\log\left(\frac{1+4e^{-\mu}}{1-4e^{-\mu}}\right).
\end{aligned}
\end{equation}
Next, we consider the transformation of
the theta functions.
Using \eqref{xiLpm}, we get
\begin{equation}
 \begin{aligned}
  \Xi_{+}^{+}(4,1)&=e^{J_{+}^{+}(4,1)}\vartheta_3\left(\frac{\xi_L'}{2}-\frac{1}{8},\frac{\tau_L'}{2}\right),~~
&\Xi_{+}^{-}(4,1)&=e^{J_{+}^{-}(4,1)}\vartheta_3\left(\frac{\xi_L'}{2}-\frac{\tau_L'}{2}+\frac{1}{8},\frac{\tau_L'}{2}\right),\\
\Xi_{-}^{+}(4,1)&=e^{J_{-}^{+}(4,1)}\vartheta_4\left(\frac{\xi_L'}{2}-\frac{1}{8},\frac{\tau_L'}{2}\right),~~
&\Xi_{-}^{-}(4,1)&=e^{J_{-}^{-}(4,1)}\vartheta_4\left(\frac{\xi_L'}{2}-\frac{\tau_L'}{2}+\frac{1}{8},\frac{\tau_L'}{2}\right),
 \end{aligned}
\end{equation}
and the product of grand partition functions can be simplified
by using the identity of theta functions \eqref{id-theta-add}
\begin{equation}
 \begin{aligned}
  &\Xi_{\pm}^{+}(4,1)\Xi_{\pm}^{-}(4,1)\\
=&e^{J_{\pm}^{+}(4,1)+J_{\pm}^{-}(4,1)}
\left[\vartheta_3\left(\xi_L'-\frac{\tau_L'}{4},\tau_L'\right)
\vartheta_3\left(\frac{\tau_L'-1}{4},\tau_L'\right)
\pm\vartheta_2\left(\xi_L'-\frac{\tau_L'}{4},\tau_L'\right)
  \vartheta_2\left(\frac{\tau_L'-1}{4},\tau_L'\right)\right].
 \end{aligned}
\end{equation}
Again, the first term is proportional to $\Xi(4,0)$ and
the $\vartheta_2$ terms are canceled by adding the two terms
 in the functional relation \eqref{rel1}.
Finally, we find that the functional relation
\eqref{rel1} is satisfied up to a factor $\mathcal{N}_4(\tau')$
\begin{equation}
 \Xi_{+}^{+}(4,1)\Xi_{+}^{-}(4,1)+\frac{e^\mu}{16}
\Xi_{-}^{+}(4,1)\Xi_{-}^{-}(4,1)=\mathcal{N}_4(\tau')\Xi(4,0),
\label{upto-rel1k4}
\end{equation}
where $\tau'=\tau_L'+1$
and $\mathcal{N}_4(\tau')$ is given by
\begin{equation}
 \mathcal{N}_4(\tau')=\exp\left[F_1'+F_1^{\text{NS}'}+\frac{\pi i\tau'}{16}+\frac{\mu}{8}
-\frac{1}{8}\log\left(\frac{1+4e^{-\mu}}{1-4e^{-\mu}}\right)\right]\vartheta_3\left(\frac{\tau'}{4},\tau'\right).
\end{equation}
Using the relation
\begin{equation}
 16e^{-2\mu}=\frac{\vartheta_2(0,\tau')^4}{\vartheta_3(0,\tau')^4},
\end{equation}
and the expression of $F_1+F_1^\text{NS}$ in \eqref{F1FNS-theta},
$\mathcal{N}_4(\tau')$ is rewritten as
\begin{equation}
 \mathcal{N}_4(\tau')=\lf[\frac{(q')^{\frac{1}{8}}}{\vartheta_3(0,\tau')^2\vartheta_4(0,\tau')^2}
\frac{2\vartheta_3(0,\tau')}{\vartheta_2(0,\tau')}\left[\frac{\vartheta_3(0,\tau')^2-\vartheta_2(0,\tau')^2}{\vartheta_3(0,\tau')^2+\vartheta_2(0,\tau')^2}\right]^\hf 
\vartheta_3\left(\frac{\tau'}{4},\tau'\right)^4\ri]^\qu,
\label{N4def}
\end{equation}
with $q'=e^{2\pi i\tau'}$.
One can show that 
$\mathcal{N}_4(\tau')=1$ (see Appendix \ref{proof:N4} for a proof), 
and the functional relation \eqref{rel1} is indeed satisfied
for $k=4$.

\subsubsection{Quantum Wronskian relation \eqref{QWrel1} for $k=4,M=2$}

Let us consider the quantum Wronskian relation \eqref{QWrel1} for the $k=4,M=2$ case:
\begin{equation}
 e^{-\frac{\pi i}{4}}\Xi_{+}^{+}(4,1)\Xi_{-}^{-}(4,1)
+e^{\frac{\pi i}{4}}\Xi_{+}^{-}(4,1)\Xi_{-}^{+}(4,1)=\rt{2}\Xi(4,2).
\label{QW1fork4M2}
\end{equation}
Here we have used the Seiberg-like duality $\Xi_\pm(4,3)=\Xi_\pm(4,1)$.
Again, we recall the exact grand partition functions
\begin{equation}
 \begin{aligned}
  \Xi(4,2)&=e^{J(4,2)}\vartheta_4\left(\xi_L'-\frac{\tau_L'}{4},\tau_L'\right),\\
\Xi_{+}(4,1)&=e^{J_{+}(4,1)}\vartheta_3\left(\frac{\xi_L}{2}-\frac{1}{8},\frac{\tau_L}{2}\right),\quad
\Xi_{-}(4,1)=e^{J_{-}(4,1)}\vartheta_4\left(\frac{\xi_L}{2}-\frac{1}{8},\frac{\tau_L}{2}\right),
 \end{aligned}
\end{equation}
and the modified grand potentials
\begin{equation}
 \begin{aligned}
  J(4,2)&=\til{A}(4,2)+\hf f_0'+F_1'+F_1^{\text{NS}'}-\frac{\mu}{4}-\frac{1}{8}\log\frac{1+4e^{-\mu}}{1-4e^{-\mu}},\\
J_\pm(4,1)&=\til{A}_\pm(4,1)+\qu f_0+F_1+F_1^\text{NS}-\frac{\mu}{16}\pm \frac{\mu}{4}.
 \end{aligned}
\end{equation}
Let us consider the first term on the left hand side of in \eqref{QW1fork4M2}.
From the transformation of $\xi_L$ and $\tau_L$ in
\eqref{xiLpm}, we find
\begin{equation}
 \Xi_{+}^{+}(4,1)=e^{J_{+}^{+}(4,1)}\vartheta_3\left(\frac{\xi_L'}{2}-\frac{1}{8},\frac{\tau_L'}{2}\right),~~
\Xi_{-}^{-}(4,1)=e^{J_{-}^{-}(4,1)}\vartheta_3\left(\frac{\xi_L'}{2}-\frac{\tau_L'}{4}
-\frac{3}{8},\frac{\tau_L'}{2}\right),
\end{equation}
and the product of $\Xi_{+}^{+}(4,1)$ and
$\Xi_{-}^{-}(4,1)$ is simplified by using \eqref{id-theta-add}
as
\begin{equation}
\begin{aligned}
 &\Xi_{+}^{+}(4,1)\Xi_{-}^{-}(4,1)\\
=&e^{J_{+}^{+}(4,1)+J_{-}^{-}(4,1)}\left[\vartheta_3\left(\xi_L'-\frac{\tau_L'}{2}
-\frac{1}{2},\tau_L'\right)\vartheta_3\left(\frac{\tau_L'+1}{4},\tau_L'\right)
+\vartheta_2\left(\xi_L'-\frac{\tau_L'}{2}
 -\frac{1}{2},\tau_L'\right)\vartheta_2\left(\frac{\tau_L'+1}{4},\tau_L'\right)\right].
\end{aligned}
\label{QWrel-1term}
\end{equation}
Here the modified grand potentials satisfy
\begin{equation}
 J_{+}^{+}(4,1)+J_{-}^{-}(4,1)-J(4,2)=\frac{\pi i}{4}-\hf\log2
+F_1'+F_1^{\text{NS}'}-\frac{1}{16}\Bigl(\del_t^2F_0(t')-2\mu'\Bigr)+\frac{1}{8}\log\left(
\frac{1+4e^{-\mu}}{1-4e^{-\mu}}\right).
\end{equation}
Also, we can rewrite the second term on the left hand side in \eqref{QW1fork4M2} in a similar form as \eqref{QWrel-1term}. 
Finally, adding the two terms in \eqref{QW1fork4M2},
we find that the functional relation \eqref{QW1fork4M2} is 
satisfied up to a factor $\til{\mathcal{N}}_4(\tau')$
\begin{equation}
  e^{-\frac{\pi i}{4}}\Xi_{+}^{+}(4,1)\Xi_{-}^{-}(4,1)
+e^{\frac{\pi i}{4}}\Xi_{+}^{-}(4,1)\Xi_{-}^{+}(4,1)=\rt{2}\til{\mathcal{N}}_4(\tau')\Xi(4,2),
\end{equation}
where the factor $\til{\mathcal{N}}_4(\tau')$ is given by
\begin{equation}
 \til{\mathcal{N}}_4(\tau')=\exp\left[F_1'+F_1^{\text{NS}'}+\frac{\pi i\tau'}{16}+\frac{\mu}{8}
+\frac{1}{8}\log\left(\frac{1+4e^{-\mu}}{1-4e^{-\mu}}\right)\right]\vartheta_4\left(\frac{\tau'}{4},\tau'\right).
\end{equation}
It turns out that this factor $\til{\mathcal{N}}_4(\tau')$ is related to 
the factor $\mathcal{N}_4(\tau')$ in \eqref{N4def}
by a shift of the argument $\tau'\to\tau'+2$:
\begin{equation}
 \til{\mathcal{N}}_4(\tau')=\mathcal{N}_4(\tau'+2).
\end{equation}
From the identity $\mathcal{N}_4(\tau)=1$ proved in Appendix \ref{proof:N4},
we find that the quantum Wronskian relation
\eqref{QWrel1} is satisfied for $k=4,M=2$.
In a similar manner, one can show  that the second
 quantum Wronskian relation \eqref{QWrel2}
is also satisfied for $k=4,M=2$.

\subsubsection{Quantum Wronskian relation \eqref{QWrel1} for $k=4,M=1$}
Next, we would like to prove the quantum Wronskian
relation \eqref{QWrel1} for $k=4,M=1$
\begin{equation}
 e^{-\frac{\pi i}{8}}\Xi_{+}^{+}(4,0)\Xi_{-}^{-}(4,2)
+e^{\frac{\pi i}{8}}\Xi_{+}^{-}(4,0)\Xi_{-}^{+}(4,2)
=2\cos\frac{\pi}{8}\Xi(4,1),
\label{QWrel1-k4M1}
\end{equation}
where the exact grand partition functions are given by
\begin{equation}
 \begin{aligned}
  \Xi(4,1)&=e^{J(4,1)}\vartheta_3\left(\xi_L+\qu,\tau_L\right),\\
\Xi_{+}(4,0)&=e^{J_{+}(4,0)}\vartheta_3\left(\frac{\xi_L'}{2}-\frac{\tau_L'}{8},\frac{\tau_L'+1}{2}\right),\\
\Xi_{-}(4,2)&=e^{J_{+}(4,2)}\vartheta_4\left(\frac{\xi_L'}{2}-\frac{\tau_L'}{8}-\frac{1}{4},\frac{\tau_L'+1}{2}\right),
 \end{aligned}
\end{equation}
and the modified grand potentials are
\begin{equation}
 \begin{aligned}
 J(4,1)&=\til{A}(4,1)+\hf f_0+F_1+F_1^\text{NS}-\frac{\mu}{8}-\frac{1}{8}\log(1+16e^{-2\mu}), \\
J_{+}(4,0)&=\til{A}_{+}(4,0)+\qu f_0'+F_1'+F_1^{\text{NS}'}+\frac{3\mu}{8}+\frac{1}{8}\log(1-4e^{-\mu}),\\
J_{-}(4,2)&=\til{A}_{+}(4,2)+\qu f_0'+F_1'+F_1^{\text{NS}'}-\frac{3\mu}{8}-\frac{3}{8}\log(1+4e^{-\mu}).
 \end{aligned}
\end{equation}
Let us consider the first term on the left hand side of 
\eqref{QWrel1-k4M1}.
Using the transformation
of $\xi_L$ and $\tau_L$ in \eqref{xiL'pm}, we find
\begin{equation}
 \Xi_{+}^{+}(4,0)=e^{J_{+}^{+}(4,0)}\vartheta_4\left(\frac{\xi_L}{2}+\frac{\tau_L}{8},\frac{\tau_L+1}{2}\right),\quad
\Xi_{-}^{-}(4,2)=e^{J_{-}^{-}(4,2)}\vartheta_4\left(\frac{\xi_L}{2}-\frac{\tau_L}{8}-\qu,\frac{\tau_L+1}{2}\right).
\end{equation}
Using the identity of theta functions \eqref{id-theta-add} again,  
the product of grand partition functions can be rewritten as
\begin{equation}
 \begin{aligned}
 & \Xi_{+}^{+}(4,0)\Xi_{-}^{-}(4,2)\\
=& e^{J_{+}^{+}(4,0)+J_{-}^{-}(4,2)}
\left[\vartheta_3\left(\xi_L-\qu,\tau_L+1\right)\vartheta_3\left(\frac{\tau}{4},\tau\right)
-\vartheta_2\left(\xi_L-\qu,\tau_L+1\right)\vartheta_2\left(\frac{\tau}{4},\tau\right)\right],
 \end{aligned}
\end{equation}
with $\tau=\tau_L+1$.
The second term on the left hand side of \eqref{QWrel1-k4M1}
can be rewritten similarly.
Finally, using the relation of the grand potentials
\begin{equation}
  J_+^+(4,0)+J_-^-(4,2)-J(4,1)=\frac{3\pi i}{8}-\log\left(2\sin\frac{\pi}{8}\right)+
F_1+F_1^\text{NS}
-\frac{1}{16}(\del_t^2F_0-2\mu)+\frac{1}{8}\log\frac{1-4ie^{-\mu}}{1+4ie^{-\mu}},
\end{equation}
one can show that the functional relation \eqref{QWrel1} holds
up to the same factor $\mathcal{N}_4(\tau)$ as \eqref{N4def}, where 
the argument is simply replaced $\tau'\to\tau$ in \eqref{N4def}.
From the identity $\mathcal{N}_4(\tau)=1$, we find that the 
functional relation \eqref{QWrel1} is indeed satisfied for $k=4,M=1$.
Also, we can prove the second quantum Wronskian relation \eqref{QWrel2} for $k=4,M=1$ in a completely parallel way.

\subsection{Comments on other $k$'s}
For other values of $k$'s we do not have a general proof of the relations
\eqref{rel1}, \eqref{QWrel1}, and \eqref{QWrel2}.
However, we can check these relations by comparing the small $\ka$
expansion on both sides
using our data of the exact values of $Z_\pm(N,k,M)$ and
the known values of the canonical partition functions $Z(N,k,M)$
of ABJ(M) theories computed in \cite{HMO2,MaMo,HO}.
We find a complete agreement for $k=3,6$.
On the other hand, for $k=8,12$ the exact values of ABJ(M)
partition functions $Z(N,k,M)$ are not known in the literature.
Rather, assuming that the functional relations are correct,
we can predict the values of $Z(N,k,M)$ with $k=8,12$ for all $0\leq M\leq k$ using the functional relations.
We have checked that those values of $Z(N,k,M)$ for $k=8,12$
obtained in this manner are all consistent with
the conjectured modified grand potential \eqref{Jconjecture}
given by the free energy of (refined) topological string
on local $\mathbb{P}^1\times\mathbb{P}^1$ \cite{HMMO}.

As shown in \eqref{82-vs-42},
the grand partition function $\Xi(8,2)$
of $(k,M)=(8,2)$ ABJ theory
is equal to $\Xi_{-}(4,2)$.
On the other hand, the functional relation \eqref{QWrel1}
predicts that $\Xi(8,2)$ is given by
\begin{equation}
 \Xi(8,2)=\frac{1}{2\cos\frac{\pi}{8}}\left[e^{-\frac{\pi i}{8}}\Xi_{+}^{+}(8,1)\Xi_{-}^{-}(8,3)
+e^{\frac{\pi i}{8}}\Xi_{+}^{-}(8,1)\Xi_{-}^{+}(8,3)\right].
\label{rel82}
\end{equation}
Plugging the small $\ka$ expansion of $\Xi_{+}(8,1)$ \eqref{Xipm81exp} 
and $\Xi_{-}(8,3)$ \eqref{Xipm83exp} into this functional relation \eqref{rel82}, we get
\begin{equation}
 \Xi(8,2)=1+\frac{4+4 \sqrt{2}-3 \pi }{32 \pi }\ka+\frac{-32-24 \left(1+\sqrt{2}\right) \pi +\left(33-8
   \sqrt{2}\right) \pi ^2}{2048 \pi ^2}\ka^2+\cdots,
\end{equation}
which indeed agrees with the expansion of $\Xi_{-}(4,2)$ in \eqref{Xipm42exp}.

\section{Conclusion}\label{sec:conclusion}
In this paper, we have studied the orientifolding
of the ABJ Fermi gas by
projecting to the even/odd sectors under
the reflection of fermion coordinate.
$\Xi_{-}(\mu,k,0)$ and
$\Xi_{+}(\mu,k,1)$ are indeed
related to the $\mathcal{N}=5$ $O(n)\times USp(n')$
theories, and we expect that
more general cases $\Xi_\pm(\mu,k,M)$ are also related
to M-theory on some orientifold background.
It would be interesting to understand the spacetime picture (if any)
for the general cases $\Xi_\pm(\mu,k,M)$
and study the precise relation between the reflection of fermion coordinate and the orientifolding of spacetime.
The relation to the orientifold is also suggested
by the appearance of non-orientable
contribution $F_\text{non-ori}$ in the constant term
$A_\pm(k,M)$ \eqref{Apm} in the perturbative part of grand potential.

The grand potential of orientifold ABJ theory receives
three types of instanton corrections \eqref{3-type}:
worldsheet instantons, membrane instantons,
and ``half-instantons''.
We find that the half-instanton
corrections come from the twisted spectral trace
$\Tr(\rho^s R)$, and we determined the first few half-instanton coefficients
in closed forms \eqref{rell} as functions of $k$ and $b=M/k$.
We also observed that the coefficients of worldsheet instanton corrections
are different from the worldsheet instanton coefficients
of ABJ theory before orientifolding.
It would be interesting to understand the structure of
worldsheet instantons in orientifold ABJ theory better,
and
see if they have some relation to the
``real  topological string'' on the
local $\mathbb{P}^1\times\mathbb{P}^1$ \cite{Walcher:2007qp,Krefl:2009mw}.

For $k=2,4,8$ we have found the closed form expression of the grand partition function of orientifold ABJ theory
in terms of the theta functions, and
we have also checked for $k=2,4$ that
the functional relations conjectured in \cite{GHM}
are indeed satisfied.
We found that the structure of grand partition function
at $k=8$
is qualitatively different from the cases
of $k=2,4$.
For $k=2,4$, the modified grand potentials are
basically determined by the genus-zero and genus-one
free energies,
while for the $k=8$ case
the modified grand potentials receives all genus corrections
in the string coupling expansion.
Nevertheless,
we have managed to find the closed form expression of the
all order resummation of worldsheet instanton corrections $G_{o}(\mu)$
in \eqref{defG} and $G_{e}(\mu)$ in \eqref{defGe}.
Since $G_{o,e}(\mu)$ 
are not invariant under the $2\pi i$-shift,
 we divided
 the periodic sum into two parts.
  In general, the worldsheet 1-instanton factor $e^{-4\mu/k}$
  is not invariant under the $2\pi i$-shift of $\mu$,
  except for the $k=1,2,4$ cases.
  Therefore, for generic $k\in\mathbb{Z}_{>0}$,
  the periodic sum \eqref{periodic-sum} should be
  taken separately according to $n$ modulo $k$ (or $n$ modulo $k/2$ for even $k$),
  and the resulting exact grand partition function
 will be written as a sum of certain building blocks
 carrying definite $\mathbb{Z}_k$-charge (or  $\mathbb{Z}_{k/2}$-charge for even $k$).
 It would be very interesting to find a closed form expression of
 the grand partition functions for general integer $k$, say
 $k=3,6$.

 As observed in \cite{CGM,GHM},
 the exact forms of the grand partition functions
 of (orientifold) ABJ theories have the same form as
 the ``non-perturbative partition functions'' proposed
 in \cite{Eynard:2008he}.
 Our exact grand partition functions have a nice modular property,
 and we expect that they are invariant under some subgroup of $SL(2,\mathbb{Z})$.
 It would be nice to understand the modular property of $\Xi_\pm(\mu,k,M)$ better.
 Interestingly, the exact form of $\Xi_\pm(\mu,k,M)$ we found
 is a holomorphic function of $\mu$ at fixed $k,M$.
 As emphasized in  \cite{Eynard:2008he},
 the holomorphic anomaly might be an artifact of
 the perturbative genus expansion and the holomorphicity
 and modularity are restored at the non-perturbative level.
 It would be very interesting to study this mechanism in
 a concrete example, and we believe that
 the (orientifold) ABJ theory serves as a good
 testing ground for this purpose.

 The exact grand partition functions of (orientifold)
  ABJ theories we obtained are reminiscent of the torus partition functions of chiral fermions in $2d$ CFT.
  It has long been suspected that the partition functions
  of matrix models have a natural interpretation in $2d$ CFT living on spectral curves (see e.g. \cite{Kostov:2010nw} and references therein).
  It would be interesting to clarify the precise relation between
  our exact grand partition functions and $2d$ CFT.
 
\acknowledgments
I would like to thank Yasuyuki Hatsuda, Masazumi Honda, and Sanefumi Moriyama
for useful discussions.

\appendix

\section{Theta functions}\label{app:theta}
In this Appendix, we summarize
some useful properties of theta functions, and
give a proof of $\mathcal{N}_2(\tau)=1$ and $\mathcal{N}_4(\tau)=1$.
Here we use the notation
\begin{equation}
 q=e^{2\pi i\tau},\qquad z=e^{2\pi iv}.
\end{equation}

\paragraph{Jacobi theta functions.}
The Jacobi theta functions $\vartheta_a(v,\tau)~(a=1,2,3,4)$ are defined by
\begin{equation}
 \begin{aligned}
  \vartheta_1(v,\tau)&=-i\sum_{n\in\mathbb{Z}}q^{\hf(n+1/2)^2}z^{n+1/2}(-1)^n
=-iq^{\frac{1}{8}}(z^{\hf}-z^{-\hf})\prod_{n=1}^\infty (1-q^n)(1-zq^n)(1-z^{-1}q^n),\\
 \vartheta_2(v,\tau)&=\sum_{n\in\mathbb{Z}}q^{\hf(n+1/2)^2}z^{n+1/2}=
q^{\frac{1}{8}}(z^{\hf}+z^{-\hf})\prod_{n=1}^\infty (1-q^n)(1+zq^n)(1+z^{-1}q^n),\\
\vartheta_3(v,\tau)&=\sum_{n\in\mathbb{Z}}q^{\hf n^2}z^{n}=
\prod_{n=1}^\infty (1-q^n)(1+zq^{n-\hf})(1+z^{-1}q^{n-\hf}),\\
\vartheta_4(v,\tau)&=\sum_{n\in\mathbb{Z}}q^{\hf n^2}(-z)^{n}=
\prod_{n=1}^\infty (1-q^n)(1-zq^{n-\hf})(1-z^{-1}q^{n-\hf}).
\label{Jtheta-def}
 \end{aligned}
\end{equation}
Under the $S$-transformation $\tau\to-1/\tau$, they behave as
\begin{equation}
 \begin{aligned}
  \vartheta_1(v,\tau)&=i(-i\tau)^{-\hf}e^{-\frac{\pi iv^2}{\tau}}\vartheta_1(v/\tau,-1/\tau),\\
\vartheta_4(v,\tau)&=(-i\tau)^{-\hf}e^{-\frac{\pi iv^2}{\tau}}\vartheta_2(v/\tau,-1/\tau),\\
\vartheta_3(v,\tau)&=(-i\tau)^{-\hf}e^{-\frac{\pi iv^2}{\tau}}\vartheta_3(v/\tau,-1/\tau).
 \end{aligned}
 \label{thetaS}
\end{equation}

\paragraph{Theta function with characteristic.}
We define the theta function with characteristic $(a,b)$ by
\begin{equation}
\vartheta\Bigl[
\begin{matrix}
 a\\b
\end{matrix}\Bigr](v,\tau)= e^{\pi i a^2\tau+2\pi ia(v+b)}\vartheta_3(v+a\tau+b,\tau)=\sum_{n\in\mathbb{Z}}e^{\pi i (n+a)^2+
2\pi i(n+a)(v+b)}.
\end{equation}
Under the $S$-transformation, this becomes
\begin{equation}
 \vartheta\Bigl[
\begin{matrix}
 a\\b
\end{matrix}\Bigr](v,\tau)=(-i\tau)^{-\hf}e^{2\pi iab-\frac{\pi iv^2}{\tau}}
 \vartheta\Bigl[
\begin{matrix}
 -b\\a
\end{matrix}\Bigr]\Bigl(\frac{v}{\tau},-\frac{1}{\tau}\Bigr).
\end{equation}
The Jacobi theta functions can be written as
\begin{equation}
 \begin{aligned}
 \vartheta_1(v,\tau)&=- \vartheta\Bigl[
\begin{matrix}
 1/2\\1/2
  \end{matrix}\Bigr](v,\tau),~~
   &\vartheta_2(v,\tau)&=\vartheta\Bigl[
\begin{matrix}
 1/2\\0
  \end{matrix}\Bigr](v,\tau),\\
    \vartheta_3(v,\tau)&=\vartheta\Bigl[
\begin{matrix}
 0\\0
  \end{matrix}\Bigr](v,\tau),~~
  &\vartheta_4(v,\tau)&=\vartheta\Bigl[
\begin{matrix}
 0\\1/2
  \end{matrix}\Bigr](v,\tau).  
 \end{aligned}
\end{equation}
 
\paragraph{Useful identities.}
The Jacobi theta functions satisfy
\begin{equation}
 \begin{aligned}
  \vartheta_3(v\pm 1/2,\tau)&=\vartheta_3(v,\tau\pm 1)=\vartheta_4(v,\tau),\\
\vartheta_2(v\pm 1/2,\tau)&=\mp \vartheta_1(v,\tau), \\
\vartheta_2(v,\tau)&=e^{\pi i(\tau/4\pm v)}\vartheta_3(v\pm\tau/2,\tau).
 \end{aligned}
\end{equation}
and
\begin{equation}
 \vartheta_1(v,\tau)^4+\vartheta_3(v,\tau)^4=\vartheta_2(v,\tau)^4+\vartheta_4(v,\tau)^4.
\label{id-abst}
\end{equation}
The Dedekind eta function
defined by
\begin{equation}
 \eta(\tau)=q^{\frac{1}{24}}\prod_{n=1}^\infty (1-q^n),
\end{equation}
is related to the Jacobi theta functions as
\begin{equation}
 2\eta(\tau)^3=\vartheta_2(0,\tau)\vartheta_3(0,\tau)\vartheta_4(0,\tau).
  \label{id-eta}
\end{equation}
The theta functions with argument $\tau$ and $2\tau$ are related by
\begin{equation}
 \vartheta_3(v,\tau)\vartheta_4(v,\tau)=\vartheta_4(0,2\tau)\vartheta_4(2v,2\tau),
\label{id-landen}
\end{equation}
and $\tau$ and $4\tau$ are related by
\begin{equation}
 \begin{aligned}
  \vartheta_3(v,\tau)&=\vartheta_3(2v,4\tau)+\vartheta_2(2v,4\tau),\\
 \vartheta_4(v,\tau)&=\vartheta_3(2v,4\tau)-\vartheta_2(2v,4\tau).
\label{id-4tau}
 \end{aligned}
\end{equation}
From \eqref{id-landen}, \eqref{id-4tau}, and the product representation of
theta functions \eqref{Jtheta-def},
one can show that
\begin{equation}
 \begin{aligned}
   \vartheta_3(0,\tau)^2-\vartheta_2(0,\tau)^2
&= \vartheta_4(0,\tau/2)^2,\\
 \vartheta_3(0,\tau)^2+\vartheta_2(0,\tau)^2
&= \vartheta_3(0,\tau/2)^2,\\
2\vartheta_2(0,\tau)\vartheta_3(0,\tau)&=\vartheta_2(0,\tau/2)^2.
 \end{aligned}
\label{id-th^2}
\end{equation}
To show the functional relations \eqref{rel1}, \eqref{QWrel1},
and \eqref{QWrel2}, we use the following identities
\begin{equation}
 \begin{aligned}
  \vartheta_3(x+y,\tau)\vartheta_3(x-y,\tau)&=\vartheta_3(2x,2\tau)\vartheta_3(2y,2\tau)
+\vartheta_2(2x,2\tau)\vartheta_2(2y,2\tau),\\
\vartheta_4(x+y,\tau)\vartheta_4(x-y,\tau)&=\vartheta_3(2x,2\tau)\vartheta_3(2y,2\tau)
-\vartheta_2(2x,2\tau)\vartheta_2(2y,2\tau).
 \end{aligned}
\label{id-theta-add}
\end{equation}

\paragraph{Elliptic modulus and theta functions.}
The elliptic modulus $k^2$ 
appearing in the complete elliptic integral of the first kind $K(k^2)$
is related to the Jacobi theta functions by
\begin{equation}
 \frac{2}{\pi}K(k^2)=\vartheta_3(0,\tau)^2,~~
\tau=i\frac{K(1-k^2)}{K(k^2)},~~
k^2=\frac{\vartheta_2(0,\tau)^4}{\vartheta_3(0,\tau)^4},~~
1-k^2=\frac{\vartheta_4(0,\tau)^4}{\vartheta_3(0,\tau)^4}.
\end{equation}

\subsection{Proof of $\mathcal{N}_2(\tau)=1$}\label{proof:N2}
In this subsection we will prove the identity $\mathcal{N}_2(\tau)=1$
in \eqref{Nid-k2}, which amounts to
showing that
\begin{equation}
 \vartheta_2(0,\tau)\vartheta_3(0,\tau)=2q^{\frac{1}{8}}\vartheta_3(\tau/2,2\tau)^2.
\label{thetaid-for-k2}
\end{equation}
As we will see below, this follows directly from the product representation of
theta functions \eqref{Jtheta-def}.
First, notice that
\begin{equation}
 \vartheta_3(\tau/2,2\tau)=\prod_{n=1}^\infty (1-q^{2n})(1+q^\hf q^{2n-1})(1+q^{-\hf}q^{2n-1})
=\prod_{n=1}^\infty (1-q^{2n})(1+q^{2n-\hf})(1+q^{2n-1-\hf}).
\end{equation}
The last two factors can be thought of as the 
product of $q^{m-\hf}$ with even $m$ $(m=2n)$ and odd $m$ $(m=2n-1)$. Thus, we find
\begin{equation}
 \vartheta_3(\tau/2,2\tau)
=\prod_{n=1}^\infty (1-q^{2n})(1+q^{n-\hf}).
\label{rhsN2}
\end{equation}
On the other hand, by multiplying $\vartheta_2(0,\tau)$
and $\vartheta_3(0,\tau)$ in
\eqref{Jtheta-def}, we find
\begin{equation}
 \vartheta_2(0,\tau)\vartheta_3(0,\tau)= 2q^{\frac{1}{8}}
\prod_{n=1}^\infty (1-q^n)^2(1+q^n)^2(1+q^{n-\hf})^2=
2q^{\frac{1}{8}}
\prod_{n=1}^\infty (1-q^{2n})^2(1+q^{n-\hf})^2.
\label{lhsN2}
\end{equation}
Comparing \eqref{rhsN2} and \eqref{lhsN2}, we find the desired
relation \eqref{thetaid-for-k2}.

\subsection{Proof of $\mathcal{N}_4(\tau)=1$}\label{proof:N4}
In this subsection, we will prove the identity $\mathcal{N}_4(\tau)=1$.
Using \eqref{id-abst}, \eqref{thetaid-for-k2}, and \eqref{id-th^2},
we can rewrite $\mathcal{N}_4(\tau)^4$ in \eqref{N4def}  as
\begin{equation}
\begin{aligned}
 \mathcal{N}_4(\tau)^4&=\frac{1}{\vartheta_3(\tau/2,2\tau)^2\vartheta_4(0,\tau)^2}
\frac{[\vartheta_3(0,\tau)^4-\vartheta_2(0,\tau)^4]^\hf}{\vartheta_3(0,\tau)^2+\vartheta_2(0,\tau)^2} 
\vartheta_3(\tau/4,\tau)^4\\
&=\frac{\vartheta_3(\tau/4,\tau)^4}{\vartheta_3(\tau/2,2\tau)^2\vartheta_3(0,\tau/2)^2}.
\end{aligned}
\end{equation}
Therefore, to prove $\mathcal{N}_4(\tau)=1$ we have to show that
\begin{equation}
 \vartheta_3(\tau/2,2\tau)\vartheta_3(0,\tau/2)=\vartheta_3(\tau/4,\tau)^2.
\label{N4amount}
\end{equation}

First, $\vartheta_3(\tau/4,\tau)$ on the right hand side of \eqref{N4amount} is 
easily obtained from \eqref{rhsN2} by replacing $q\to q^\hf$
\begin{equation}
 \vartheta_3(\tau/4,\tau)=\prod_{n=1}^\infty(1-q^n)(1+q^{\frac{n}{2}-\qu}).
\label{rhsN4}
\end{equation}
Next consider the the left hand side of \eqref{N4amount}.
From \eqref{rhsN2},  we can rewrite $\vartheta_3(\tau/2,2\tau)$  as
\begin{equation}
 \vartheta_3(\tau/2,2\tau)=\prod_{n=1}^\infty(1-q^n)(1+q^n)(1+q^{n-\hf})=
\prod_{n=1}^\infty(1-q^n)(1+q^{\frac{n}{2}}).
\label{th3-1}
\end{equation}
In the last step we have used the same reasoning to derive \eqref{rhsN2}.
From \eqref{Jtheta-def}$,
\vartheta_3(0,\tau/2)$ is given by
\begin{equation}
 \vartheta_3(0,\tau/2)=\prod_{n=1}^\infty(1-q^{\frac{n}{2}})(1+q^{\hf(n-\hf)})^2.
\label{th3-2}
\end{equation}
Then, multiplying \eqref{th3-1} and \eqref{th3-2} we find
\begin{equation}
 \vartheta_3(\tau/2,2\tau)\vartheta_3(0,\tau/2)=\prod_{n=1}^\infty
(1-q^n)^2(1+q^{\frac{n}{2}-\qu})^2.
\label{lhsN4}
\end{equation}
Comparing \eqref{lhsN4} and \eqref{rhsN4},
we find the desired relation \eqref{N4amount}.

\section{Instanton expansion of $J_\pm(k,M)$}\label{app:Jpm-inst}
In this Appendix, we summarize the non-perturbative corrections
to the grand potential
$J_\pm^{\text{np}}(k,M)$ obtained from
the exact values of $Z_\pm(N,k,M)$.
\footnotesize
\subsection{$k=2$}
\begin{equation}
  \begin{aligned}
J_+^\text{np}(2,0)&=\left[\frac{4\mu^2+2\mu+1}{2\pi^2}+2\right]e^{-2\mu}
+\left[-\frac{52\mu^2+\mu+9/4}{4\pi^2}-14\right]e^{-4\mu}\\
&+\left[\frac{736\mu^2-304/3\mu+154/9}{6\pi^2}+\frac{416}{3}\right]e^{-6\mu}
+\left[-\frac{2701\mu^2-13949/24\mu+11291/192}{2\pi^2}-1582\right]e^{-8\mu}\\
&+\left[\frac{161824\mu^2-624244/15\mu+285253/75}{10\pi^2}+\frac{97472}{5}\right]e^{-10\mu}\\
J_-^\text{np}(2,0)&=\left[\frac{4\mu^2+2\mu+1}{2\pi^2}-2\right]e^{-2\mu}
+\left[-\frac{52\mu^2+\mu+9/4}{4\pi^2}+18\right]e^{-4\mu}
\\
&+\left[\frac{736\mu^2-304/3\mu+154/9}{6\pi^2}-\frac{608}{3}\right]e^{-6\mu}
+\left[-\frac{2701\mu^2-13949/24\mu+11291/192}{2\pi^2}+2514\right]e^{-8\mu}
\\
&+\left[\frac{161824\mu^2-624244/15\mu+285253/75}{10\pi^2}-\frac{164672}{5}\right]e^{-10\mu}
\end{aligned}
\label{Jpm20}
\end{equation}

\begin{equation}
 \begin{aligned}
  J_{+}^\text{np}(\mu,2,1)&=e^{-\mu}-\frac{4\mu^2+2\mu+1}{2\pi^2}e^{-2\mu}
+\frac{16}{3}e^{-3\mu}
+\left[-\frac{52\mu^2+\mu+9/4}{4\pi^2}+2\right]e^{-4\mu}+\frac{256}{5}e^{-5\mu}\\
&+\left[-\frac{736\mu^2-304/3\mu+154/9}{6\pi^2}+32\right]e^{-6\mu}+\frac{4096}{7}e^{-7\mu}\\
&+\left[-\frac{2701\mu^2-13949/24\mu+11291/192}{2\pi^2}+466\right]e^{-8\mu}\\
J_{-}^\text{np}(\mu,2,1)&=-e^{-\mu}-\frac{4\mu^2+2\mu+1}{2\pi^2}e^{-2\mu}-\frac{16}{3}e^{-3\mu}
+\left[-\frac{52\mu^2+\mu+9/4}{4\pi^2}+2\right]e^{-4\mu}-\frac{256}{5}e^{-5\mu}\\
&+\left[-\frac{736\mu^2-304/3\mu+154/9}{6\pi^2}+32\right]e^{-6\mu}-\frac{4096}{7}e^{-7\mu}\\
&+\left[-\frac{2701\mu^2-13949/24\mu+11291/192}{2\pi^2}+466\right]e^{-8\mu}
 \end{aligned}
\label{Jpm21}
\end{equation}

\subsection{$k=4$}
\begin{equation}
 \begin{aligned}
  J_+^\text{np}(4,0)&=-\hf e^{-\mu}+\left[-\frac{4\mu^2+2\mu+1}{4\pi^2}-1\right]e^{-2\mu}-\frac{8}{3}e^{-3\mu}
 +\left[-\frac{52\mu^2+\mu+9/4}{8\pi^2}-6\right]e^{-4\mu}\\
 &\quad-\frac{128}{5}e^{-5\mu}
  +\left[-\frac{736\mu^2-304/3\mu+154/9}{12\pi^2}-\frac{160}{3}\right]e^{-6\mu}-\frac{2048}{7}e^{-7\mu}\\
&\quad+\left[-\frac{2701\mu^2-13949/24\mu+11291/192}{4\pi^2}-558\right]e^{-8\mu}
-\frac{32768}{9}e^{-9\mu}\\
&\quad+\left[-\frac{161824\mu^2-624244/15\mu+285253/75}{20\pi^2}-\frac{31936}{5}\right]e^{-10\mu}\\
  J_-^\text{np}(4,0)&=\frac{3}{2}e^{-\mu}
  +\left[-\frac{4\mu^2+2\mu+1}{4\pi^2}+3\right]e^{-2\mu}
  +8e^{-3\mu}
  +\left[-\frac{52\mu^2+\mu+9/4}{8\pi^2}+26\right]e^{-4\mu}\\
  &\quad+\frac{384}{5}e^{-5\mu}
  +\left[-\frac{736\mu^2-304/3\mu+154/9}{12\pi^2}+288\right]e^{-6\mu}
+\frac{6144}{7}e^{-7\mu}\\
&\quad+\left[-\frac{2701\mu^2-13949/24\mu+11291/192}{4\pi^2}+3538\right]e^{-8\mu}
+\frac{32768}{3}e^{-9\mu}\\
&\quad+\left[-\frac{161824\mu^2-624244/15\mu+285253/75}{20\pi^2}+\frac{230208}{5}\right]e^{-10\mu}
 \end{aligned}
\label{Jpm40}
\end{equation}

\begin{equation}
 \begin{aligned}
  J_+^\text{np}(4,1)&=\frac{4\mu^2+2\mu+1}{4\pi^2}e^{-2\mu}
+\left[-\frac{52\mu^2+\mu+9/4}{8\pi^2}+2\right]e^{-4\mu}
+\left[\frac{736\mu^2-304/3\mu+154/9}{12\pi^2}-32\right]e^{-6\mu}\\
&+\left[-\frac{2701\mu^2-13949/24\mu+11291/192}{4\pi^2}+466\right]e^{-8\mu}\\
J_-^\text{np}(4,1)&=\frac{4\mu^2+2\mu+1}{4\pi^2}e^{-2\mu}
+\left[-\frac{52\mu^2+\mu+9/4}{8\pi^2}+2\right]e^{-4\mu}
+\left[\frac{736\mu^2-304/3\mu+154/9}{12\pi^2}-32\right]e^{-6\mu}\\
&+\left[-\frac{2701\mu^2-13949/24\mu+11291/192}{4\pi^2}+466\right]e^{-8\mu}
 \end{aligned}
\label{Jpm41}
\end{equation}

\begin{equation}
 \begin{aligned}
  J_+^\text{np}(4,2)&=\hf e^{-\mu}+\left[-\frac{4\mu^2+2\mu+1}{4\pi^2}-1\right]e^{-2\mu} +\frac{8}{3}e^{-3\mu}
  +\left[-\frac{52\mu^2+\mu+9/4}{8\pi^2}-6\right]e^{-4\mu}\\
 &\quad+\frac{128}{5}e^{-5\mu}
  +\left[-\frac{736\mu^2-304/3\mu+154/9}{12\pi^2}-\frac{160}{3}\right]e^{-6\mu}
+\frac{2048}{7}e^{-7\mu}\\
&\quad+\left[-\frac{2701\mu^2-13949/24\mu+11291/192}{4\pi^2}-558\right]e^{-8\mu}
+\frac{32768}{9}e^{-9\mu}\\
&\quad+\left[-\frac{161824\mu^2-624244/15\mu+285253/75}{20\pi^2}-\frac{31936}{5}\right]e^{-10\mu}\\
  J_-^\text{np}(4,2)&=-\frac{3}{2} e^{-\mu}
  +\left[-\frac{4\mu^2+2\mu+1}{4\pi^2}+3\right]e^{-2\mu}
  -8e^{-3\mu}+\left[-\frac{52\mu^2+\mu+9/4}{8\pi^2}+26\right]e^{-4\mu}\\
  &\quad-\frac{384}{5}e^{-5\mu}
  +\left[-\frac{736\mu^2-304/3\mu+154/9}{12\pi^2}+288\right]e^{-6\mu}
-\frac{6144}{7}e^{-7\mu}\\
&\quad+\left[-\frac{2701\mu^2-13949/24\mu+11291/192}{4\pi^2}+3538\right]e^{-8\mu}
-\frac{32768}{3}e^{-9\mu}\\
&\quad+\left[-\frac{161824\mu^2-624244/15\mu+285253/75}{20\pi^2}+\frac{230208}{5}\right]e^{-10\mu}
 \end{aligned}
\label{Jpm42}
\end{equation}

\subsection{$k=8$}
For odd $M$, we find
\begin{equation}
 \begin{aligned}
  J_{+}^\text{np}(8,1)&=\frac{1}{\rt{2}}e^{-\frac{\mu}{2}}-\frac{4}{3\rt{2}}e^{-\frac{3\mu}{2}}
+\frac{4\mu^2+2\mu+1}{8\pi^2}e^{-2\mu}
-\frac{16}{5\rt{2}}e^{-\frac{5\mu}{2}}
+\frac{64}{7\rt{2}}e^{-\frac{7\mu}{2}}\\
&+\left[-\frac{52\mu^2+\mu+9/4}{16\pi^2}+2\right]e^{-4\mu}
+\frac{256}{9\rt{2}}e^{-\frac{9\mu}{2}}
-\frac{1024}{11\rt{2}}e^{-\frac{11\mu}{2}}\\
&+\left[\frac{736\mu^2-304/3\mu+154/9}{24\pi^2}-32\right]e^{-6\mu}
-\frac{4096}{13\rt{2}}e^{-\frac{13\mu}{2}}
+\frac{4^7}{15\rt{2}}e^{-\frac{15\mu}{2}}\\
&+\left[-\frac{2701\mu^2-13949/24\mu+11291/192}{8\pi^2}+466\right]e^{-8\mu}\\
 \end{aligned}
\label{Jpm81}
\end{equation}

\begin{equation}
 \begin{aligned}
  J_{+}^\text{np}(8,3)&=-\frac{1}{\rt{2}}e^{-\frac{\mu}{2}}
+\frac{4}{3\rt{2}}e^{-\frac{3\mu}{2}}
+\frac{4\mu^2+2\mu+1}{8\pi^2}e^{-2\mu}
+\frac{16}{5\rt{2}}e^{-\frac{5\mu}{2}}
-\frac{64}{7\rt{2}}e^{-\frac{7\mu}{2}}\\
&+\left[-\frac{52\mu^2+\mu+9/4}{16\pi^2}+2\right]e^{-4\mu}
-\frac{256}{9\rt{2}}e^{-\frac{9\mu}{2}}
+\frac{1024}{11\rt{2}}e^{-\frac{11\mu}{2}}\\
&+\left[\frac{736\mu^2-304/3\mu+154/9}{24\pi^2}-32\right]e^{-6\mu}
+\frac{4096}{13\rt{2}}e^{-\frac{13\mu}{2}}
-\frac{4^7}{15\rt{2}}e^{-\frac{15\mu}{2}}\\
&+\left[-\frac{2701\mu^2-13949/24\mu+11291/192}{8\pi^2}+466\right]e^{-8\mu}
 \end{aligned}
\label{Jpm83}
\end{equation}
We find that the instanton corrections of $J_{-}(8,M)$ are
equal to $J_{+}^\text{np}(8,M)$ for $M=1,3$
\begin{equation}
 J_{-}^\text{np}(8,M)=J_{+}^\text{np}(8,M)\qquad(M=1,3).
\end{equation}

For even $M$, we find
\begin{align}
 \begin{aligned}
 J_{+}(8,0)&=e^{-\frac{\mu}{2}}+\frac{3}{4}e^{-\mu}-\frac{2}{3}e^{-\frac{3\mu}{2}}
+\left[-\frac{4\mu^2+2\mu+1}{8\pi^2}+\frac{1}{2}\right]e^{-2\mu}+\frac{6}{5}e^{-\frac{5\mu}{2}}
+4e^{-3\mu}-\frac{20}{7}e^{-\frac{7\mu}{2}}\\
&+\left[-\frac{52\mu^2+\mu+9/4}{16\pi^2}+6\right]e^{-4\mu}+\frac{70}{9}e^{-\frac{9\mu}{2}}
+\frac{192}{5}e^{-5\mu}-\frac{252}{11}e^{-\frac{11\mu}{2}}\\
&+\left[-\frac{736\mu^2-304/3\mu+154/9}{24\pi^2}+\frac{224}{3}\right]e^{-6\mu}
+\frac{924}{13}e^{-\frac{13\mu}{2}}+\frac{3072}{7}e^{-7\mu}-\frac{3432}{15}e^{-\frac{15\mu}{2}}
\\
&+\left[-\frac{2701\mu^2-13949/24\mu+11291/192}{8\pi^2}+978\right]e^{-8\mu}
+\frac{12870}{17}e^{-\frac{17\mu}{2}}+\frac{16384}{3}e^{-9\mu}-\frac{48620}{19}e^{-\frac{19\mu}{2}}\\
  J_{-}(8,0)&=e^{-\frac{\mu}{2}}-\frac{5}{4}e^{-\mu}
-\frac{2}{3}e^{-\frac{3\mu}{2}}
+\left[-\frac{4\mu^2+2\mu+1}{8\pi^2}+\frac{9}{2}\right]e^{-2\mu}
+\frac{6}{5}e^{-\frac{5\mu}{2}}
-\frac{20}{3}e^{-3\mu}-\frac{20}{7}e^{-\frac{7\mu}{2}}\\
&+\left[-\frac{52\mu^2+\mu+9/4}{16\pi^2}+38\right]e^{-4\mu}
+\frac{70}{9}e^{-\frac{9\mu}{2}}
-64e^{-5\mu}-\frac{252}{11}e^{-\frac{11\mu}{2}}\\
&+\left[-\frac{736\mu^2-304/3\mu+154/9}{24\pi^2}+416\right]e^{-6\mu}+\frac{924}{13}e^{-\frac{13\mu}{2}}-\frac{5120}{7}e^{-7\mu}-\frac{3432}{15}e^{-\frac{15\mu}{2}}\\
&+\left[-\frac{2701\mu^2-13949/24\mu+11291/192}{8\pi^2}+5074\right]e^{-8\mu}
+\frac{12870}{17}e^{-\frac{17\mu}{2}}-\frac{81920}{9}e^{-9\mu}
-\frac{48620}{19}e^{-\frac{19\mu}{2}}
 \end{aligned}
\label{Jpm80}
\end{align}

\begin{align}
 \begin{aligned}
J_{+}(8,2)&=-\frac{3}{4}e^{-\mu}+\left[-\frac{4\mu^2+2\mu+1}{8\pi^2}+\frac{1}{2}\right]e^{-2\mu}
-4e^{-3\mu}+\left[-\frac{52\mu^2+\mu+9/4}{16\pi^2}+6\right]e^{-4\mu}\\
&-\frac{192}{5}e^{-5\mu}+\left[-\frac{736\mu^2-304/3\mu+154/9}{24\pi^2}+\frac{224}{3}\right]e^{-6\mu}
-\frac{3072}{7}e^{-7\mu}\\
&+\left[-\frac{2701\mu^2-13949/24\mu+11291/192}{8\pi^2}+978\right]e^{-8\mu}
-\frac{16384}{3}e^{-9\mu}\\
  J_{-}(8,2)&=\frac{5}{4}e^{-\mu}+\left[-\frac{4\mu^2+2\mu+1}{8\pi^2}+\frac{9}{2}\right]e^{-2\mu}+\frac{20}{3}e^{-3\mu}+\left[-\frac{52\mu^2+\mu+9/4}{16\pi^2}+38\right]e^{-4\mu}\\
&+64e^{-5\mu}+\left[\frac{736\mu^2-304/3\mu+154/9}{24\pi^2}+416\right]e^{-6\mu}+\frac{5120}{7}e^{-7\mu}\\
&+\left[-\frac{2701\mu^2-13949/24\mu+11291/192}{8\pi^2}+5074\right]e^{-8\mu}+\frac{81920}{9}e^{-9\mu}
 \end{aligned}
\label{Jpm82}
\end{align}

\begin{align}
 \begin{aligned}
J_{+}(8,4)&=-e^{-\frac{\mu}{2}}+\frac{3}{4}e^{-\mu}+\frac{2}{3}e^{-\frac{3\mu}{2}}
+\left[-\frac{4\mu^2+2\mu+1}{8\pi^2}+\frac{1}{2}\right]e^{-2\mu}-\frac{6}{5}e^{-\frac{5\mu}{2}}
+4e^{-3\mu}+\frac{20}{7}e^{-\frac{7\mu}{2}}\\
&+\left[-\frac{52\mu^2+\mu+9/4}{16\pi^2}+6\right]e^{-4\mu}-\frac{70}{9}e^{-\frac{9\mu}{2}}
+\frac{192}{5}e^{-5\mu}+\frac{252}{11}e^{-\frac{11\mu}{2}}\\
&+\left[-\frac{736\mu^2-304/3\mu+154/9}{24\pi^2}+\frac{224}{3}\right]e^{-6\mu}
-\frac{924}{13}e^{-\frac{13\mu}{2}}+\frac{3072}{7}e^{-7\mu}+\frac{3432}{15}e^{-\frac{15\mu}{2}}
\\
&+\left[-\frac{2701\mu^2-13949/24\mu+11291/192}{8\pi^2}+978\right]e^{-8\mu}
-\frac{12870}{17}e^{-\frac{17\mu}{2}}+\frac{16384}{3}e^{-9\mu}+\frac{48620}{19}e^{-\frac{19\mu}{2}}\\
  J_{-}(8,4)&=-e^{-\frac{\mu}{2}}-\frac{5}{4}e^{-\mu}
+\frac{2}{3}e^{-\frac{3\mu}{2}}+\left[-\frac{4\mu^2+2\mu+1}{8\pi^2}+\frac{9}{2}\right]e^{-2\mu}
-\frac{6}{5}e^{-\frac{5\mu}{2}}-\frac{20}{3}e^{-3\mu}+\frac{20}{7}e^{-\frac{7\mu}{2}}\\
&+\left[-\frac{52\mu^2+\mu+9/4}{16\pi^2}+38\right]e^{-4\mu}
-\frac{70}{9}e^{-\frac{9\mu}{2}}-64e^{-5\mu}+\frac{252}{11}e^{-\frac{11\mu}{2}}\\
&+\left[-\frac{736\mu^2-304/3\mu+154/9}{24\pi^2}+416\right]e^{-6\mu}-\frac{924}{13}e^{-\frac{13\mu}{2}}-\frac{5120}{7}e^{-7\mu}+\frac{3432}{15}e^{-\frac{15\mu}{2}}\\
&+\left[-\frac{2701\mu^2-13949/24\mu+11291/192}{8\pi^2}+5074\right]e^{-8\mu}
-\frac{12870}{17}e^{-\frac{17\mu}{2}}-\frac{81920}{9}e^{-9\mu}
+\frac{48620}{19}e^{-\frac{19\mu}{2}}
 \end{aligned}
\label{Jpm84}
\end{align}
\subsection{$k=3$}\label{app:J3}
\begin{equation}
 \begin{aligned}
    J_+^\text{np}(3,0)&=-\frac{1}{\rt{2}}e^{-\mu}+
  \frac{2}{3}e^{-\frac{4\mu}{3}}+\frac{4}{3\rt{2}}e^{-3\mu}
  +\left[\frac{4\mu^2+\mu+1/4}{6\pi^2}-\frac{8}{9}\right]e^{-4\mu}
  +\frac{16}{5\rt{2}}e^{-5\mu}-\frac{17}{9}e^{-\frac{16\mu}{3}}\\
&\quad+\frac{2}{15}e^{-\frac{20\mu}{3}}-\frac{64}{7\rt{2}}e^{-7\mu}
+\left[-\frac{52\mu^2+\mu/2+9/16}{12\pi^2}+\frac{88}{9}\right]e^{-8\mu}
-\frac{256}{9\rt{2}}e^{-9\mu}+\frac{2776}{189}e^{-\frac{28\mu}{3}}\\
J_-^\text{np}(3,0)&=\frac{1}{\rt{2}}e^{-\mu}+
  \frac{2}{3}e^{-\frac{4\mu}{3}}-\frac{4}{3\rt{2}}e^{-3\mu}
+\left[\frac{4\mu^2+\mu+1/4}{6\pi^2}-\frac{8}{9}\right]e^{-4\mu}
-\frac{16}{5\rt{2}}e^{-5\mu}-\frac{17}{9}e^{-\frac{16\mu}{3}}
\\
&\quad+\frac{2}{15}e^{-\frac{20\mu}{3}}+\frac{64}{7\rt{2}}e^{-7\mu}
+\left[-\frac{52\mu^2+\mu/2+9/16}{12\pi^2}+\frac{88}{9}\right]e^{-8\mu}
  +\frac{256}{9\rt{2}}e^{-9\mu}+\frac{2776}{189}e^{-\frac{28\mu}{3}}\\
  &\quad  -\frac{31}{18}e^{-\frac{32\mu}{3}}-\frac{2^{10}}{11\rt{2}}e^{-11\mu}+\left[\frac{736\mu^2-152/3\mu+77/18}{18\pi^2}-\frac{3200}{27}\right]e^{-12\mu}-
  \frac{2^{12}}{13\rt{2}}e^{-13\mu}
 \end{aligned}
\end{equation}

\begin{equation}
 \begin{aligned}
  J_{+}^\text{np}(3,1)&=\frac{1}{\rt{2}}e^{-\mu}
-\frac{1}{3}e^{-\frac{4\mu}{3}}
+\hf e^{-\frac{8\mu}{3}}-\frac{4}{3\rt{2}}e^{-3\mu}
+\left[\frac{4\mu^2+\mu+1/4}{6\pi^2}-\frac{8}{9}\right]e^{-4\mu}
-\frac{16}{5\rt{2}}e^{-5\mu}\\
&+\frac{25}{36}e^{-\frac{16\mu}{3}}-\frac{12}{5}e^{-\frac{20\mu}{3}}
+\frac{64}{7\rt{2}}e^{-7\mu}
+\left[-\frac{52\mu^2+\mu/2+9/16}{12\pi^2}+\frac{88}{9}\right]e^{-8\mu}
+\frac{256}{9\rt{2}}e^{-9\mu}-\frac{947}{189}e^{-\frac{28\mu}{3}}\\
&+\frac{1489}{72}e^{-\frac{32\mu}{3}}-\frac{1024}{11\rt{2}}e^{-11\mu}
+\left[\frac{736\mu^2-152/3\mu+77/18}{18\pi^2}-\frac{3200}{27}\right]e^{-12\mu}
-\frac{2^{12}}{13\rt{2}}e^{-13\mu}\\
&+\frac{59119}{1215}e^{-\frac{40\mu}{3}}\\
J_{-}^\text{np}(3,1)&=-\frac{1}{\rt{2}}e^{-\mu}-\frac{1}{3}e^{-\frac{4\mu}{3}}
+\hf e^{-\frac{8\mu}{3}}+\frac{4}{3\rt{2}}e^{-3\mu}
+\left[\frac{4\mu^2+\mu+1/4}{6\pi^2}-\frac{8}{9}\right]e^{-4\mu}
+\frac{16}{5\rt{2}}e^{-5\mu}\\
&+\frac{25}{36}e^{-\frac{16\mu}{3}}-\frac{12}{5}e^{-\frac{20\mu}{3}}
-\frac{64}{7\rt{2}}e^{-7\mu}
+\left[-\frac{52\mu^2+\mu/2+9/16}{12\pi^2}+\frac{88}{9}\right]e^{-8\mu}
-\frac{256}{9\rt{2}}e^{-9\mu}-\frac{947}{189}e^{-\frac{28\mu}{3}}\\
&+\frac{1489}{72}e^{-\frac{32\mu}{3}}+\frac{1024}{11\rt{2}}e^{-11\mu}
+\left[\frac{736\mu^2-152/3\mu+77/18}{18\pi^2}-\frac{3200}{27}\right]e^{-12\mu}
+\frac{2^{12}}{13\rt{2}}e^{-13\mu}\\
&+\frac{59119}{1215}e^{-\frac{40\mu}{3}}
 \end{aligned}
\end{equation}

\subsection{$k=6$}

\begin{equation}
 \begin{aligned}
  J_+^\text{np}(6,0)&=\frac{2}{3}e^{-\frac{2\mu}{3}}
+\left[\frac{4\mu^2+2\mu+1}{6\pi^2}+\frac{10}{9}\right]e^{-2\mu}
-\frac{17}{9}e^{-\frac{8\mu}{3}}+\frac{2}{15}e^{-\frac{10\mu}{3}}+\left[-\frac{52\mu^2+\mu+9/4}{12\pi^2}-\frac{56}{9}\right]e^{-4\mu}\\
&+\frac{2776}{189}e^{-\frac{14\mu}{3}}-\frac{31}{18}e^{-\frac{16\mu}{3}}
+\left[\frac{736\mu^2-304/3\mu+154/9}{18\pi^2}+\frac{1408}{27}\right]e^{-6\mu}
  -\frac{35938}{243}e^{-\frac{20\mu}{3}}
  +\frac{6508}{297}e^{-\frac{22\mu}{3}}\\
   &+\left[-\frac{2701\mu^2-13949/24\mu+11291/192}{6\pi^2}
  -\frac{4648}{9}\right]e^{-8\mu}
+\frac{15932974}{13\cdot3^6}e^{-\frac{26\mu}{3}}
-\frac{69622}{3^5}e^{-\frac{28\mu}{3}}\\
J_-^\text{np}(6,0)&=\frac{2}{3}e^{-\frac{2\mu}{3}}
+\left[\frac{4\mu^2+2\mu+1}{6\pi^2}-\frac{26}{9}\right]e^{-2\mu}
-\frac{17}{9}e^{-\frac{8\mu}{3}}
+\frac{2}{15}e^{-\frac{10\mu}{3}}
+\left[-\frac{52\mu^2+\mu+9/4}{12\pi^2}+\frac{232}{9}\right]e^{-4\mu}\\
&+\frac{2776}{189}e^{-\frac{14\mu}{3}}-\frac{31}{18}e^{-\frac{16\mu}{3}}
+\left[\frac{736\mu^2-304/3\mu+154/9}{18\pi^2}-\frac{7808}{27}\right]e^{-6\mu}
  -\frac{35938}{243}e^{-\frac{20\mu}{3}}
  +\frac{6508}{297}e^{-\frac{22\mu}{3}}\\
  &+\left[-\frac{2701\mu^2-13949/24\mu+11291/192}{6\pi^2}
  +\frac{32216}{9}\right]e^{-8\mu}
+\frac{15932974}{13\cdot3^6}e^{-\frac{26\mu}{3}}
-\frac{69622}{3^5}e^{-\frac{28\mu}{3}}
 \end{aligned}
 \label{Jpm60}
\end{equation}

\begin{equation}
 \begin{aligned}
  J_{+}^\text{np}(6,1)&=\frac{1}{3}e^{-\frac{2\mu}{3}}-e^{-\mu}
+\hf e^{-\frac{4\mu}{3}}+\left[-\frac{4\mu^2+2\mu+1}{6\pi^2}+\frac{8}{9}\right]e^{-2\mu}
+\frac{25}{36}e^{-\frac{8\mu}{3}}-\frac{16}{3}e^{-3\mu}
+\frac{12}{5}e^{-\frac{10\mu}{3}}\\
&+\left[-\frac{52\mu^2+\mu+9/4}{12\pi^2}+\frac{88}{9}\right]e^{-4\mu}
+\frac{947}{189}e^{-\frac{14\mu}{3}}-\frac{256}{5}e^{-5\mu}
+\frac{1489}{72}e^{-\frac{16\mu}{3}}\\
&+\left[-\frac{736\mu^2-304/3\mu+154/9}{18\pi^2}+\frac{3200}{27}\right]e^{-6\mu}
+\frac{59119}{1215}e^{-\frac{20\mu}{3}}
-\frac{2^{12}}{7}e^{-7\mu}
+\frac{193825}{891}e^{-\frac{22\mu}{3}}\\
&+\left[-\frac{2701\mu^2-13949/24\mu+11291/192}{6\pi^2}+\frac{13784}{9}\right]e^{-8\mu}
+\frac{5126663}{13\cdot 3^6}e^{-\frac{26\mu}{3}}
-\frac{2^{16}}{9}e^{-9\mu}\\
 J_{-}^\text{np}(6,1)&=\frac{1}{3}e^{-\frac{2\mu}{3}}+e^{-\mu}+\hf e^{-\frac{4\mu}{3}}
+\left[-\frac{4\mu^2+2\mu+1}{6\pi^2}+\frac{8}{9}\right]e^{-2\mu}
+\frac{25}{36}e^{-\frac{8\mu}{3}}+\frac{16}{3}e^{-3\mu}
+\frac{12}{5}e^{-\frac{10\mu}{3}}\\
&+\left[-\frac{52\mu^2+\mu+9/4}{12\pi^2}+\frac{88}{9}\right]e^{-4\mu}
+\frac{947}{189}e^{-\frac{14\mu}{3}}+\frac{256}{5}e^{-5\mu}
+\frac{1489}{72}e^{-\frac{16\mu}{3}}\\
&+\left[-\frac{736\mu^2-304/3\mu+154/9}{18\pi^2}+\frac{3200}{27}\right]e^{-6\mu}
+\frac{59119}{1215}e^{-\frac{20\mu}{3}}
+\frac{2^{12}}{7}e^{-7\mu}
+\frac{193825}{891}e^{-\frac{22\mu}{3}}\\
&+\left[-\frac{2701\mu^2-13949/24\mu+11291/192}{6\pi^2}+\frac{13784}{9}\right]e^{-8\mu}
+\frac{5126663}{13\cdot 3^6}e^{-\frac{26\mu}{3}}
+\frac{2^{16}}{9}e^{-9\mu}
 \end{aligned}
\end{equation}

\begin{equation}
 \begin{aligned}
  J_{+}^\text{np}(6,2)&=-\frac{1}{3}e^{-\frac{2\mu}{3}}+\hf e^{-\frac{4\mu}{3}}
+\left[\frac{4\mu^2+2\mu+1}{6\pi^2}+\frac{10}{9}\right]e^{-2\mu}
+\frac{25}{36}e^{-\frac{8\mu}{3}}-\frac{12}{5}e^{-\frac{10\mu}{3}}
\\
&+\left[-\frac{52\mu^2+\mu+9/4}{12\pi^2}-\frac{56}{9}\right]e^{-4\mu}
-\frac{947}{189}e^{-\frac{14\mu}{3}}
+\frac{1489}{72}e^{-\frac{16\mu}{3}}\\
&+\left[\frac{736\mu^2-304/3\mu+154/9}{18\pi^2}+\frac{1408}{27}\right]e^{-6\mu}
  +\frac{59119}{1215}e^{-\frac{20\mu}{3}}
  -\frac{193825}{891}e^{-\frac{22\mu}{3}}\\
  &+\left[-\frac{2701\mu^2-13949/24\mu+11291/192}{6\pi^2}-\frac{4648}{9}\right]e^{-8\mu}
-\frac{5126663}{13\cdot 3^6}e^{-\frac{26\mu}{3}}\\
 J_{-}^\text{np}(6,2)&=-\frac{1}{3}e^{-\frac{2\mu}{3}}+\hf e^{-\frac{4\mu}{3}}
+\left[\frac{4\mu^2+2\mu+1}{6\pi^2}-\frac{26}{9}\right]e^{-2\mu}
+\frac{25}{36}e^{-\frac{8\mu}{3}}-\frac{12}{5}e^{-\frac{10\mu}{3}}\\
&+\left[-\frac{52\mu^2+\mu+9/4}{12\pi^2}+\frac{232}{9}\right]e^{-4\mu}
-\frac{947}{189}e^{-\frac{14\mu}{3}}+\frac{1489}{72}e^{-\frac{16\mu}{3}}\\
&+\left[\frac{736\mu^2-304/3\mu+154/9}{18\pi^2}-\frac{7808}{27}\right]e^{-6\mu}
  +\frac{59119}{1215}e^{-\frac{20\mu}{3}}
   -\frac{193825}{891}e^{-\frac{22\mu}{3}}\\
  &+\left[-\frac{2701\mu^2-13949/24\mu+11291/192}{6\pi^2}+\frac{32216}{9}\right]e^{-8\mu}
-\frac{5126663}{13\cdot 3^6}e^{-\frac{26\mu}{3}}
 \end{aligned}
\end{equation}

\begin{equation}
 \begin{aligned}
  J_{+}^\text{np}(6,3)&=-\frac{2}{3}e^{-\frac{2\mu}{3}}
+e^{-\mu}
+\left[-\frac{4\mu^2+2\mu+1}{6\pi^2}+\frac{8}{9}\right]e^{-2\mu}
-\frac{17}{9}e^{-\frac{8\mu}{3}}
+\frac{16}{3}e^{-3\mu}-\frac{2}{15}e^{-\frac{10\mu}{3}}\\
&+\left[-\frac{52\mu^2+\mu+9/4}{12\pi^2}+\frac{88}{9}\right]e^{-4\mu}
-\frac{2776}{189}e^{-\frac{14\mu}{3}}
+\frac{256}{5}e^{-5\mu}-\frac{31}{18}e^{-\frac{16\mu}{3}}\\
&+\left[-\frac{736\mu^2-304/3\mu+154/9}{18\pi^2}+\frac{3200}{27}\right]e^{-6\mu}\\
J_{-}^\text{np}(6,3)&=-\frac{2}{3}e^{-\frac{2\mu}{3}}-e^{-\mu}
+\left[-\frac{4\mu^2+2\mu+1}{6\pi^2}+\frac{8}{9}\right]e^{-2\mu}
-\frac{17}{9}e^{-\frac{8\mu}{3}}
-\frac{16}{3}e^{-3\mu}-\frac{2}{15}e^{-\frac{10\mu}{3}}\\
&+\left[-\frac{52\mu^2+\mu+9/4}{12\pi^2}+\frac{88}{9}\right]e^{-4\mu}
-\frac{2776}{189}e^{-\frac{14\mu}{3}}-\frac{256}{5}e^{-5\mu}-\frac{31}{18}e^{-\frac{16\mu}{3}}\\
&+\left[-\frac{736\mu^2-304/3\mu+154/9}{18\pi^2}+\frac{3200}{27}\right]e^{-6\mu}
 \end{aligned}
\end{equation}

\subsection{$k=12$}
For odd $M$, we find
\begin{equation}
 \begin{aligned}
  J_{+}^\text{np}(12,1)&=\rt{3}e^{-\frac{\mu}{3}}
-\frac{7}{6}e^{-\frac{2\mu}{3}}
+\frac{9}{4}e^{-\frac{4\mu}{3}}
-\frac{16\rt{3}}{5}e^{-\frac{5\mu}{3}}
+\left[\frac{4\mu^2+2\mu+1}{12\pi^2}+\frac{74}{9}\right]e^{-2\mu}
-\frac{185\rt{3}}{21}e^{-\frac{7\mu}{3}}\\
&+\frac{1057}{72}e^{-\frac{8\mu}{3}}
-\frac{611}{15}e^{-\frac{10\mu}{3}}
+\frac{2203\rt{3}}{33}e^{-\frac{11\mu}{3}}
+\left[-\frac{52\mu^2+\mu+9/4}{24\pi^2}-\frac{1864}{9}\right]e^{-4\mu}
+\frac{23201\rt{3}}{117}e^{-\frac{13\mu}{3}}
 \end{aligned}
\end{equation}

\begin{equation}
 \begin{aligned}
  J_{+}^\text{np}(12,3)&=-\frac{2}{3}e^{-\frac{2\mu}{3}}
-e^{-\frac{4\mu}{3}}
+\left[\frac{4\mu^2+2\mu+1}{12\pi^2}-\frac{34}{9}\right]e^{-2\mu}
+\frac{25}{18}e^{-\frac{8\mu}{3}}+
\frac{68}{15}e^{-\frac{10\mu}{3}}\\
&+\left[-\frac{52\mu^2+\mu+9/4}{24\pi^2}+\frac{296}{9}\right]e^{-4\mu}
-\frac{1894}{189}e^{-\frac{14\mu}{3}}
-\frac{2766}{72}e^{-\frac{16\mu}{3}}\\
&+\left[\frac{736\mu^2-304/3\mu+154/9}{36\pi^2}-\frac{9838}{27}\right]e^{-6\mu}
 \end{aligned}
\end{equation}

\begin{equation}
 \begin{aligned}
  J_{+}^\text{np}(12,5)&=-\rt{3}e^{-\frac{\mu}{3}}
-\frac{7}{6}e^{-\frac{2\mu}{3}}
+\frac{9}{4}e^{-\frac{4\mu}{3}}
+\frac{16\rt{3}}{5}e^{-\frac{5\mu}{3}}
+\left[\frac{4\mu^2+2\mu+1}{12\pi^2}+\frac{74}{9}\right]e^{-2\mu}
+\frac{185\rt{3}}{21}e^{-\frac{7\mu}{3}}\\
&+\frac{1057}{72}e^{-\frac{8\mu}{3}}
-\frac{611}{15}e^{-\frac{10\mu}{3}}
-\frac{2203\rt{3}}{33}e^{-\frac{11\mu}{3}}
+\left[-\frac{52\mu^2+\mu+9/4}{24\pi^2}-\frac{1864}{9}\right]e^{-4\mu}
-\frac{23201\rt{3}}{117}e^{-\frac{13\mu}{3}}
 \end{aligned}
\end{equation}
We find that the instanton corrections for $J_{-}(12,M)$ are
equal to $J_{+}^\text{np}(12,M)$ for all $M=1,3,5$
\begin{equation}
 J_{-}^\text{np}(12,M)=J_{+}^\text{np}(12,M)\qquad(M=1,3,5).
\end{equation}

For even $M$, we find
\begin{align}
 \begin{aligned}
  J_{+}(12,0)&=2e^{-\frac{\mu}{3}}-\frac{4}{3}e^{-\frac{2\mu}{3}}-\frac{5}{6}e^{-\mu}
+3 e^{-\frac{4\mu}{3}}-\frac{38}{5}e^{-\frac{5\mu}{3}}
+\left[-\frac{4\mu^2+2\mu+1}{12\pi^2}+\frac{127}{9}\right]e^{-2\mu}\\
&\quad-\frac{344}{21}e^{-\frac{7\mu}{3}}+\frac{265}{18}e^{-\frac{8\mu}{3}}
-\frac{40}{9}e^{-3\mu}-\frac{514}{15}e^{-\frac{10\mu}{3}}
+\frac{3196}{33}e^{-\frac{11\mu}{3}}
+\left[-\frac{52\mu^2+\mu+9/4}{24\pi^2}-\frac{1552}{9}\right]e^{-4\mu}\\
&\quad+\frac{32050}{13\cdot3^2}e^{-\frac{13\mu}{3}}
-\frac{50030}{7\cdot3^3}e^{-\frac{14\mu}{3}}
-\frac{384}{9}e^{-5\mu}
+\frac{28313}{36}e^{-\frac{16\mu}{3}}-\frac{353752}{17\cdot3^2}e^{-\frac{17\mu}{3}}\\
&\quad+\left[-\frac{736\mu^2-304/3\mu+154/9}{36\pi^2}+\frac{126850}{27}\right]e^{-6\mu}\\
J_{-}(12,0)&=2e^{-\frac{\mu}{3}}-\frac{4}{3}e^{-\frac{2\mu}{3}}+\frac{7}{6}e^{-\mu}
+3 e^{-\frac{4\mu}{3}}-\frac{38}{5}e^{-\frac{5\mu}{3}}
+\left[-\frac{4\mu^2+2\mu+1}{12\pi^2}+\frac{163}{9}\right]e^{-2\mu}\\
&\quad-\frac{344}{21}e^{-\frac{7\mu}{3}}+\frac{265}{18}e^{-\frac{8\mu}{3}}
+\frac{56}{9}e^{-3\mu}-\frac{514}{15}e^{-\frac{10\mu}{3}}
+\frac{3196}{33}e^{-\frac{11\mu}{3}}
+\left[-\frac{52\mu^2+\mu+9/4}{24\pi^2}-\frac{1264}{9}\right]e^{-4\mu}\\
&\quad+\frac{32050}{13\cdot3^2}e^{-\frac{13\mu}{3}}
-\frac{50030}{7\cdot3^3}e^{-\frac{14\mu}{3}}
+\frac{896}{15}e^{-5\mu}
+\frac{28313}{36}e^{-\frac{16\mu}{3}}
-\frac{353752}{17\cdot3^2}e^{-\frac{17\mu}{3}}\\
&\quad+\left[-\frac{736\mu^2-304/3\mu+154/9}{36\pi^2}+\frac{136066}{27}\right]e^{-6\mu}
 \end{aligned}
\end{align}

\begin{align}
 \begin{aligned}
  J_{+}(12,2)&=e^{-\frac{\mu}{3}}-\frac{5}{6}e^{-\frac{2\mu}{3}}+\frac{5}{6}e^{-\mu}
+\qu e^{-\frac{4\mu}{3}}-\frac{4}{5}e^{-\frac{5\mu}{3}}
+\left[-\frac{4\mu^2+2\mu+1}{12\pi^2}+\frac{19}{9}\right]e^{-2\mu}\\
&\quad+\frac{17}{21}e^{-\frac{7\mu}{3}}-\frac{143}{72}e^{-\frac{8\mu}{3}}
+\frac{40}{9}e^{-3\mu}+e^{-\frac{10\mu}{3}}
-\frac{151}{33}e^{-\frac{11\mu}{3}}
+\left[-\frac{52\mu^2+\mu+9/4}{24\pi^2}+\frac{176}{9}\right]e^{-4\mu}\\
J_{-}(12,2)&=e^{-\frac{\mu}{3}}-\frac{5}{6}e^{-\frac{2\mu}{3}}-\frac{7}{6}e^{-\mu}
+\qu e^{-\frac{4\mu}{3}}-\frac{4}{5}e^{-\frac{5\mu}{3}}
+\left[-\frac{4\mu^2+2\mu+1}{12\pi^2}+\frac{55}{9}\right]e^{-2\mu}\\
&\quad+\frac{17}{21}e^{-\frac{7\mu}{3}}-\frac{143}{72}e^{-\frac{8\mu}{3}}
-\frac{56}{9}e^{-3\mu}+e^{-\frac{10\mu}{3}}-\frac{151}{33}e^{-\frac{11\mu}{3}}
+\left[-\frac{52\mu^2+\mu+9/4}{24\pi^2}+\frac{464}{9}\right]e^{-4\mu}
 \end{aligned}
\end{align}

\begin{align}
 \begin{aligned}
  J_{+}(12,4)&=-e^{-\frac{\mu}{3}}-\frac{5}{6}e^{-\frac{2\mu}{3}}-\frac{5}{6}e^{-\mu}
+\qu e^{-\frac{4\mu}{3}}+\frac{4}{5}e^{-\frac{5\mu}{3}}
+\left[-\frac{4\mu^2+2\mu+1}{12\pi^2}+\frac{19}{9}\right]e^{-2\mu}\\
&\quad-\frac{17}{21}e^{-\frac{7\mu}{3}}-\frac{143}{72}e^{-\frac{8\mu}{3}}
-\frac{40}{9}e^{-3\mu}+e^{-\frac{10\mu}{3}}
+\frac{151}{33}e^{-\frac{11\mu}{3}}
+\left[-\frac{52\mu^2+\mu+9/4}{24\pi^2}+\frac{176}{9}\right]e^{-4\mu}\\
J_{-}(12,4)&=-e^{-\frac{\mu}{3}}-\frac{5}{6}e^{-\frac{2\mu}{3}}+\frac{7}{6}e^{-\mu}
+\qu e^{-\frac{4\mu}{3}}+\frac{4}{5}e^{-\frac{5\mu}{3}}
+\left[-\frac{4\mu^2+2\mu+1}{12\pi^2}+\frac{55}{9}\right]e^{-2\mu}\\
&\quad-\frac{17}{21}e^{-\frac{7\mu}{3}}-\frac{143}{72}e^{-\frac{8\mu}{3}}
+\frac{56}{9}e^{-3\mu}+e^{-\frac{10\mu}{3}}+
\frac{151}{33}e^{-\frac{11\mu}{3}}
+\left[-\frac{52\mu^2+\mu+9/4}{24\pi^2}+\frac{464}{9}\right]e^{-4\mu}
 \end{aligned}
\end{align}

\begin{align}
 \begin{aligned}
  J_{+}(12,6)&=-2e^{-\frac{\mu}{3}}-\frac{4}{3}e^{-\frac{2\mu}{3}}+\frac{5}{6}e^{-\mu}
+3 e^{-\frac{4\mu}{3}}+\frac{38}{5}e^{-\frac{5\mu}{3}}
+\left[-\frac{4\mu^2+2\mu+1}{12\pi^2}+\frac{127}{9}\right]e^{-2\mu}\\
&\quad+\frac{344}{21}e^{-\frac{7\mu}{3}}+\frac{265}{18}e^{-\frac{8\mu}{3}}
+\frac{40}{9}e^{-3\mu}-\frac{514}{15}e^{-\frac{10\mu}{3}}
-\frac{3196}{33}e^{-\frac{11\mu}{3}}
+\left[-\frac{52\mu^2+\mu+9/4}{24\pi^2}-\frac{1552}{9}\right]e^{-4\mu}\\
  J_{-}(12,6)&=-2e^{-\frac{\mu}{3}}-\frac{4}{3}e^{-\frac{2\mu}{3}}-\frac{7}{6}e^{-\mu}
+3 e^{-\frac{4\mu}{3}}+\frac{38}{5}e^{-\frac{5\mu}{3}}
+\left[-\frac{4\mu^2+2\mu+1}{12\pi^2}+\frac{163}{9}\right]e^{-2\mu}\\
&\quad+\frac{344}{21}e^{-\frac{7\mu}{3}}+\frac{265}{18}e^{-\frac{8\mu}{3}}
-\frac{56}{9}e^{-3\mu}-\frac{514}{15}e^{-\frac{10\mu}{3}}
-\frac{3196}{33}e^{-\frac{11\mu}{3}}
+\left[-\frac{52\mu^2+\mu+9/4}{24\pi^2}-\frac{1264}{9}\right]e^{-4\mu}
 \end{aligned}
\end{align}

\normalsize


\end{document}